\documentclass[journal]{IEEEtran}
\usepackage{mathrsfs}
\usepackage{amssymb}
\usepackage[mathcal]{euscript}
\usepackage{cite}
\usepackage{graphicx}
\usepackage{psfrag}
\usepackage{subfigure}
\usepackage{url}
\usepackage{stfloats}
\usepackage{amsmath}
\usepackage{float}
\usepackage{array}
\usepackage[]{hyperref}
\usepackage{amsthm}
\usepackage{booktabs} 
\usepackage{algorithm} 
\usepackage{algorithmic} 
\usepackage{amsmath,amsfonts,amssymb,amsbsy,bm,paralist,theorem,ifthen,color}
\usepackage{CJK}

\usepackage{cases}
\usepackage{mathrsfs}
\usepackage{rotating}
\usepackage{makecell}
\usepackage{multirow}

\usepackage{colortbl}
\usepackage{arydshln}
\usepackage{multicol}

\ifCLASSINFOpdf
\else
\fi
\begin{document}
%
\title{RF-based Energy Harvesting: Nonlinear Models, Applications and Challenges}


\author{Ruihong Jiang 



%
\thanks{R. Jiang is with the State Key Laboratory of Networking and Switching Technology, Beijing University of Posts and Telecommunications, Beijing 100876, China (e-mail: rhjiang@bupt.edu.cn).}
}
\maketitle

\begin{abstract}

So far, various aspects associated with wireless energy harvesting (EH) have been investigated from diverse perspectives, including energy sources and models, usage protocols, energy scheduling and optimization, and EH implementation in different wireless communication systems. However, a comprehensive survey specifically focusing on models of radio frequency (RF)-based EH behaviors has not yet been presented. To address this gap, this article provides an overview of the mainstream mathematical models that capture the nonlinear behavior of practical EH circuits, serving as a valuable handbook of mathematical models for EH application research. Moreover, we summarize the application of each nonlinear EH model, including the associated challenges and precautions. We also analyze the impact and advancements of each EH model on RF-based EH systems in wireless communication, utilizing artificial intelligence (AI) techniques. Additionally, we highlight emerging research directions in the context of nonlinear RF-based EH. This article aims to contribute to the future application of RF-based EH in novel communication research domains to a significant extent.


\end{abstract}


\section*{Abbreviations}
The list of abbreviations and definitions used in this paper are summarized in Table \ref{tb_abbreviations}.

\begin{table}[t!]
\renewcommand{\arraystretch}{1.24716}
\centering 
 \caption{List of abbreviation}
 \begin{tabular}{c|c}
 \hline
 \hline
 \textbf{Full name} & \textbf{Abbreviation} \\
 \hline
 Energy harvesting & EH \\
 \cdashline{1-2}[0.8pt/1pt]
 Wireless device & WD \\
 \cdashline{1-2}[0.8pt/1pt]
 Electromagnetic & EM \\
 \cdashline{1-2}[0.8pt/1pt]
 Radio frequency & RF \\
 \cdashline{1-2}[0.8pt/1pt]
 Wireless power transfer & WPT\\
 \cdashline{1-2}[0.8pt/1pt]
 Federal Communications Commission & FCC \\
 \cdashline{1-2}[0.8pt/1pt]
 Solar Power Satellite & SPS \\
 \cdashline{1-2}[0.8pt/1pt]
 Fixed High Altitude Relay Platform & SHARP \\
 \cdashline{1-2}[0.8pt/1pt]
 Massachusetts Institute of Technology & MIT \\
 \cdashline{1-2}[0.8pt/1pt]
 Wireless sensor networks & WSN \\
 \cdashline{1-2}[0.8pt/1pt]
 Internet of Things & IoT \\
 \cdashline{1-2}[0.8pt/1pt]
 Simultaneous wireless information and power transfer & SWIPT \\
 \cdashline{1-2}[0.8pt/1pt]
 Wireless information transmission & WIT \\
 \cdashline{1-2}[0.8pt/1pt]
 Information receiver & IR \\
 \cdashline{1-2}[0.8pt/1pt]
 Energy receiver & ER \\
 \cdashline{1-2}[0.8pt/1pt]
 Wireless powered communication network & WPCN \\
 \cdashline{1-2}[0.8pt/1pt]
 Hyper access point & H-AP \\
 \cdashline{1-2}[0.8pt/1pt]
 Information decoding & ID \\
 \cdashline{1-2}[0.8pt/1pt]
 Power splitting & PS \\
 \cdashline{1-2}[0.8pt/1pt]
 Time switching & TS \\
 \cdashline{1-2}[0.8pt/1pt]
 RF-Direct Current & RF-DC \\
 \cdashline{1-2}[0.8pt/1pt]
 Multi-input multi-output & MIMO \\
 \cdashline{1-2}[0.8pt/1pt]
 Multiple input single output & MISO \\
 \cdashline{1-2}[0.8pt/1pt]
 Time-division multiple access & TDMA \\
 \cdashline{1-2}[0.8pt/1pt]
 Non-orthgonal multiple access & NOMA \\
 \cdashline{1-2}[0.8pt/1pt]
 Orthogonal frequency division multiplexing & OFDM \\
 \cdashline{1-2}[0.8pt/1pt]
 Non-orthogonal multiple access & NOMA \\
 \cdashline{1-2}[0.8pt/1pt]
 Rate-splitting multiple access & RSMA \\ 
 \cdashline{1-2}[0.8pt/1pt]
 Cognitive radio & CR \\
 \cdashline{1-2}[0.8pt/1pt]
 Device-to-device & D2D \\
 \cdashline{1-2}[0.8pt/1pt]
 Unmanned aerial vehicle & UAV \\
 \cdashline{1-2}[0.8pt/1pt]
 Point-to-Point & P2P \\
 \cdashline{1-2}[0.8pt/1pt]
 Energy efficiency & EE \\
 \cdashline{1-2}[0.8pt/1pt]
 Spectral efficiency & SE \\ 
 \cdashline{1-2}[0.8pt/1pt]
 Intelligent reflecting surface & IRS \\ 
 \cdashline{1-2}[0.8pt/1pt]
 Mobile edge computing & MEC \\ 
 \cdashline{1-2}[0.8pt/1pt]
 Channel state information & CSI \\
 \cdashline{1-2}[0.8pt/1pt]
 Decode and forward & DF \\
 \cdashline{1-2}[0.8pt/1pt]
 Amplify and forward & AF \\
 \cdashline{1-2}[0.8pt/1pt]
 Full Duplex & FD \\
 \cdashline{1-2}[0.8pt/1pt]
 Rate/Information-energy & R/I-E \\
 \cdashline{1-2}[0.8pt/1pt]
 Semidefinite relaxation SDR \\
 \cdashline{1-2}[0.8pt/1pt]
 Alternating direction method of multipliers & ADMM \\
 \cdashline{1-2}[0.8pt/1pt]
 Second-order cone program & SOCP \\
 \cdashline{1-2}[0.8pt/1pt]
 Successive convex approximation & SCA \\
 \cdashline{1-2}[0.8pt/1pt]
 Artificial Intelligence & AI \\
 \cdashline{1-2}[0.8pt/1pt]
 Machine learning & ML \\
 \cdashline{1-2}[0.8pt/1pt]
 Deep learning & DL \\
 \cdashline{1-2}[0.8pt/1pt]
 Deinforcement learning & RL \\
 \cdashline{1-2}[0.8pt/1pt]
 Deep RL & DRL \\
 \cdashline{1-2}[0.8pt/1pt]
 Deep deterministic policy gradient & DDPG \\
 \cdashline{1-2}[0.8pt/1pt]
 Inverse RL & IRL \\
 \cdashline{1-2}[0.8pt/1pt]
 Lifelong learning & LL \\
 \cdashline{1-2}[0.8pt/1pt]
 Integrated sensing and communication & ISAC \\
 \cdashline{1-2}[0.8pt/1pt]
 Integrated sensing, computing and communication & ISCAC \\
 \cdashline{1-2}[0.8pt/1pt]
 Semantic communication & SemCom \\

 \hline
 \hline
 \end{tabular}
 \label{tb_abbreviations}
\end{table}

\section{\textcolor[rgb]{0.00,0.00,0.00}{Introduction}}


In recent years, there has been a remarkable surge in the number of smart devices and the emergence of Internet of Everything (IoE) applications, which has resulted in an overwhelming burden on wireless networks, including 5G, B5G, and upcoming 6G \cite{6g_1, du2022semantic, du2023attention}. Meanwhile, the integration of the Internet of Things (IoT) with cutting-edge technologies such as artificial intelligence (AI), blockchain, cloud computing, and big data has also been accelerated. IoT terminals now possess enhanced sensing capabilities, while application platforms exhibit improved data processing prowess, resulting in elevated levels of intelligence \cite{zhang2023generative, du2023ai, du2022exploring}. Furthermore, the application scenarios of IoT has expanded extensively in critical domains like smart cities, digital villages, transportation, agriculture, manufacturing, construction, and homes. The rapid pace of IoT technology advancements has given rise to novel technologies, products, and models, including AIoT—an integration of IoT, AI, and cloud computing.

By now, billions of IoT devices heavily rely on battery power to sustain their operations. Depending on their computational demands and the nature of their battery chemistry, these devices can either operate steadily for short durations or persist for several decades \cite{eh_2023app_2, mao_2023_1}. However, there exists a category of IoT devices that possess the capability to either harvest energy autonomously or tap into externally collected energy sources, granting them the ability to function almost indefinitely. Within the realm of IoT, energy harvesting (EH) has emerged as a promising solution to reduce or even eliminate the reliance on batteries \cite{eh_2023app_1, jiang2015outage, jiang2017outage}. This holds particular advantages for devices positioned in challenging environments, such as livestock sensors, smart buildings, remote monitoring systems, as well as wearable electronics and mobile asset tracking. Despite its immense potential, the utilization of EH in IoT has remained somewhat restricted thus far \cite{se_eh}. Hence, the development of efficient and compact energy harvesting solutions remains an essential requirement for IoT devices, especially when they are deployed in remote or inaccessible locations that pose challenges for battery replacement.

The EH technology is gaining momentum, with an expanding array of silicon products available from leading companies like ADI, Atmosic, EnOcean, Metis Microsystems, ONiO, Powercast, Renesas Electronics, STMicroelectronics, and Texas Instruments \cite{product_eh}. Simultaneously, AI advancements are driving smaller, lighter, smarter, and more energy-efficient products. With continuous innovation, EH technology is maturing, unlocking limitless possibilities for the future. The exponential growth in the number of radio transmitters has made radio frequency (RF)-based EH a ubiquitous technology, garnering significant attention and research efforts in academia and industry over the past decade. Global mobile phone users have exceeded 5 billion, with over 1 billion utilizing mobile broadband, alongside an abundance of Wi-Fi routers and wireless devices like laptops. Even from a modest Wi-Fi router with a transmit power of 50mW to 100mW, small amounts of energy can be harvested over short distances. For harvesting RF energy from mobile base stations and radio towers across longer distances, longer antennas with higher gain are required.



\begin{figure*}[t!]
\centering
\includegraphics[width=0.995\textwidth]{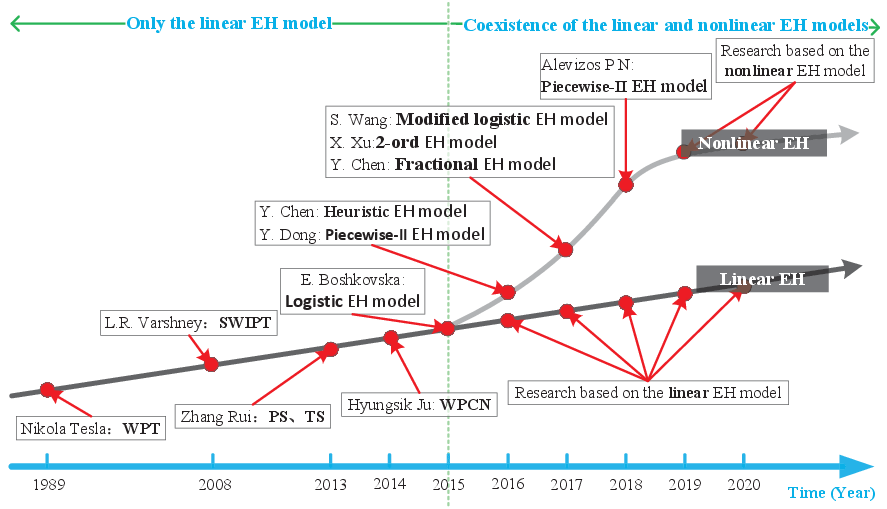}
\caption{The value of RF-based EH in typical 6G scenarios}
\label{fg_surveys}
\end{figure*}

\textcolor[rgb]{0.00,0.00,0.00}{On the other hand, the RF-based EH technology also holds great promise for applications in typical 6G scenarios, i.e., massive connectivity, immersive communication, ultra-reliable low-latency communication, AI integration with communication, fusion of sensing and communication, and ubiquitous connectivity. 
By harnessing ambient RF energy, RF-based EH can enable sustainable and self-sufficient power sources for the multitude of connected devices, ensuring uninterrupted and efficient operation in the next generation of wireless communication and sensing systems.}

\begin{itemize}

\item \textcolor[rgb]{0.00,0.00,0.00}{ \textit{Ubiquitous Connectivity:} 6G aims to achieve comprehensive connectivity across the physical, machine, human, and digital domains. This scenario is designed to enhance connectivity and bridge the digital divide by enabling interoperability with other systems. Whether in urban or rural settings, or even in diverse environments such as air, space, land, and sea, 6G networks will establish ubiquitous coverage, delivering high-speed and stable communication services to users. The RF-based EH can provide power to ensure the sustainable operation of these connected users. This technology allows IoT devices such as smart thermostats, security cameras, and voice assistants to be charged wirelessly, reducing the need for battery replacements and increasing device autonomy.}



\item \textcolor[rgb]{0.00,0.00,0.00}{ \textit{Massive communication:} This scenario includes typical use cases in smart cities, transportation, logistics, healthcare, energy, environmental monitoring, agriculture, and both extending and introducing new applications. It demands support for high device density, including a massive number of IoT devices. 6G enables the simultaneous connection of billions of devices. The RF-based EH can play a crucial role by providing sustainable energy sources for these devices, reducing the need for battery replacements or frequent recharging. For example, in a smart city scenario, streetlights could be equipped with RF-based EH capability, enabling them to capture ambient RF signals from 6G networks and convert them into electrical energy. This harvested energy can then be used to power LED lights, sensors, and communication modules embedded in the streetlights.}


\item\textcolor[rgb]{0.00,0.00,0.00}{ \textit{Hyper-reliable and low-latency Communication:} This scenario encompasses typical use cases in industrial environments that demand high-performance communication to achieve full automation in control and operations, including robotic interactions, emergency services, remote healthcare, as well as monitoring power transmission and distribution. The RF-based EH can offer significant potential by providing sustainable power, enhancing reliability, enabling low-latency communication, and supporting precise positioning. For example, in smart factory automation scenario, the collaborative robots equipped with RF-based EH capability can harvest energy from nearby RF sources, such as Wi-Fi, 5G and 6G networks. This harvested RF energy can be used to supplement or even replace traditional power sources for the robots.}


\item \textcolor[rgb]{0.00,0.00,0.00}{ \textit{Integrated AI and communication:} This scenario includes assisting autonomous driving, enabling autonomous collaboration among medical assistance devices, offloading intensive computing tasks across devices and networks, creating and predicting digital twins, and facilitating collaborative robots. Through the deep integration of artificial intelligence technologies, the 6G communication system will possess the capabilities of intelligent sensing, decision-making, and optimization, providing users with personalized and intelligent communication services. The RF-based EH can provide power to AI devices, enabling distributed sensing and decision-making. For example, in autonomous agriculture with AI integration, some autonomous drones can be equipped with RF-based EH function to harvest energy from ambient RF signals, such as those from 6G networks or satellite communications. The harvested RF energy is used to power the drone's sensors, AI processors, and communication modules.}


\item \textcolor[rgb]{0.00,0.00,0.00}{ \textit{Integrated sensing and communication:} This scenario includes assisting navigation, detecting and tracking activities and movements, environmental monitoring, and providing sensor data/information about the surrounding environment for AI, XR, and digital twin applications. Through 6G's intelligent sensing capabilities, it provides efficient solutions for smart cities and environmental monitoring. The RF-based EH can power sensor networks, supporting data collection and transmission, particularly facilitating efficient solutions for smart cities and environmental monitoring. In regions prone to natural disasters such as earthquakes, tsunamis, or forest fires, the RF-based EH can be employed to power a network of environmental sensors deployed in disaster-prone areas for disaster management. These sensors can capture ambient RF energy from 6G networks and convert it into electrical power to keep the sensors operational.}

\item \textcolor[rgb]{0.00,0.00,0.00}{ \textit{Immersive communication:} Immersive communication is a key feature of 6G, promising high-quality virtual and augmented reality experiences. The RF-based EH can power head-mounted displays, virtual reality goggles, and augmented reality applications, extending usage time and enhancing the user experience. This is critical for immersive XR communication, remote multi-sensory presentation, and holographic communication. Immersive remote healthcare scenario, the RF-based EH can be used to power medical devices such as high-resolution cameras, 3D scanners, vital signs monitoring equipment, etc., which are crucial for immersive remote healthcare.} 

\end{itemize}

\textcolor[rgb]{0.00,0.00,0.00}{In these scenarios, RF-based EH technology's key advantage is its ability to capture RF energy from the environment and convert it into usable electricity without relying on traditional batteries or grid power. This contributes to achieving more sustainable, self-sustaining, and efficient communication and sensing systems. Note that as 6G technology continues to evolve, specific application scenarios and integration methods may change. It's necessary to closely monitor the latest research and development trends in the 6G field to better understand the potential applications of RF-based EH. Additionally, for specific application scenarios, in-depth engineering and technical research may be required to optimize RF-based EH systems to meet specific needs.}

\subsection{\textcolor[rgb]{0.00,0.00,0.00}{Motivation}}
With the growing availability of RF-based EH, it not only eliminates the need for extensive cabling but also provides a resilient system that is immune to environmental conditions and hazardous substances. In the case of mesh networks, the integration of RF-based EH with sophisticated device-to-device (D2D) communications enables various applications like wireless sensor networks (WSN), wearable devices, wireless charging, and IoT \cite{Sharma21}. By reducing the reliance on batteries, RF-based EH contributes positively to the environment. However, it is crucial that power-free RF receivers or RF-based EH devices, such as Powercast's P2110 Powerharvester receiver \cite{cast2019power}, have a sensitivity greater than or equal to -11 dBm. Enhanced RF sensitivity allows RF-to-DC (RF/DC) power conversion over longer distances from the RF energy source. Nevertheless, as the distance increases, available power decreases, and charging time increases. Additionally, \emph{\textbf{practical EH circuits exhibit nonlinear characteristics, necessitating proper modeling, design, and optimization of EH behavior, whether in a linear or nonlinear state. This aspect holds significant importance in RF-based EH networks, particularly in scenarios involving large-scale applications, low efficiency, and low energy density}} \cite{clex2019nonlinear, Brono_2016, hujie_eh}.



\subsection{\textcolor[rgb]{0.00,0.00,0.00}{Comparison to Existing Surveys}}

\begin{table*}[t!]
\renewcommand{\arraystretch}{1.40}
\centering 
 \caption{Existing surveys on EH (WPT, EH, SWIPT, WPCN) and improvement of our survey}
 \begin{tabular}{m{2.5cm}|m{6.5cm}|m{7.5cm}}
 \hline
 \hline
 \textbf{\makecell[c]{Publication}} & \textbf{\makecell[c]{Topic}} & \textbf{\makecell[c]{Main contribution}} \\
 \hline
 Sudevalayam, 2011 \cite{eh_2011} & Sensor networks, energy-aware systems, EH & Survey on aspects of EH sensor systems \\
 \cdashline{1-3}[0.8pt/1pt]
 Prasad, 2014 \cite{eh_2014} & Ambient EH, WSN, energy storage, harvested network protocols & Study of various types of energy harvesting techniques \\
 \cdashline{1-3}[0.8pt/1pt]
 Huang, 2015 \cite{eh_2015} & CR, spectrum efficiency, energy efficiency, EH & Survey on EE CR techniques and the optimization of green-energy-powered wireless networks \\
 \cdashline{1-3}[0.8pt/1pt]
 Lu, 2015 \cite{ref_eh_2} & RF-powered CR network, SWIPT, receiver operation policy, beamforming, communication protocols & Review on the research progresses in wireless networks with RF-based EH capability \\
 \cdashline{1-3}[0.8pt/1pt]
 Ku, 2016 \cite{ref_eh_1} & EH, cooperative/CR/multi-user/cellular networks & Overview of EH communications and networks \\
 \cdashline{1-3}[0.8pt/1pt]
 Lu, 2016 \cite{ref_eh_5} & WPT, inductive coupling, resonance coupling, RF/Microwave radiation, SWIPT, energy beamforming, WPCN & Overview of wireless charging techniques (WPT, SWIPT, WPCN, etc.), the developments in technical standards, and the advances in network applications \\
 \cdashline{1-3}[0.8pt/1pt]
 Van Huynh, 2018 \cite{eh_2018_1} & Ambient backscatter, IoT, wireless EH, backscatter and low-power communications & Review on the ambient backscatter communications on the architectures, protocols, and applications \\
 \cdashline{1-3}[0.8pt/1pt]
 Alsaba, 2018 \cite{alsaba2018beamforming} & EH, beamforming, WPCN, SWIPT, WPT, physical layer security & Survey on exploiting beamforming technique in EH-enabled wireless communication and its physical layer security application \\
 \cdashline{1-3}[0.8pt/1pt]
 Ponnimbaduge Perera, 2018 \cite{ref_eh_3} & RF, WPT, SWIPT, interference exploitation, RF EH & Survey on the issues associated with SWIPT and WPT assisted technologies \\
 \cdashline{1-3}[0.8pt/1pt]
 Quintero, 2019 \cite{eh_2019} & MAC protocols, energy efficiency, EH, battery model, state-of-charge & Focused on the techniques of EE developed in the medium access control layer \\
 \cdashline{1-3}[0.8pt/1pt]
 Tedeschi, 2020 \cite{eh_2020_1} & EH security, green communications security, IoT security, physical-layer security & Survey of security issues, applications, techniques, and challenges arising in wireless EH networks \\
 \cdashline{1-3}[0.8pt/1pt]
 Ma, 2020 \cite{eh_2020_2} & EH communications, IoT, sensing, intermittent computing & Survey on advances in EH-IoTs from the design of sensing, computing and communications \\
 \cdashline{1-3}[0.8pt/1pt]
 Kaswan, 2022 \cite{eh_2022_1} & Wireless rechargeable sensor networks, mobile charging techniques, periodic/on-demand charging & Survey on mobile charging techniques in wireless rechargeable sensor networks \\
 \cdashline{1-3}[0.8pt/1pt]
 Mao, 2020 \cite{eh_2022_2} & 6G, green communications, artificial intelligence (AI), EH & Overview of the related research on AI-based green communications \\
 \cdashline{1-3}[0.8pt/1pt]
 Sandhu, 2021 \cite{iot_2021} & EH, energy prediction, IoT, sensing, task scheduling, ubiquitous computing, wearables & Survey on task scheduling algorithms to to minimize the energy consumption EH-based sensors in IoT. \\
 \cdashline{1-3}[0.8pt/1pt]
 Zhao, 2017 \cite{access_2017} & Beamforming optimization, interference alignment and management, SWIPT, EH & Survey on exploiting interference for wireless EH \\
 \cdashline{1-3}[0.8pt/1pt]
 Hossain, 2019 \cite{access_2019} & Cooperative relay, SWIPT, EH, 5G, resource allocation, relay selection & Review on the combination of cooperative relay and SWIPT \\
 \cdashline{1-3}[0.8pt/1pt]
 Williams, 2021 \cite{access_2021_1} & EH, industry 4.0, WSN & Survey on EH technology for small-scale WSNs including EH methods, energy storage technologies, EH system architectures, and optimization considerations \\
 \cdashline{1-3}[0.8pt/1pt]
 Ashraf, 2021 \cite{access_2021_2} & Cooperative relaying, 5G, MIMO, IoT, SWIPT, WPT & Review of SWIPT technology with cooperative relaying networks for 5G and B5G mobile networks \\
 \cdashline{1-3}[0.8pt/1pt]
 Padhy, 2021 \cite{access_2021_3} & CR, cooperative sensing, EH, IoT, next generation wireless networks, spectrum harvesting. & Review on EH and SH technologies and protocols for next-generation wireless networks \\
 \cdashline{1-3}[0.8pt/1pt]
 Ojukwu, 2022 \cite{access_2022} & Metasurface, WPT, EH, WPCN, SWIPT, mmWave, intelligent surface intelligent surface (RIS) & Outline of WPT and wireless EH systems with the metasurface technology and its applications \\
 \cdashline{1-3}[0.8pt/1pt]
 Lee, 2023 \cite{eh_2023_1} & Multiband RF EH, power conversion efficiency, & Outline of multiband RF-based EH systems for power-efficient IoT applications \\
 \cdashline{1-3}[0.8pt/1pt]
 Cai, 2023 \cite{eh_2023_2} & Battery-free, coverage, IoT, networking, sensor networks & Survey on the algorithms for battery-free WSNs \\
 \cdashline{1-3}[0.8pt/1pt]
 Jiang, 2023 \cite{eh_2023_3} & Backscatter, battery-free, hardware implementation & Overview of battery-free IoT and backscatter communication integration \\
 \cdashline{1-3}[0.8pt/1pt]
 Halimi, 2023 \cite{eh_2023_4} & RF EH, WPT, rectenna, RF energy harvester & Overview of dielectric resonator-based sensing elements in RF-based EH and WPT systems \\
 \cdashline{1-3}[0.8pt/1pt]
 Our work & EH, RF, SWIPT, WPCN, linear and nonlinear EH, AI & Survey on the RF-based EH models, applications and challenges \\
 \hline
 \hline
 \end{tabular}
 \label{tb_surveys}
\end{table*}

Over the past decade, numerous surveys have been conducted on the topic of RF-based EH. For the readers' convenience, we provide Table \ref{tb_surveys}, which summarizes the concerned survey papers.

Specifically, in \cite{eh_2011, eh_2014, eh_2019} and \cite{access_2021_1}, the authors summarized various aspects of EH sensor systems and various types of EH techniques, respectively. In \cite{ref_eh_1}, the authors offered an overview of EH communications and networks, covering topics such as energy sources and models, EH and usage protocols, energy scheduling and optimization, as well as the implementation of EH in cooperative, cognitive radio, multiuser and cellular networks, among others. With regards to different EH sources, in \cite{ref_eh_5, eh_2020_1} and \cite{access_2021_3}, the authors overviewed wireless charging techniques, security issues, energy and spectrum harvesting technologies, respectively. By utilizing ambient RF signals without requiring active RF transmission, the authors in \cite{eh_2018_1} outlined the ambient backscatter communications on the architectures, protocols, and applications. Meanwhile, the energy efficient (EE) techniques and optimization for green-energy-powered cognitive radio (CR) networks using readily available ambient energy sources were reviewed in \cite{eh_2015}.

For EH-enabled IoT, in \cite{eh_2020_2} and \cite{iot_2021}, the authors surveyed sensing, computing and communications, and the task scheduling algorithms of sensor nodes, respectively. With the RF-based EH, several works \cite{ref_eh_2, alsaba2018beamforming, ref_eh_3, access_2017} provided comprehensive literature reviews, discussed beamforming techniques, the issues associated with simultaneous wireless information and power transfer (SWIPT) and wireless power transfer (WPT) technologies, interference for wireless EH. Given the benefits of relay technology, the authors in \cite{access_2019} and \cite{access_2021_2} outlined SWIPT with cooperative relays for next-generation wireless networks. For the mobile charging issues, the work in \cite{eh_2022_1} provided an overview of mobile charging techniques in empowered wireless rechargeable sensor networks, including the network model, various WPT techniques, system design issues and performance metrics.

Furthermore, considering the emergence of new technologies, in \cite{eh_2022_2} and \cite{access_2022}, the authors presented comprehensive surveys on research related to green communications with AI techniques, and wireless WPT/wireless EH systems with the metasurface technology, respectively. In \cite{eh_2023_1}, the authors reviewed the potential of multiband RF-based EH systems for power-efficient IoT applications, covering circuitry, antenna, rectifier, and performance improvement. In \cite{eh_2023_2}, the survey summarized and analyzed existing algorithms for battery-free WSNs, including energy management, networking, data acquisition, and specific applications. In \cite{eh_2023_3}, this article comprehensively explored the integration of battery-free IoT with backscatter communication, highlighting key components, prototypes, and fundamental issues for practical applications. In \cite{eh_2023_4}, the authors presented an overview of dielectric resonator-based sensing elements and their applications in RF-based EH and WPT systems, highlighting performance enhancements and research gaps.

Above all, in \cite{eh_2011, eh_2014, eh_2019, access_2021_1, ref_eh_1, ref_eh_5, eh_2020_1, access_2021_3, eh_2018_1, eh_2015, eh_2020_2, iot_2021, ref_eh_2, alsaba2018beamforming, ref_eh_3, access_2017, access_2019, access_2021_2, eh_2022_1, eh_2022_2, access_2022, eh_2023_1, eh_2023_2, eh_2023_3, eh_2023_4}, the inherent challenges of (RF-based) EH were presented from several perspectives, such as energy sources and models, usage protocols, energy scheduling and optimization, implementation of EH in WSN, cooperative, CR, multi-user, multi-antenna, and so on. \emph{\textbf{However, a comprehensive survey that offers a complete overview of EH models, encompassing their behaviors, methods, and associated applications/challenges, has not been conducted yet.}}

\begin{figure}[t!]
\centering
\includegraphics[width=0.425\textwidth]{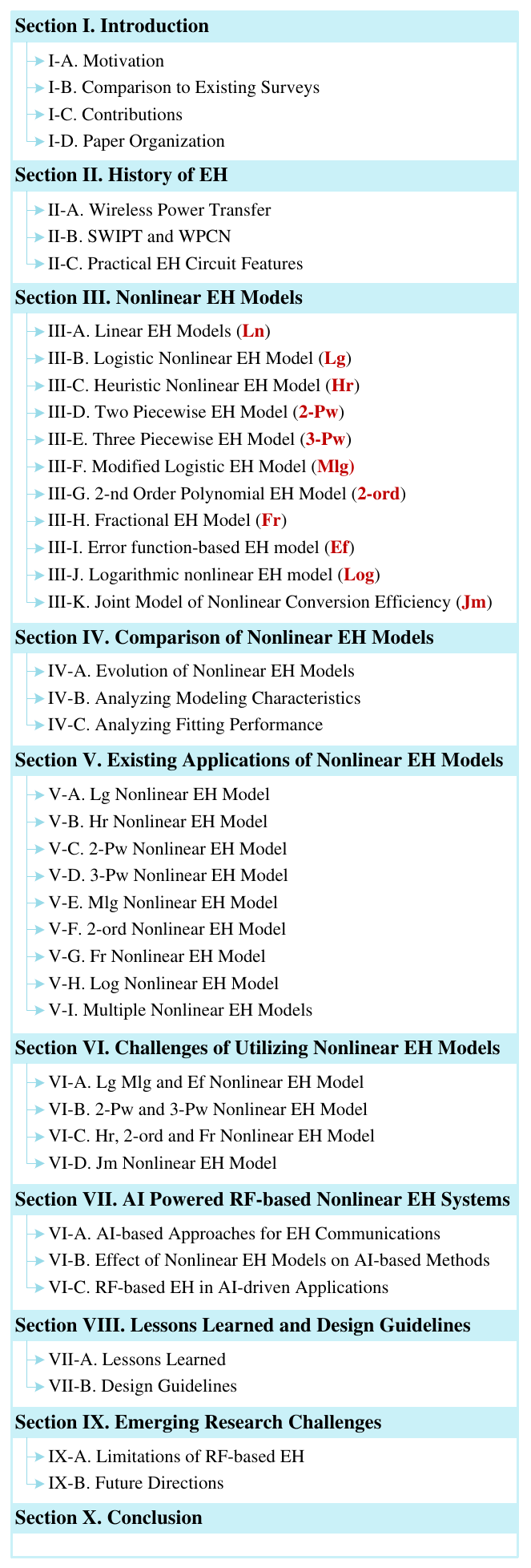}
\caption{Outline of this survey paper}
\label{fg_surveys}
\end{figure}

\subsection{\textcolor[rgb]{0.00,0.00,0.00}{Contributions}}

\textit{Considering the aforementioned analysis and to the best of our knowledge, none of the existing surveys has provided an overview of models on the EH behaviors, methods, and applications/challenges of EH models.} To fill the gap, the key contributions of this survey are summarized as follows.

\begin{itemize}
\item In Section \ref{eh_history}, we outline some mainstream mathematical EH models that characterize the nonlinear behavior of practical EH circuit, which is a handbook of mathematical models for the application research of EH.

\item In Section \ref{section_eh_models}, we provides a brief history of EH, covering WPT, SWPIT, and WPCN.

\item In Section \ref{section_comparison_eh}, we discusses RF-DC circuit features and compares existing EH models.

\item In Section \ref{section_apps_ehmodels}, we explores applications of nonlinear EH models in various scenarios.

\item In Section \ref{section_challenge_apps_ehmodels}, we examines challenges of utilizing each nonlinear EH model.

\item In Section \ref{section_AI_ehmodels}, we analyzes the impact and development of each model on wireless communication EH systems using AI.  we give an analysis of the impact and development of each EH model on wireless communication RF-based EH systems with the help of AI. At last, we underline some emerging research directions about nonlinear RF-based EH.

\item In Section \ref{section_lessons_eh}, We present the lessons learned and design guidelines, facilitating the creation of efficient and dependable RF-based EH systems through the utilization of precise nonlinear models.

\item In Section \ref{section_direction_ehmodels} presents emerging research directions in RF-based EH. we give an analysis of the impact and development of each EH model on wireless communication RF-based EH systems with the help of AI. At last, we underline some emerging research directions about nonlinear RF-based EH.

\item In Section \ref{section_conclusion_ehmodels} summarizes the article. We give an analysis of the impact and development of each EH model on wireless communication RF-based EH systems with the help of AI. At last, we underline some emerging research directions about nonlinear RF-based EH.

\end{itemize}

\subsection{Paper Organization}
The rest of this article is organized as follows. Section \ref{eh_history} provides a brief history of EH, covering WPT, SWPIT, and WPCN. Section \ref{section_eh_models} discusses RF-DC circuit features and compares existing EH models. Section \ref{section_apps_ehmodels} explores applications of nonlinear EH models in various scenarios. Section \ref{section_challenge_apps_ehmodels} examines challenges of utilizing each nonlinear EH model. Section \ref{section_AI_ehmodels} analyzes the impact and development of each model on wireless communication EH systems using AI. Section \ref{section_direction_ehmodels} presents emerging research directions in RF-based EH, while Section \ref{section_conclusion_ehmodels} summarizes the article. Fig. \ref{fg_surveys} provides a detailed outline for easier navigation and reading.



\section{\textcolor[rgb]{0.00,0.00,0.00}{History of EH}}\label{eh_history}

\begin{table*}[t!] 
\renewcommand{\arraystretch}{1.40}
\centering 
\caption{Development of EH} 
\label{tab:nL_0}
\begin{tabular}{c|m{16.2cm}}
\hline
\hline
 \textbf{\makecell[c]{Year}} & \textbf{Main event} \\
	 \hline
 1880 & Heinrich Hertz proved the existence and propagation of electromagnetic waves in free space. \\
 \cdashline{1-2}[0.8pt/1pt]
	 1890 & Nikola Tesla conducts WPT experiment for the first time. \\
	 \cdashline{1-2}[0.8pt/1pt]
 1964 & William C. Brown invented the silicon rectifier diode antenna. \\
 \cdashline{1-2}[0.8pt/1pt]
	 1968 & Peter Glaser proposed the concept of SPS. \\
	 \cdashline{1-2}[0.8pt/1pt]
 1987 & Canada demonstrated its first free-flying wireless-powered aircraft. \\
 \cdashline{1-2}[0.8pt/1pt]
 1992 & Japan applied electronic scanning phased array transmitter to WPT for the first time in the MILAX experiment. \\
 \cdashline{1-2}[0.8pt/1pt]
 1994 & Academician Weiqian Lin introduced microwave power transmission technology to domestic scholars for the first time.\\
 \cdashline{1-2}[0.8pt/1pt]
 2001 & In the Reunion Island project, France realized the transmission of 10 kilowatts of electricity to a remote village. \\
 \cdashline{1-2}[0.8pt/1pt]
 2007 & MIT in the United States has realized wireless energy transmission between two coils that are 2.13 m apart. \\
 \cdashline{1-2}[0.8pt/1pt]
 2008 & Electrical energy was successfully wirelessly transmitted 148 kilometers between the two islands of Hawaii. \\
 \cdashline{1-2}[0.8pt/1pt]
 2015 & Japan successfully transmitted 1.8 kilowatts of power to a small receiving device 55 m away. \\
 \cdashline{1-2}[0.8pt/1pt]
 2017 & The wireless charging technology of two American companies, PowerCast and Energous, has passed the FCC certification. \\
 \cdashline{1-2}[0.8pt/1pt]
 \makecell[c]{\multirow{2}{0.6cm}{2018}} & PowerCast has realized the function of wireless charging for wireless devices up to 80 feet away. \\
 \cdashline{2-2}[0.8pt/1pt]
 & Energous and Dialog semiconductor company (European) produced wireless charging devices at a distance of about 4.5 meters. \\
 \cdashline{1-2}[0.8pt/1pt]
 2019 & MIT realized that it harvests enough energy from indoor WiFi signals to light up mobile phones and activate related chips. \\
 \cdashline{1-2}[0.8pt/1pt]
 \multicolumn{1}{c|}{\multirow{2}{*}{2020}} & The new Mophile wireless charging board was released, which can charge four electronic devices at the same time. \\
 \cdashline{2-2}[0.8pt/1pt]
 & Xiaomi mobile phone announced the first 80 watt wireless second charging, setting a new global mobile phone wireless charging record. \\
 \cdashline{1-2}[0.8pt/1pt]
 \multicolumn{1}{c|}{\multirow{6}{*}{2021}} & The new Mophile wireless charging board was released, which can charge four electronic devices at the same time. \\
 \cdashline{2-2}[0.8pt/1pt]
 & Powercast won the BIG Innovation Awards 2021, enabling convenient and secure monitoring of employee temperatures in Covid-19 protocols. \\
 \cdashline{2-2}[0.8pt/1pt]
 & Atmosic and Energous Corporation claimed ``first interoperability energy harvesting'' for wireless charging from up to 2 meters away. \\
 \cdashline{2-2}[0.8pt/1pt]
 & Belgian company, e-peas introduced two PMICs to market, focused on providing more performant EH for IoT. \\
 \cdashline{1-2}[0.8pt/1pt]
 \multicolumn{1}{c|}{\multirow{2}{*}{2022}} & Atmosic creates EH solutions integrated directly into the system-on- a-chip (SoC) for IoT devices directly. \\
 \cdashline{2-2}[0.8pt/1pt]
 & researchers from the University of South Florida create ``perfect EM absorption'' rectenna for RF-based EH. \\
 \cdashline{1-2}[0.8pt/1pt]
 2023 & Ossia's Cota Real Wireless Power enables wirelessly powered IoT devices without the need for batteries or wiring, used in over 62 countries. \\
 
\hline
\hline
 \end{tabular}
\end{table*}

As wireless data services experience unprecedented growth, the power demands placed on wireless devices (WDs) continue to rise, resulting in a depletion problem. To address this, EH has emerged as a promising solution, allowing WDs to capture energy from various sources such as solar/light, thermoelectric power, mechanical motion, and electromagnetic (EM) radiation \cite{ref_eh_5, ref_eh_1, ref_eh_2}. Particularly, EM radiation sources can be categorized into near-field (ambient harvesting) and far-field (dedicated harvesting) based on the distance applications. In this context, antennas of the receivers capture EM radiation, such as RF/microwave signals, which are subsequently converted into DC power by rectifier circuits.

\subsection{\textcolor[rgb]{0.00,0.00,0.00}{Wireless Power Transfer (WPT)}}

In fact, the concept of realizing wireless EH through electromagnetic radiation is not novel, as demonstrated by early-stage WPT technology. The history of WPT could be traced back to Heinrich Hertz's early research in 1880, which aimed to prove the existence and propagation of electromagnetic waves in free space \cite{hertz1990h}. In the experiments, a spark gap transmitter was employed to generate and detect high frequencies at the receiver, resembling a WPT system.
 In 1890, Nikola Tesla, a famous electrical engineer (physicist), built a large wireless transmission station for information, telephone and wireless power supply in the Wardenclyffe tower project, illuminating a neon lamp 25 m away without the need for wires, thereby validating the concept of wireless power transfer (WPT).\cite{steer2006ieee}. 
 In 1934, the Federal Communications Commission (FCC) of the United States (US) reserved the 2.4-2.5 GHz frequency band for industrial, scientific, and medical purposes, promoting significant scientific research and enabling the progress of WPT. During World War II, the technology of using magnetrons to convert electric energy into microwaves was successfully developed, while the discovery of the method to convert microwaves back into electric current was not made until 1964.

In 1964, William C. Brown invented the silicon rectifier diode antenna and successfully verified the idea of converting microwave into electric current \cite{steer2006ieee}. In 1968, Peter Glaser proposed Solar Power Satellite (SPS), using geostationary satellites to collect and transmit solar energy via microwave beams, addressing energy shortage and greenhouse gas emissions, captivating research for over half a century \cite{Gla1986ieee}. 
In 1987, Canada demonstrated its first free flying wireless powered aircraft, called the Fixed High Altitude Relay Platform (SHARP), marking a breakthrough in the International Aviation Alliance by demonstrating the ability of a small aircraft to sustain flight using RF beam-provided energy \cite{mou2015wireless}.
In 1992, Japan achieved a milestone by utilizing electronic scanning phased array transmitter in the MILAX (Microwave Lift-off Aircraft Experiment) experiment, enabling controlled microwave beam tracking and wireless power supply for a mobile fuel-free airplane model \cite{shi2013beam}.
In 1994, Weigan Lin, the pioneer of electromagnetic field and microwave technology in China, introduced microwave power transmission technology to domestic scholars, laying a solid foundation for subsequent research and development of radio transmission technology in China \cite{wei1994power}.
In 2001, G. Pignolet successfully used microwave wireless power transmission to illuminate a 200 W light bulb 40 m away in Reunion Island, France. Subsequently, in 2003, a 10 kW experimental microwave power transmission device was deployed on the island, enabling a point-to-point (P2P) wireless power supply at 2.45 GHz to the nearby Grand-Bassin village, located 1 km away \cite{luk2013ptop}. 
In 2007, Marin Soljacic of Massachusetts Institute of Technology (MIT) and his team achieved wireless energy transmission between two coils 2.13 m apart through electromagnetic resonance, successfully powering a 60 W bulb \cite{kurs1994wireless}. In 2008, electricity was wirelessly transmitted over a distance of 148 km between two islands in Hawaii, showcasing a significantly larger power transmission range compared to previous experiments, even though only 20 W of power was received. \cite{whi2008wired}. In 2015, Japan achieved a significant milestone by successfully transmitting 1.8 kW of power to a small receiving device located 55 m away \cite{zeng2017communications}.

RF-based EH technology has garnered significant attention as an appealing and promising solution to extend the lifespan of energy-limited networks (e.g., WSN, IoT) and electronic devices (e.g., sensors, low-power mobile devices) due to its controllable and predictable nature compared to other EH sources \cite{ref_eh_3, ref_eh_4, ref_eh_5}. Thanks to the small size and low deployment cost of the RF energy conversion circuit module, it is suitable for installation on mobile device terminals and sensor nodes. As a result, RF-based EH technology has been widely considered as an effective means of providing continuous and stable power for low-power devices like IoT ones \cite{alsaba2018beamforming}.

In late 2017, PowerCast and Energous, two US companies, obtained FCC certification for their wireless charging technology. 
In 2018, PowerCast unveiled a long-range RF-based wireless charging system capable of charging devices up to 80 feet away at the Consumer Electronics Show in Las Vegas, which offered wearable devices, smartphones, and smart accessory manufacturers an opportunity to harness energy through PowerCast's wireless charging solutions \cite{cast2019power}. 
In addition, Energous and Dialog Semiconductor collaborated to develop an RF-based wireless charging system capable of wirelessly charging devices up to approximately 4.5 m away, eliminating the need for coils and backplanes \cite{dia2019eng}. In 2019, MIT Microtechnology Laboratory announced the development of a new type of silicon rectifier, which can collect enough energy from indoor WiFi signals to light up mobile phones and activate related chips \cite{zhang2019wifi}. In 2020, the new Mophile wireless charging board was released, which can charge four electronic devices at the same time \cite{mop2020charge}. Furthermore, Xiaomi mobile phone announced the first 80 watt wireless second charging, charging 100\% in 19 minutes, setting a new global record for wireless charging speed \cite{xiaomi2020charge}.

In early 2021, Powercast was awarded the BIG Innovation Awards 2021 by the Business Intelligence Group for their innovative wirelessly powered RFID temperature scanning system, enabling convenient and secure monitoring of employee temperatures to aid in Covid-19 protocols \cite{eh_ai_apps}. Later, Atmosic and Energous Corporation claimed ``first interoperability energy harvesting'' for wireless charging, enabling charging from up to 2 meters away \cite{whatup_app}. In the same year, Belgian company, e-peas Semiconductor introduced two PMICs for IoT, providing enhanced EH capabilities with input voltages ranging from 100 mV to 4.5V and output voltages up to 3.3V, supporting 60 mA in high power mode \cite{Belgian_1}.
In 2022, Atmosic integrates EH into the system-on-a-chip (SoC) for low-power IoT devices directly, drawing only 0.7 mA \cite{Atmosic_1}. Later, Univ. of South Florida researchers created "perfect EM absorption" rectenna using metamaterials for RF-based EH, capturing low-intensity RF waves (100 $\mu$W) emitted by cell phone towers \cite{Florida_1, Florida_2}.
In 2023, Ossia's Cota Real Wireless Power enables wirelessly powered IoT devices, such as electronic shelf labels and asset tracking systems, without the need for batteries or wiring, approved in over 62 countries \cite{Ossia_1}.

Stated thus, we summarize the major historical development milestones of WPT by Table \ref{tab:nL_0} in chronological order.


\subsection{\textcolor[rgb]{0.00,0.00,0.00}{SWIPT and WPCN}}

\begin{figure}[t!]
\centering
\includegraphics[width=0.475\textwidth]{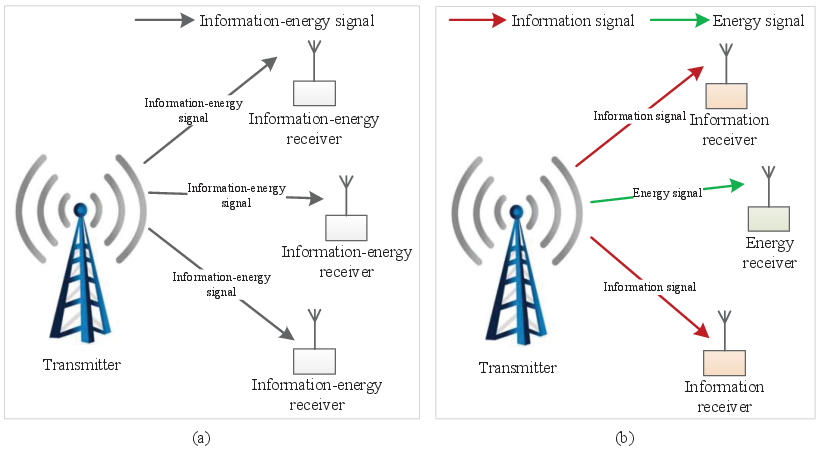}
\caption{SWIPT systems}
\label{fig_swipt}
\end{figure}

\begin{figure}[t!]
\centering
\includegraphics[width=0.475\textwidth]{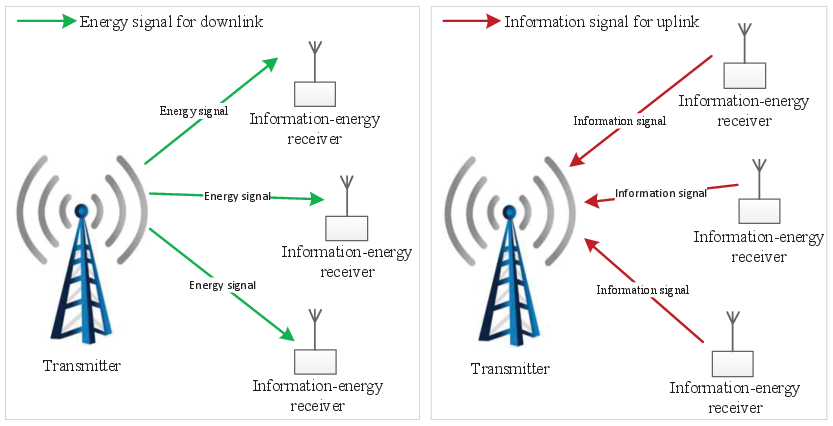}
\caption{WPCN systems}
\label{fig_wpcn}
\end{figure}

In wireless communication systems, RF-based EH has formed two popular applications. The first application is SWIPT, as shown in Fig. \ref{fig_swipt}, where WPT and wireless information transmission (WIT) are realized by transmitting information and energy simultaneously from one or multiple transmitters to one or multiple receivers \cite{ref_swipt_0}. The receivers in SWIPT systems can either be co-located or separated. In the case of separated receivers, the energy receiver (ER) and the information receiver (IR) are distinct devices. The ER is a low-power device designed to receive and store energy, while the IR is responsible for receiving information. In the case of co-located receivers, each receiver is a single low-power device that simultaneously receives information and charges its battery. The second application is WPCN \cite{ref_wpcn_0}, where a hyper access point (H-AP) transmits power to WDs in the downlink WPT using RF signals carrying energy, and then the WDs utilize the harvested energy to transmit information in the uplink WIT using RF signals carrying information.


\begin{figure}[t!]
\centering
\includegraphics[width=0.475\textwidth]{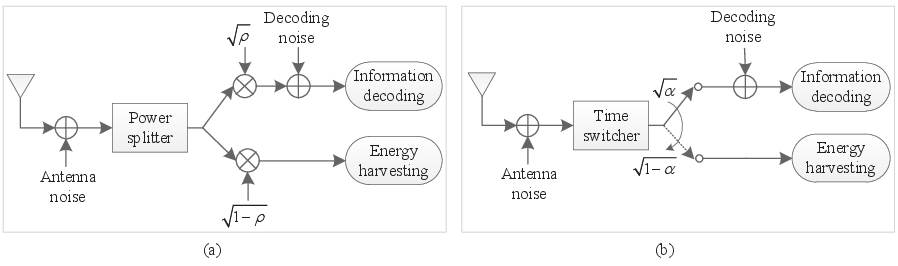}
\caption{The PS and TS receiver architectures for SWIPT}
\label{fig_swipt_pts}
\end{figure}

Particularly, in SWIPT systems, the power sensitivity of practical circuits for EH and ID from RF signals differs significantly (e.g., -10 dBm for EH versus -60 dBm for ID). This discrepancy makes it impossible to directly decode the carried information and power of the RF signals. In order to enable SWIPT technology, the authors in \cite{ref_pst} proposed two practical receiver architectures: power splitting (PS) and time switching (TS) receiver architectures, as illustrated in Fig. \ref{fig_swipt_pts}. In the PS architecture, the received RF signals are divided into two separate streams using a power splitter, with a predetermined PS ratio. One stream is directed towards an energy circuit for harvesting power, while the other stream is forwarded to an information decoder for decoding the transmitted information. In essence, the PS receiver performs simultaneous EH and ID. In the PS architecture, the received RF signals are alternately processed by either an energy circuit for EH or an information decoder for ID, based on a predefined TS ratio. In this case, the TS receiver periodically switches between EH and ID operations.



Although theoretical possibilities exist for EH in SWIPT systems using PS and TS receiver architectures, practical implementation becomes challenging when it comes to harvesting energy from the same received RF signals. In contrast, WPCN systems are relatively easier to implement since they separate EH and information decoding (ID) functions at dedicated receivers, without being constrained by the circuit sensitivity of EH and ID. However, both SWIPT and WPCN have been extensively studied in various wireless communication networks, such as WSN \cite{ref_sw_wsn_0, ref_sw_wsn_1}, cooperative relay \cite{ref_sw_relay_1, ref_sw_relay_2}, multi-input multi-output (MIMO) \cite{ref_sw_mimo_1, ref_sw_mimo_2}, orthogonal frequency division multiplexing (OFDM) \cite{ref_sw_ofdm_1}, non-orthogonal multiple access (NOMA) \cite{ref_sw_noma_1}, CR \cite{ref_sw_cr_1, ref_sw_cr_2}, device-to-device (D2D) \cite{ref_sw_d2d_1, ref_sw_d2d_2}, and unmanned aerial vehicle (UAV) systems \cite{ref_sw_uav_1, ref_sw_uav_2}.

\subsection{\textcolor[rgb]{0.00,0.00,0.00}{Practical EH Circuit Features}} \label{section_rf_dc}

\begin{figure}[t!]
\centering
\includegraphics[width=0.475\textwidth]{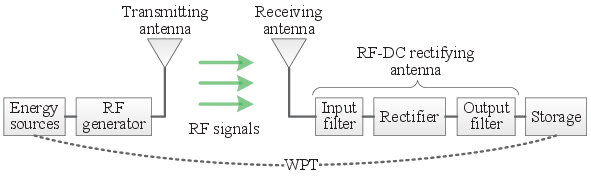}
\caption{Practical EH process: RF-DC receiver structure diagram}
\label{fig_rf_dc}
\end{figure}

In practice, wireless transmission signals experience a certain degree of distortion when passing through the RF channel\footnote{The so-called RF channel here refers to the RF transceiver information channel, excluding the spatial fading channel.}, where the distortion can be categorized into linear distortion and nonlinear distortion. The actual EH process is shown in Fig. \ref{fig_rf_dc}, where the transmitter transmits RF signals to the receiver, and then converts the power of the received RF signals into direct current (i.e., RF-DC) through a rectifier antenna for storage or use. The rectifier is mainly composed of nonlinear circuit components such as diodes, transistors connected with diodes, etc., in which the linear distortion primarily arises from some passive components like filters, whereas the nonlinear distortion mainly stems from some active devices such as amplifiers and frequency mixers. Consequently, the energy harvested by EH circuit tends to exhibit nonlinear characteristics rather than linear ones due to the influence of electronic circuit components \cite{clex2019nonlinear}.
Furthermore, due to the saturation effect of practical EH circuit, the amount of harvested energy at the receiver does not always increase proportionally with the input power. Instead, it initially increases but eventually reaches a point of saturation, becoming stable.

Initially, it was commonly assumed that the harvested energy could be linearly increased by increasing the power of received RF signals. However, it was later discovered that the linear EH model was too idealized, and practical EH circuits exhibit nonlinear behavior instead of linear behavior. Using the linear EH model may result in mismatching the practical systems, yielding fault design output and causing system performance loss and the inefficiency of SWIPT, WPCN, WPT, and EH-enabled networks. As a result, various EH models have been proposed by scholars in recent years to address these limitations.
\emph{\textbf{As the EH models play a crucial role in the design of RF-based EH wireless powered systems and have different features in the application, this article provides a comprehensive survey of these models, aiming to offer valuable insights for their applications in the design of WPT, SWIPT, WPCN and EH-enabled systems.}}



\section{\textcolor[rgb]{0.00,0.00,0.00}{Nonlinear EH Models}} \label{section_eh_models}



\textit{Notation:} $Q(\cdot)$ is the function of the output/harvested power. $P_{\rm{in}}$ is the input/received RF power. $P_{\rm{sen}}$ is the harvester's sensitivity. $P_{\rm{sat}}$ is the saturation threshold of EH. ${\rm Q}_{\max}$ is the maximum harvested power when practical EH circuit reaches saturation. $\eta/\widetilde{\eta}( \cdot )$ is the energy conversion efficiency/function.


\subsection{Linear EH Model (Ln)}
In traditional linear EH model, the harvested energy ${\rm Q}_{\rm Ln}$ is modeled as a linear function in terms of the input RF power $P_{\rm in}$\cite{eh_2011}, which is given by
\begin{flalign} \label{lin_md}
{\rm Q}_{\rm Ln} (P_{\rm in}) = \eta P_{\rm in},
\end{flalign}
where $\eta$ denotes the energy conversion efficiency, such that $0 \le \eta \le 1$. It is seen that $\eta$ is independent of $P_{\rm in}$, and ${\rm Q}_{\rm Ln} (P_{\rm in})$ is increment with $P_{\rm in}$. However, the linear EH model ignores the following: (i) the dependence of RF harvesting efficiency on input power, (ii) the harvester cannot operate below the sensitivity threshold, and (iii) the harvested power saturates when the input power level is above a power threshold. 



As mentioned in section \ref{section_rf_dc}, one of the significant advances in wireless EH is the rectifier, which is able to convert RF power into DC voltage. Due to the presence of nonlinear components such as diodes, inductors, and capacitors in practical EH circuits, the conversion of harvested energy often shows nonlinear characteristics. Specifically, the amount of harvested energy at the EH receiver first increases with the increment of the input RF power $P_{\rm in}$, reaching a maximum value, and then decrease. This behavior can be likened to an "S" curve \cite{phdthesis}. Therefore, the conventional linear EH model is inadequate to capture the nonlinear features inherent in practical EH circuits employed in wireless communication networks.


\subsection{Logistic Nonlinear EH Model (Lg)}
Due to the limitations of traditional linear EH model, E. Boshkovska, et al. \cite{ref_mod_lg1_0} presented a nonlinear EH model based on empirical data obtained from practical EH circuit measurements. This model incorporates a pseudo-concave logic function (i.e., ``S" shape) to approximately capture the nonlinear behavior exhibited by EH circuits. The corresponding mathematical description is given by
\begin{flalign}
{\rm Q}_{\rm Lg} (P_{\rm in}) &= \frac{\frac{{\rm Q}_{\max}}{1+e^{-a(P_{\rm in} - b)}} - \frac{{\rm Q}_{\max}}{1+e^{a b}}} {1-\frac{1}{1+e^{a b}}},
\label{mod_Logi}
\end{flalign}
where $P_{\rm in}$ is the input RF power with a unit of Watt, ${\rm Q}_{\max}$, $a$ and $b$ are constants. Particularly, ${\rm Q}_{\max}$ is the maximum harvested power when practical EH circuit reaches saturation. $a$ and $b$ are determined by EH circuit properties. By reviewing the existing literature, we summarize the available parameters for ${\rm Q}_{\max}$, $a$, and $b$ in Table \ref{tab1}.


\begin{table}[t!]
\renewcommand{\arraystretch}{1.40}
 \centering
 \caption{Parameters for the Lg nonlinear EH model}
 \begin{tabular}{c|c|c|c}
 \hline
 \hline
 ${\rm Q}_{\max}$ (mW) & a & b & Ref. \\
 \hline
 20 & 6400 & 0.003 & \cite{ref_mod_lg1_0, ref_lg1_app_1, ref_lg1_app_5} \\
 \cdashline{1-4}[0.8pt/1pt]
 24 & 1500 & 0.0014 & \cite{ref_lg1_app_2} \\
 \cdashline{1-4}[0.8pt/1pt]
 24 & 150 & 0.0014 & \cite{ref_lg1_app_23} \\
 \cdashline{1-4}[0.8pt/1pt]
 24 & 1500 & 0.0022 & \cite{ref_lg1_app_4} \\
 \cdashline{1-4}[0.8pt/1pt]
 24 & 150 & 0.014 & \cite{ref_lg1_app_6} \\
 \cdashline{1-4}[0.8pt/1pt]
 9.097 (uW) & 47083 & 2.9uW & \cite{ref_lg1_app_13} \\
 \hline
 \hline
 \end{tabular}
 \label{tab1}
\end{table}

\subsection{Heuristic Nonlinear EH Model (Hr)}
In traditional linear EH model, the conversion efficiency, denoted as $\eta$, is typically assumed to be constant \cite{ref_sw_wsn_1}-\cite{ref_sw_uav_2}. while, $\eta$ generally varies with $P_{\rm in}$ in practice. To account for this variability, a heuristic EH model (Hr) was proposed in \cite{ref_mod_hr_0}, wherein $\eta$ was modeled as a function of $P_{\rm in}$. That is

\begin{flalign}
{\rm Q}_{\rm Hr} (P_{\rm in}) &= \widetilde{\eta} ( P_{\rm in} ) P_{\rm in}, \nonumber \\
&=\frac{p_2 P_{\rm in}^2 + p_1 P_{\rm in} + p_0}{q_3 P_{\rm in}^3 + q_2 P_{\rm in}^2 + q_1 P_{\rm in} + q_0} P_{\rm in}, \nonumber \\
&=\frac{p_2 P_{\rm in}^3 + p_1 P_{\rm in}^2 + p_0 P_{\rm in}}{q_3 P_{\rm in}^3 + q_2 P_{\rm in}^2 + q_1 P_{\rm in} + q_0},
\label{mod_duo}
\end{flalign}
where $\widetilde{\eta}( \cdot )$ is the conversion efficiency function, i.e., 
\begin{flalign}
\widetilde{\eta}( P_{\rm in} ) = \frac{p_2 P_{\rm in}^2 + p_1 P_{\rm in} + p_0}{q_3 P_{\rm in}^3 + q_2 P_{\rm in}^2 + q_1 P_{\rm in} + q_0}, 
\end{flalign}
where $P_{\rm in}$ measures in milliwatt, the parameters $p_0, p_1, p_2, q_0, q_1, q_2$ and $q_3$ vary for different RF-DC circuits (i.e., different EH receiver). In \cite[Table II]{ref_mod_hr_0}, certain fitting parameters for $p_0, p_1, p_2, q_0, q_1$ and $q_2$ were provided, where $q_3$ is normalized to 1.

\subsection{Two Piecewise EH Model (2-Pw)}
As the mathematical expressions of the logistic (Lg) and heuristic (Hr) EH models, i.e., \eqref{lin_md} and \eqref{mod_duo}, are highly complex, nonlinear, and non-convex functions, they pose challenges in using them to analyze optimization problems in wireless communication networks and derive theoretical analysis results. To address this issue and effectively capture the nonlinear characteristics of practical EH circuits, the authors in \cite{ref_mod_pw1_0} presented a two piecewise linear EH model (2-Pw) without considering the sensitive voltage characteristics of EH circuit, which is

\begin{flalign}
{\rm Q}_{\textrm{2-Pw}} (P_{\rm in}) & =
\begin{cases}
\eta P_{\rm in}, & P_{\rm in} \le P_{\textrm{sat}} \\
{\rm Q}_{\max}, & P_{\rm in} > P_{\textrm{sat}}
\end{cases},
\label{mod_pw1}
\end{flalign}
where $P_{\textrm{sat}}$ denotes the saturation threshold of EH, set to be 10 dBm in \cite{ref_mod_pw1_0}. In this model, the harvested energy ${\rm Q}_{\textrm{2-Pw}} (P_{\rm in})$ is a constant calculated by $\eta P_{\textrm{sat}}$ when the input power $P_{\rm in}$ exceeds the threshold $P_{\textrm{sat}}$.

\subsection{Three Piecewise EH Model (3-Pw)}
Considering both the sensitivity and saturation characteristics of practical EH circuits, the authors in \cite{ref_mod_pw2_0} proposed a model that captures the EH behavior as an arbitrary nonlinear, continuous, and non-decreasing function with respect to the input power, i.e.,
\begin{flalign}
{\rm Q}_{\textrm{3-Pw}} (P_{\rm in}) & =
\begin{cases}
 0, & 0 \le P_{\rm in} \le P_{\textrm{sen}} \\
\widetilde{\eta} (P_{\rm in}) P_{\rm in} & P_{\textrm{sen}} \le P_{\rm in} \le P_{\textrm{sat}} \\
\widetilde{\eta} (P_{\textrm{sat}}) P_{\textrm{sat}} & P_{\rm in} \ge P_{\textrm{sat}} \\
\end{cases}.
\label{mod_pw3_1}
\end{flalign}
where $\widetilde{\eta} (\cdot)$ is modeled as a high-order polynomial in the dBm scale, offering higher granularity over the very small input power values. i.e.,
\begin{flalign}
\widetilde{\eta} (x) = w_0 + \sum_{i=1}^W w_i \big(10 \log_{10} (x) \big)^i, x \in P_{\rm in},
\label{mod_pw3_2}
\end{flalign}
with $\widetilde{\eta} (x)$ being parametrized by $(W+1)$ real numbers (the coefficients of the polynomial). Here, $W$ represents the degree of the polynomial. Additionally, the model in \eqref{mod_pw3_1} can be utilized to evaluate a simplified piecewise linear approximation in the following,
\begin{flalign}
{\rm Q}_{\textrm{3-Pw}} (P_{\rm in}) & =
\begin{cases}
 0, & P_{\rm in} < P_{\textrm{sen}} \\
\eta (P_{\rm in} - P_{\rm{sen}}) & P_{\textrm{sen}} \le P_{\rm in} \le P_{\textrm{sat}} \\
{\rm Q}_{\max} & P_{\rm in} > P_{\textrm{sat}} \\
\end{cases}.
\label{mod_pw2}
\end{flalign}

In order to better capture the actual EH circuit characteristics and reduce errors, a more refined form of 3-Pw EH model is presented in \cite{ref_mod_pw2_0,ref_mod_pw2_1,ref_pw2_app_1}, which is given by
\begin{flalign}
{\rm Q}_{\textrm{3-Pw}} (P_{\rm in}) &=
\begin{cases}
0, & P_{\rm in} < P_{\textrm{th}}^1 \\
a_i P_{\rm in} + b_i & P_{\textrm{th}}^i \le P_{\rm in} \le P_{\textrm{th}}^{i+1}, \\ & (i = 1, ..., N-1) \\
{\rm Q}_{\max}, & P_{\rm in} > P_{\textrm{th}}^{N} \
\end{cases},
\label{mod_pw3}
\end{flalign}
where $P_{\rm th} = \{P_{\rm th}^i \vert 1 \le i \le N \}$ (in mW) are thresholds on $P_{\rm in}$ for $(N+1)$ linear segments. $a_i$ and $b_i$ (in mW) are the scope and the intercept for the $i$-th linear pieces, respectively. Each linear segment is obtained through linear regression, aiming to minimize the deviation from practical EH circuit.

\subsection{Modified Logistic EH Model (Mlg)}
This model, proposed by \cite{ref_mod_lg2_0}, also expresses the input and output power as a nonlinear function, and points out that all any nonlinear EH model should satisfy the following properties:
\begin{itemize}
\item[($i$)] When $P_{\textrm{in}}$ falls below the sensitivity threshold $P_{\rm{sen}}$ of EH circuit, the circuit does not work and the output power is zero, i.e., ${\rm Q} (P_{\rm in})=0$.
\item[($ii$)] ${\rm Q} (P_{\rm in})$ is a monotonically increasing function of $P_{\rm in}$.
\item[($iii$)] With the increment of $P_{\rm in}$, the EH efficiency $\frac{{\rm Q} (P_{\rm in})}{P_{\rm in}}$ first increases to its maximum then decreases.
\item[($iv$)] When EH circuit reaches saturation, ${\rm Q} (P_{\rm in}) \le {\rm Q}_{\max}$ for all $P_{\rm in}$.
\end{itemize}
Although the Lg model, i.e., \eqref{mod_Logi}, was presented in \cite{ref_mod_lg1_0}, it does not take into account the sensitive characteristics of EH circuit and cannot satisfy the property ($i$). To address this limitation, the authors in \cite{ref_mod_lg2_0} proposed a modified logistic EH model, which is
\begin{flalign}
{\rm Q}_{\rm Mlg} (P_{\rm in}) &= \left[ \frac{{\rm Q}_{\max}}{e^{-\tau P_{\rm sen} + \nu}} \big( \frac{1+e^{\tau P_{\rm sen} + \nu}}{1 + e^{-\tau P_{\rm in} + \nu}} -1 \big) \right]^+,
\label{mod_Logi2}
\end{flalign}
where $\tau$ and $\nu$ reflect the steepness of this model, enabling it to capture the nonlinear dynamics of EH circuit. The values of these parameters can be found in Table \ref{tab3}.

\begin{table}[t!]
\renewcommand{\arraystretch}{1.40}
\centering 
 \caption{Parameters for the Mlg nonlinear EH model}
 \begin{tabular}{c|c|c|c|c}
 \hline
 \hline
 ${\rm Q}_{\max}$ (mW) & $\tau$ & $\nu$ & $P_{\rm{sen}}$ & Ref. \cite{ref_lg2_app_1} \\
 \hline
 4.927 & 274 & 0.29 & 0.064 & Powercast P2110 \\
 \cdashline{1-5}[0.8pt/1pt]
 8.2 & 411 & 2.2 & 0.08 & Avago HSMS-286x \\
 \cdashline{1-5}[0.8pt/1pt]
 37.5 & 116 & 2.3 & 0.08 & Avago HSMS-286x \\
 \cdashline{1-5}[0.8pt/1pt]
 $0.11 \times 10^{-3}$ & 47 & 2.4 & 0.08 & Avago HSMS-286x \\
 \hline
 \hline
 \end{tabular}
 \label{tab3}
\end{table}

\subsection{2-nd Order Polynomial EH Model (2-ord)}
This model was proposed in \cite{ref_mod_2or_0}, which could achieve a good fitting effect in microwatt through data fitting. It is given by
\begin{flalign}
{\rm Q}_{\textrm{2-ord}} (P_{\rm in})= \alpha_1 P_{\rm in}^2 + \alpha_2 P_{\rm in} + \alpha_3,
\label{mod_2or}
\end{flalign}
where $\alpha_1 \le 0$ provides a good match for EH measurements. Typically, ${\rm Q}_{\textrm{2-ord}} (P_{\rm in})=0$ has one root in close proximity to 0 and another positive root with a larger absolute value, indicating that $\alpha_2 > 0$. Furthermore, it is likely that EH circuit does not work until the input power exceeds a threshold, hence $\alpha_3 \le 0$. The values of the parameters depend on the actual antenna design and circuitry, which can be determined by data fitting tools.

\subsection{Fractional EH Model (Fr)}
Since the logistic and heuristic (Lg and Hr) EH models cannot be mathematically tractable to theoretically derive the probability density function (PDF) and the cumulative distribution function (CDF) of the average output power of EH circuit, a simpler fractional EH model (Fr) was presented in \cite{ref_mod_frac_0}, i.e.,
\begin{flalign}
{\rm Q}_{\rm Fr} (P_{\rm in}) &= \frac{a P_{\rm in} + b}{P_{\rm in} + c} - \frac{b}{c},
\label{mod_frac}
\end{flalign}
where $a, b$ and $c$ are constants determined by standard curve fitting. Also, the term $\frac{b}{c}$ is included to ensure that the output power is zero when the input power is zero. The fitted parameters of the Fr model are presented in Table \ref{tab4} according to \cite{ref_mod_frac_0}.

\begin{table}[t!]
\renewcommand{\arraystretch}{1.40}
\centering 
 \caption{Parameters for the Fr nonlinear EH model}
 \begin{tabular}{c|c|c|c|c}
 \hline
 \hline
 a & b & c & RMSE \footnotemark[1] & Ref. \\
 \hline
 2.463 & 1.635 & 0.826 & 0.009737 & \makecell[c]{\multirow{2}{0.6cm}{\cite{ref_mod_frac_0}}} \\
 \cdashline{1-4}[0.8pt/1pt]
 0.3929 & 0.01675 & 0.04401 & 0.0003993 & \\
 \hline
 \hline
 \end{tabular}
 \label{tab4}
\end{table}
\footnotetext[1]{RMSE is short for the root mean squared error.}

\subsection{Error function-based EH model (Ef)}
Similar to EH models mentioned above, considering two limitations of practical EH circuit, i.e., ($i$) when the input power is relatively large, the output power reaches a saturation state; ($ii$) when the input power is below the sensitivity level, the output power drops to zero, a nonlinear EH model was proposed in \cite{erf_mod_2020_0}. This model effectively captures the saturation and sensitivity features of practical EH circuit and is given by
\begin{flalign}
{\rm Q}_{\rm Ef} (P_{\rm in}) &= {\rm Q}_{\max} \left[ \frac{\rm{erf}(a((P_{\rm{in}} - P_{\rm{sen}} ) +b)) -\rm{erf}(a b)}{1 - \rm{erf}(ab)} \right]^{+},
\label{mod_ef_0}
\end{flalign}
where $a > 0$, $b > 0$, $[x]^+ = \max(x, 0)$, and $\rm{erf}(x) = \frac{2}{\sqrt{\pi}} \int_0^x e^{-t^2} \textit{dt}$ is the error function. Besides, the parameters $a$, $b$ and ${\rm Q}_{\max}$ can be determined via a best-fit match with experimental data, i.e., $a = 0.0086$, $b = 11.8689 \mu$W, ($a$ and $b$ are best fit \cite[21]{erf_mod_2020_0}), and ${\rm Q}_{\max} = 10.219 \mu$W.
Meanwhile, when the input power is large, a asymptotic model is given by
\begin{flalign}
{\rm Q}_{\rm Ef} (P_{\rm in}) &= {\rm Q}_{\max} \left[1 - e^{-\kappa (P_{\rm{in}} - P_{\rm{sen}})} \right]^{+},
\label{mod_ef_1}
\end{flalign}
where $\kappa = 2 \frac{a e^{-a^2 b^2}}{\sqrt{\pi} (1 - \rm{erf}(a b) )}$.

\subsection{Logarithmic nonlinear EH model (Log)}
In the 900 MHz frequency band, the authors \cite{han2020joint} presented a logarithmic nonlinear EH model (Log). This model offers the benefit of simplifying the optimization problem while providing a more accurate representation of the received power, which is
\begin{flalign}
{\rm Q}_{\rm Log} (x) &= a \log (1 + bx), \,\, \rm{s.t.} \,\, 0 \le x \le c
\label{mod_log}
\end{flalign}
where $a$, $b$ and $c$ are determined by detailed circuit. Specifically, $a$ and $b$ are obtained by minimizing the sum of squared error between the rectifier model and the data extracted using Engauge Digitizer \cite[18]{han2020joint}, $c$ is determined by examining an extracted data that represents the operational limit of the rectifier \cite{han2020joint}. In the specific case of \cite[27]{han2020joint}, the values are $a = 0.0319$ and $b = 3.6169$. For \cite[20]{han2020joint}, the values are $a = 0.2411$ and $b = 0.4566$. In both cases, $c$ is set to 3 mW.

\subsection{Joint Model of Nonlinear Conversion Efficiency (Jm)}
Typically, the energy conversion efficiency, i.e., $\eta$, is commonly modeled as a function of the input power, disregarding the influence of the operating frequency. Therefore, to account for the impact of both input power and operating frequency, a joint model of the nonlinear conversion efficiency (Jm) was introduced in \cite{Joint_mod_0}. That is,
\begin{flalign}
\eta^{\rm{Jm}} (P_{\rm{in}}, f) = \frac{1}{\eta_P^{\rm{Jm}}(P_{\rm{r}})} \eta_P^{\rm{Jm}}(P_{\rm{in}}) \eta_f^{\rm{Jm}}(f),
\label{mod_jm}
\end{flalign}
where $\eta_P^{\rm{Jm}}(P_{\rm{in}})$ and $\eta_f^{\rm{Jm}}(f)$ are the conversion efficiency functions in terms of the input power and frequency respectively. Especially, $\eta_P^{\rm{Jm}}(P_{\rm{in}})$ adopts the 2-ord model in \eqref{mod_2or}; $\eta_f^{\rm{Jm}}(f)$ for Type-I harvester is given by $\eta_f^{\rm{Jm}}(f) = \sum_{i=1}^n a_i e^{[-(\frac{f-b_i}{c_i})^2]}$, where $f$ is the frequency in hertz, $\eta_f^{\rm{Jm}}$ is the efficiency in percentage, $n$ is the order of the Gaussian model, and the parameters of $a_1, a_2, ..., a_n$, $b_1, b_2, ..., b_n$, and $c_1, c_2, ..., c_n$ are different for different harvesters, which can be determined from the experimental data, see \cite[Table V]{Joint_mod_0}; $\eta_f^{\rm{Jm}}(f)$ for Type-II harvester is given by $\eta_f^{\rm{Jm}}(f) = a_0 + \sum_{i=1}^n ( a_i \cos(ifw) + b_i \sin(ifw) )$, where $a_0, a_2, ..., a_n$, $b_1, b_2, ..., b_n$ and $w$ are the parameters to be fit for different harvesters, see \cite[Table VIII]{Joint_mod_0}. Besides, $\eta_P^{\rm{Jm}}(P_{\rm{r}})$ is the value of $\eta$ at $P_{\rm{r}}$ with fixed frequency $f_{\rm{r}}$.

Similar to the Hr model, the harvested power via the Jm model can be given by
\begin{flalign} 
{\rm Q}_{\rm Jm} (P_{\rm in}) = \eta^{\rm{Jm}} (P_{\rm{in}}, f) P_{\rm in}.
\label{mod_jm_eh}
\end{flalign}

\section{\textcolor[rgb]{0.00,0.00,0.00}{Comparison of Nonlinear EH Models} \label{section_comparison_eh}}

\subsection{\textcolor[rgb]{0.00,0.00,0.00}{Evolution of Nonlinear EH Models}}

\begin{figure*}[t!]
\centering
\includegraphics[width=0.98\textwidth]{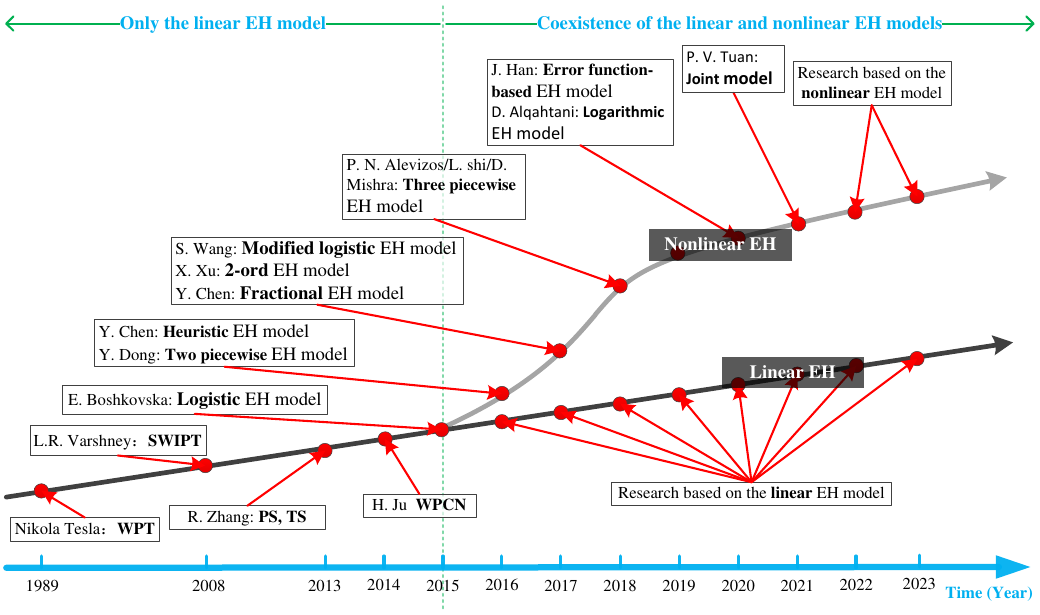}
\caption{The linear EH model vs. the nonlinear EH model}
\label{fig_two_lines}
\end{figure*}

The linear EH model, which is often used for initial research, fails to capture the nonlinear behavior exhibited by practical EH circuits. Meanwhile, utilizing results obtained with the linear EH model for the design of wireless communication systems would lead to some serious bias in terms of performance evaluations. Currently, two main branches of RF-based EH models exist. One is the linear model, while the other is the nonlinear model, as depicted in Fig. \ref{fig_two_lines}. In 2015, the logistic nonlinear EH model was introduced, opening up a new research path for RF-based EH. However, the linear EH model continues to receive attention. Although the logistic EH model aligns better with the practical EH circuit's characteristics, it possesses a more complex form. The model function is typically non-convex and non-concave, posing significant challenges and difficulties in analyzing the performance of wireless communication systems. These challenges encompass optimization problems and deriving theoretical results. Consequently, various approximate forms of nonlinear EH models have been proposed. For instance, in 2016, the two-piecewise EH model was introduced, followed by the three-piecewise EH model in 2018. These models aimed to better capture the nonlinear characteristics and account for the sensitive nature of practical EH circuits. Furthermore, a modified logistic EH model, similar to the logistic EH model, was proposed in 2017. Additionally, several other models have been presented, such as the heuristic, 2nd order polynomial, fractional, error function-based, Joint model, and logarithmic EH models. These models were proposed between 2016 and 2023, respectively.

\begin{table*}[t!]
\renewcommand{\arraystretch}{1.26}
\centering 
 \caption{The summarization and comparison of nonlinear EH models}
 \begin{tabular}{m{3.0cm}<{\centering}|m{8.5cm}|m{1.25cm}<{\centering}|m{1.25cm}<{\centering}|m{1.15cm}<{\centering}}
 \hline
 \hline
 \multicolumn{5}{c}{\multirow{4}{17.0cm}{\textbf{Parameters in each EH model}: $Q(\cdot)$ is the function of the output/harvested power. $P_{\rm{in}}$ is the input/received RF power. $P_{\rm{sen}}$ and $P_{\rm{sat}}$ are the sensitivity and saturation thresholds of practical EH circuit, respectively. ${\rm Q}_{\max}$ is the maximum harvested power when EH circuit reaches saturation. $\eta/\widetilde{\eta}( \cdot )$ is the energy conversion efficiency/function.}} \\
 \multicolumn{5}{c}{} \\
 \multicolumn{5}{c}{} \\
 \multicolumn{5}{c}{} \\

 \hline
 
 \textbf{EH models} & \textbf{\makecell[c]{Formulations}} & \textbf{\makecell[c]{Sensitivity}} & \textbf{Saturation} & \textbf{\makecell[c]{$\eta$}} \\ 
 \hline
 Linear (Ln) \cite{eh_2011} & \makecell[c]{${\rm Q}_{\rm Ln} (P_{\rm in}) = \eta P_{\rm in}, \,\, \eqref{lin_md}$} & $\times$ & $\times$ & constant \\ 
 
 \cdashline{1-5}[0.8pt/1pt]
 Logistic (Lg) \cite{ref_mod_lg1_0} & $${\rm Q}_{\rm Lg} (P_{\rm in}) = \frac{\frac{{\rm Q}_{\max}}{1+e^{-a(P_{\rm in} - b)}} - \frac{{\rm Q}_{\max}}{1+e^{a b}}} {1-\frac{1}{1+e^{a b}}}, \,\, \eqref{mod_Logi},$$ where $a$ and $b$ are constants determined by EH circuit. & $\times$ & $\surd$ & -- \\ 
 
 \cdashline{1-5}[0.8pt/1pt]
 Heuristic (Hr) \cite{ref_mod_hr_0} & $${\rm Q}_{\rm Hr} (P_{\rm in}) = \widetilde{\eta} ( P_{\rm in} ) P_{\rm in} = \frac{p_2 P_{\rm in}^3 + p_1 P_{\rm in}^2 + p_0 P_{\rm in}}{q_3 P_{\rm in}^3 + q_2 P_{\rm in}^2 + q_1 P_{\rm in} + q_0}, \,\, \eqref{mod_duo}$$ where $p_0, p_1, p_2, q_0, q_1, q_2, q_3$ are constants. & $\times$ & $\surd$ & $\widetilde{\eta}( P_{\rm in} )$ \\ 
 
 \cdashline{1-5}[0.8pt/1pt]
 Two Piecewise (2-Pw) \cite{ref_mod_pw1_0} & \makecell[c]{${\rm Q}_{\textrm{2-Pw}} (P_{\rm in}) = \begin{cases}
\eta P_{\rm in}, & P_{\rm in} \le P_{\textrm{sat}} \\
{\rm Q}_{\max}, & P_{\rm in} > P_{\textrm{sat}} \end{cases}, \,\, \eqref{mod_pw1}$} & $\times$ & $\surd$ & constant \\ 


 \cdashline{1-5}[0.8pt/1pt] 
 \multirowcell{12}{Three Piecewise (3-Pw) \\ \cite{ref_mod_pw2_0,ref_mod_pw2_1,ref_pw2_app_1}} &
$${\rm Q}_{\textrm{3-Pw}} (P_{\rm in}) =
\begin{cases}
 0, & 0 \le P_{\rm in} \le P_{\textrm{sen}} \\
\widetilde{\eta} (P_{\rm in}) P_{\rm in} & P_{\textrm{sen}} \le P_{\rm in} \le P_{\textrm{sat}} \\
\widetilde{\eta} (P_{\textrm{sat}}) P_{\textrm{sat}} & P_{\rm in} \ge P_{\textrm{sat}} \\
\end{cases}, \,\, \eqref{mod_pw3_1}$$ where $\widetilde{\eta} (x) = w_0 + \sum_{i=1}^W w_i \big(10 \log_{10} (x) \big)^i, x \in P_{\rm in}, \,\,\eqref{mod_pw3_2}$. & $\surd$ & $\surd$ & $\widetilde{\eta} (x)$ \\ 

 \cdashline{2-5}[0.8pt/1pt] 
 
 ~ & $${\rm Q}_{\textrm{3-Pw}} (P_{\rm in}) = \begin{cases}
 0, & P_{\rm in} < P_{\textrm{sen}} \\
\eta P_{\rm in} & P_{\textrm{sen}} \le P_{\rm in} \le P_{\textrm{sat}} \\
{\rm Q}_{\max} & P_{\rm in} > P_{\textrm{sat}} \ \end{cases}, \,\, \eqref{mod_pw2}$$
$${\rm Q}_{\textrm{3-Pw}} (P_{\rm in}) =
\begin{cases}
0, & P_{\rm in} < P_{\textrm{th}}^1 \\
a_i P_{\rm in} + b_i & P_{\textrm{th}}^i \le P_{\rm in} \le P_{\textrm{th}}^{i+1}, \\ & (i = 1, ..., N-1) \\
{\rm Q}_{\max}, & P_{\rm in} > P_{\textrm{th}}^{N} \
\end{cases}, \,\, \eqref{mod_pw3}$$
where $P_{\rm th} = \{P_{\rm th}^i \vert 1 \le i \le N \}$ mW are thresholds on $P_{\rm in}$ for $(N+1)$ linear segments. $a_i$ and $b_i$ mW are the scope and the intercept for the $i$-th linear pieces, respectively.
 & $\surd$ & $\surd$ & constant \\


 \cdashline{1-5}[0.8pt/1pt]
 Modify Logistic (Mlg) \cite{ref_mod_lg2_0} & $${\rm Q}_{\rm Mlg} (P_{\rm in}) = \left[ \frac{{\rm Q}_{\max}}{e^{-\tau P_{\rm sen} + \nu}} \big( \frac{1+e^{\tau P_{\rm sen} + \nu}}{1 + e^{-\tau P_{\rm in} + \nu}} -1 \big) \right]^+, \,\, \eqref{mod_Logi2}$$ where $\tau$ and $\nu$ are constants. & $\surd$ & $\surd$ & -- \\ 
 
 \cdashline{1-5}[0.8pt/1pt]
 2-nd order polynomial (2-ord) \cite{ref_mod_2or_0} & $${\rm Q}_{\textrm{2-ord}} (P_{\rm in})= \alpha_1 P_{\rm in}^2 + \alpha_2 P_{\rm in} + \alpha_3, \,\, \eqref{mod_2or}$$ where $\alpha_1 \le 0$, $\alpha_2 > 0$ and $\alpha_3 \le 0$. & $\times$ & $\surd$ & -- \\ 
 
 \cdashline{1-5}[0.8pt/1pt]
 Fractional (Fr) \cite{ref_mod_frac_0} & $${\rm Q}_{\rm Fr} (P_{\rm in}) = \frac{a P_{\rm in} + b}{P_{\rm in} + c} - \frac{b}{c}, \,\, \eqref{mod_frac}$$ where $a, b$, and $c$ are constants. & $\times$ & $\surd$ & -- \\ 
 
 \cdashline{1-5}[0.8pt/1pt]
 Error function (Ef) \cite{erf_mod_2020_0} & $${\rm Q}_{\rm Ef} (P_{\rm in}) = {\rm Q}_{\max} \left[ \frac{\rm{erf}(a((P_{\rm{in}} - P_{\rm{sen}} ) +b)) -\rm{erf}(a b)}{1 - \rm{erf}(ab)} \right]^{+}, \,\, \eqref{mod_ef_0}$$ where $a > 0$, $b > 0$, and $\rm{erf}(x) = \frac{2}{\sqrt{\pi}} \int_0^x e^{-t^2} \textit{dt}$ is the error function. & $\surd$ & $\surd$ & -- \\ 
 
 \cdashline{1-5}[0.8pt/1pt]
 Logarithmic (Log) \cite{han2020joint} & $${\rm Q}_{\rm Log} (x) = a \log (1 + bx), \,\, \rm{s.t.} \,\, 0 \le x \le c, \,\, \eqref{mod_log}$$
where $a$, $b$ and $c$ are determined by detailed circuit. & $\times$ & $\surd$ & -- \\ 
 
 \cdashline{1-5}[0.8pt/1pt]
 Joint Model (Jm) \cite{Joint_mod_0} & 
 $${\rm Q}_{\rm Jm} (P_{\rm in}) = \widetilde{\eta}^{\rm{Jm}} (P_{\rm{in}}, f) P_{\rm in}, \,\, \eqref{mod_jm_eh}$$
 where $\widetilde{\eta}^{\rm{Jm}} (P_{\rm{in}}, f) = \frac{1}{\widetilde{\eta}_P^{\rm{Jm}}(P_{\rm{r}})} \widetilde{\eta}_P^{\rm{Jm}}(P_{\rm{in}}) \widetilde{\eta}_f^{\rm{Jm}}(f) \eqref{mod_jm}$, $\widetilde{\eta}_P^{\rm{Jm}}(P_{\rm{in}})$ and $\widetilde{\eta}_f^{\rm{Jm}}(f)$ are the conversion efficiency functions of power and frequency, respectively, $\widetilde{\eta}_P^{\rm{Jm}}(P_{\rm{r}})$ is the value of $\eta$ at $P_{\rm{r}}$ with fixed frequency $f_{\rm{r}}$. & $\surd$ & $\surd$ & $\widetilde{\eta}^{\rm{Jm}} (P_{\rm{in}}, f)$ \\ 
 
 \hline
 \hline
 \end{tabular}
 \label{tb_mds_0}
\end{table*}

\subsection{\textcolor[rgb]{0.00,0.00,0.00}{Analyzing Modeling Characteristics}}
For comparison, the aforementioned EH models are summarized in Table \ref{tb_mds_0} with following conclusions:
\begin{itemize}
\item[-] \textit{The sensitivity of EH circuit:} Among the existing nonlinear EH models, the 3-Pw, Mlg, Ef and Jm EH models take into account the sensitivity of practical EH circuit\footnote{Since the Jm EH model is formed with both power and frequency, it can adopt any EH nonlinear model in terms of the power (input power).}, others do not.

\item[-] \textit{The saturation of EH circuit:} The saturation features should be the primary nonlinear characteristic of practical EH circuits, so all nonlinear models reflect this point except for the Ln EH model.

\item[-] \textit{The energy conversion efficiency $\eta$:} The Lg, Mlg, 2-ord, Fr, Ef, Log EH models do not consider the characterization of $\eta$. In the Ln, 2-Pw and 3-Pw EH models, $\eta$ is usually constructed as a constant. In order to characterize the nonlinearity of practical EH circuit, a variant $\eta$ is described by the power or the frequency in the Hr, 3-Pw and Jm EH models, respectively. For example, $\eta$ in the Hr and 3-Pw EH models is related to the power, and $\eta$ in the JM EH model is affected by both power and frequency.
\end{itemize}

\subsection{\textcolor[rgb]{0.00,0.00,0.00}{Analyzing Fitting Performance}}

\begin{figure}[t!]
\centering
\includegraphics[width=0.475\textwidth]{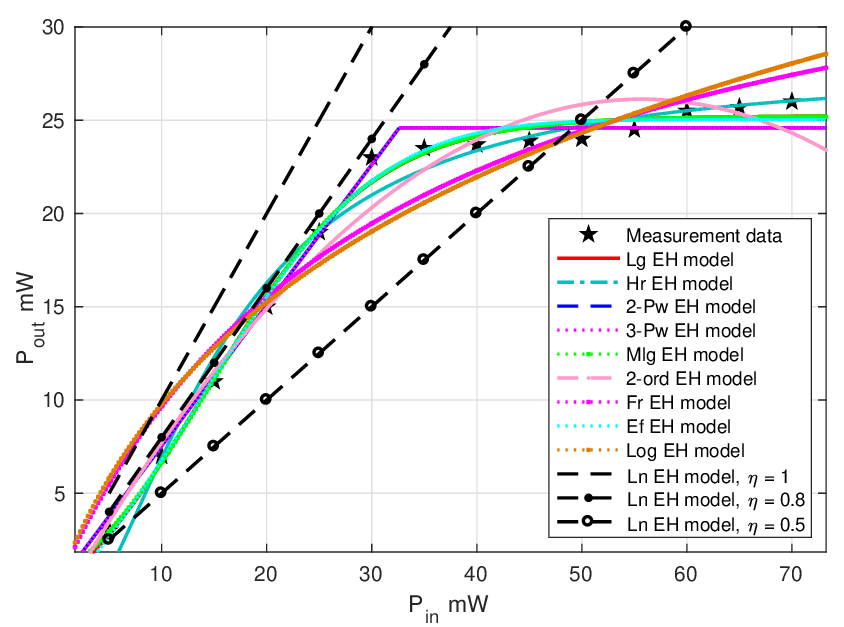}
\caption{Comparison of current EH models in milliwatt domain}
\label{fig_fit_mw}
\end{figure}

\begin{figure}[t!]
\centering
\includegraphics[width=0.475\textwidth]{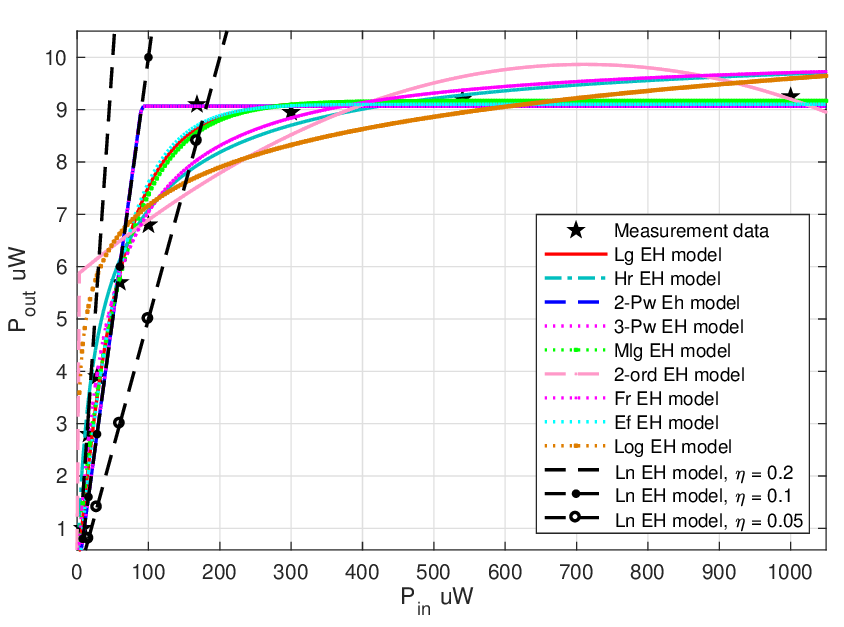}
\caption{Comparison of current EH models in microwatt domain}
\label{fig_fit_uw}
\end{figure}

To better understand and analyze each EH model, we fit the relationship between the input and output of EH in milliwatt and microwatt domains respectively for all EH models\footnote{MATLAB is used for fitting.}, according to the measurement data in \cite{fit_data}. The corresponding results are shown in Fig. \ref{fig_fit_mw} and Fig. \ref{fig_fit_uw}, respectively. It can be observed that
\begin{itemize}
\item[-] When the input power $P_{\textrm{in}}$ is relatively low (about $0 \sim 20$ mW, or $0 \sim 10$ $\mu$W), the output power $P_{\textrm{out}}$ can be approximately linearly increased. That is, it can be obtained by using the Ln EH model with an appropriate $\eta$.

\item[-] When the input power $P_{\textrm{in}}$ is relatively large, practical EH circuit enters a saturation state where the output power $P_{\textrm{out}}$ either increases slowly or remains constant. At the moment, the error caused by the Ln EH model is relatively so large, which cannot reflect the saturation characteristics of EH circuit. That is, the results obtained by the Ln EH model are not imprecise, which may lead to an incorrect evaluation of the system performance.

\item[-] The Lg, Mlg and Ef (and 3-Pw) EH models can achieve relatively better performance compared with others\footnote{The 3-Pw EH model in \eqref{mod_pw3} also can obtain good fitting results. For a simple description, we just compare the 3-Pw EH model in \eqref{mod_pw2} here.}. The Hr EH model follows closely but has a more complex form. Although the remaining models bring some errors to some extent, they may offer certain benefits in studying different problems. For example, a simplified form of the EH model can reduce the analytical complexity of a problem and provide approximate solutions to the situations that are otherwise difficult to analyze.

\end{itemize}


\section{\textcolor[rgb]{0.00,0.00,0.00}{The Road to study the Nonlinear EH models}} \label{section_apps_ehmodels}


\begin{figure}[t!]
\centering
\includegraphics[width=0.475\textwidth]{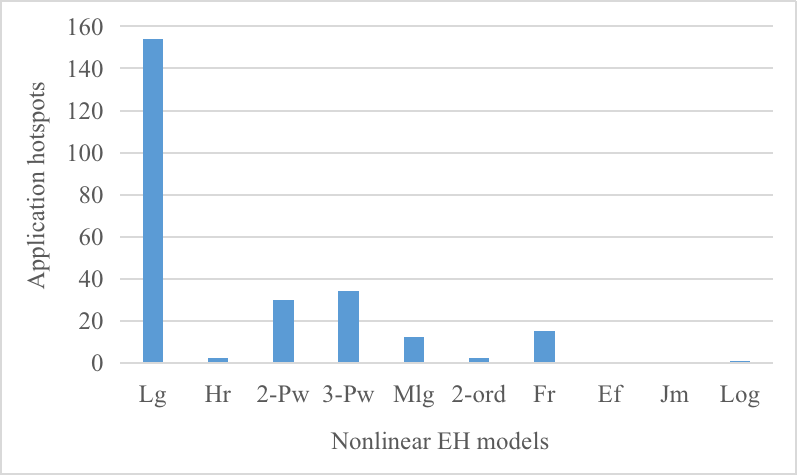}
\caption{The application hotspots of existing nonlinear EH models}
\label{fig_works_nons}
\end{figure}

\begin{figure*}[t!]
\centering
\includegraphics[width=0.99\textwidth]{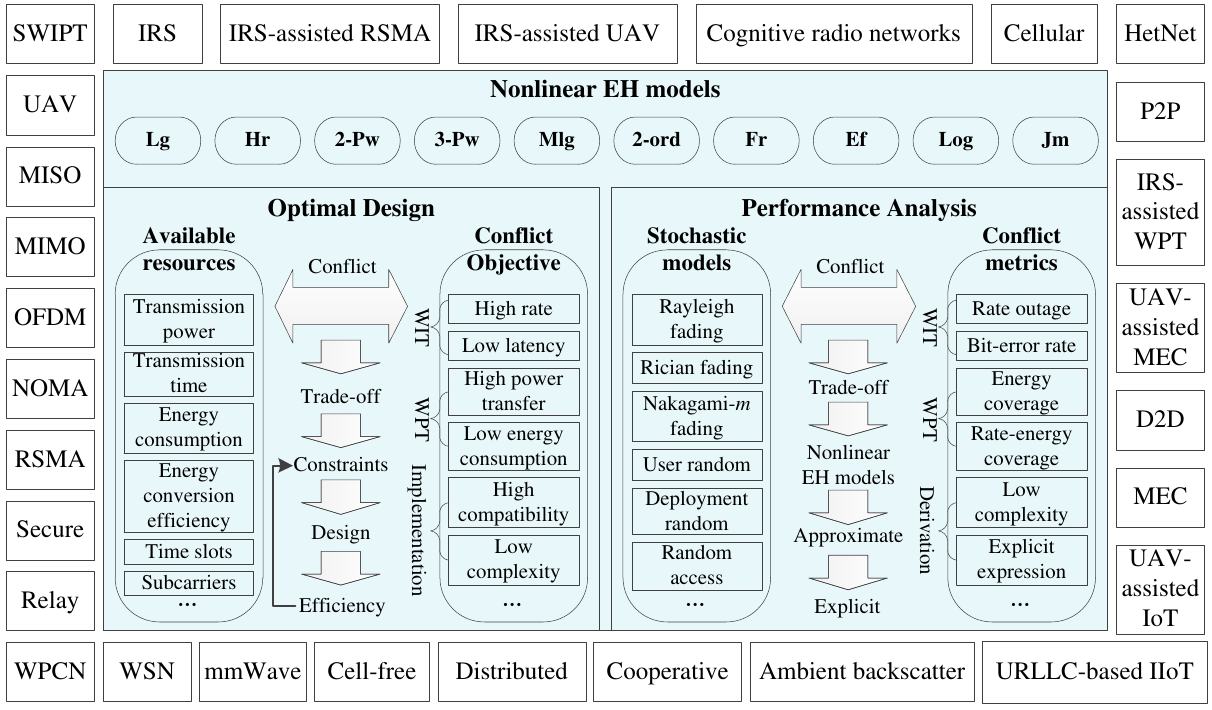}
\caption{Key techniques and scenarios for the optimal design and performance analysis of the RF-based EH systems under the nonlinear EH models}
\label{fg_scenaros_ehs}
\end{figure*}

This section provides an overview of previous studies utilizing the aforementioned nonlinear EH models, as illustrated in Fig. \ref{fig_works_nons}. Almost all existing nonlinear EH models have been employed and investigated extensively. Moreover, we summarized the key techniques and scenarios for the optimal design and performance analysis of the RF-based EH systems under the nonlinear EH models as shown in Fig. \ref{fg_scenaros_ehs}.

%
%
%
%
%

In the sequel, we delve into the specific applications of each model for further analysis and discussion.




\begin{table*}[t!] 
\renewcommand{\arraystretch}{1.46}
\centering 
\caption{The applications of the Lg and Hr nonleianr EH model} \label{tab_lg_app}
\begin{tabular}{c|c|m{8.1cm}|m{2.5cm}<{\centering}}
\hline
\hline

\multicolumn{4}{c}{} \\
\multicolumn{4}{c}{\textbf{Lg nonleianr EH model}} \\
\hline

 \textbf{Systems} & \textbf{Scenario} & \textbf{\makecell[c]{Goal}} & \textbf{Method} \\
	 \hline
 \makecell[c]{\multirow{37}{0.6cm}{SWIPT}} & MISO & Harvested power/EH efficiency \cite{ref_lg1_app_1, ref_lg1_app_5, ref_lg1_app_2, lg1_swipt_2021app5, lg1_swipt_2021app23, lg1_swipt_2021app30, ref_lg1_app_36}, transmit power \cite{ref_lg1_app_6, ref_lg1_app_11, ref_lg1_app_12, ref_lg1_app_17, ref_lg1_app_42, ref_lg1_app_44, lg1_swipt_2021app3, ref_lg1_app_43}, multi-objective optimization \cite{ref_lg1_app_29}, outage \cite{ref_lg1_app_51}, artificial noise power \cite{ref_lg1_app_46}, secrecy EE \cite{ref_lg1_app_55, lg1_swipt_2021app19}, secrecy rate \cite{lg1_swipt_2021app29}, throughput, energy cost and interference \cite{ref_lg1_app_22} & \multirow{27}{2.5cm}{Convex optimization theory, SDR, SCA, S-procedure, ADMM, Lagrangian, probability theory, mathematic} \\
 
 \cdashline{2-3}[0.8pt/1pt]
 & MIMO & EH efficiency \cite{ref_lg1_app_7}, R-E region \cite{ref_lg1_app_9}, outage-constrained secrecy rate \cite{ref_lg1_app_21}, SE and EE \cite{ref_lg1_app_50}, max-min harvested energy \cite{lg1_swipt_2021app24, lg_2021_swipt8}, harvested energy \cite{lg1_swipt_2021app26} & \\
 
 \cdashline{2-3}[0.8pt/1pt]
 & Relay & Achievable rate \cite{ref_hr_lg_app_1}, transmit power \cite{lg1_swipt_2021app11}, secrecy outage \cite{ref_lg1_app_38, lg_2021_swipt3}, residual energy \cite{ref_lg1_app_57}, sum rate \cite{lg1_swipt_2021app1}, throughput \cite{lg1_swipt_2021app12}, rate fairness \cite{lg1_swipt_2021app14}, ergodic secrecy capacity \cite{lg1_swipt_2021app20}, EE \cite{lg1_swipt_2021app21}, outage and ergodic capacity \cite{lg_2021_swipt2}, outage \cite{lg_2021_swipt13} & \\
 
 \cdashline{2-3}[0.8pt/1pt]
 & P2P & Rate \cite{ref_lg1_app_8, ref_lg1_app_30}, harvested energy \cite{ref_lg1_app_31}, R-E region \cite{ref_lg1_app_32, ref_lg1_app_52, lg1_swipt_2021app17}, I-E region \cite{lg1_swipt_2021app6, lg1_swipt_2021app7}, & \\ 
 
 \cdashline{2-3}[0.8pt/1pt]
 & Multi-user & R-E region \cite{lg1_swipt_2021app4, lg1_swipt_2021app2}, outage \cite{ref_lg1_app_35, ref_lg1_app_35_1}, SINR \cite{lg1_swipt_2021app27}, transmit power minimization and the sum harvested energy maximization \cite{lg_2020_swipt1}, & \\
 
 \cdashline{2-3}[0.8pt/1pt]
 & OFDM & Achievable sum rate \cite{lg_2022_swipt14}, power efficiency \cite{lg_2023app6} & \\
 
 \cdashline{2-3}[0.8pt/1pt]
 & NOMA & Harvested energy \cite{ref_lg1_app_19, ref_lg1_app_20}, transmit power \cite{ref_lg1_app_53, ref_lg1_app_48, ref_lg1_app_48_0}, SE/EE \cite{lg1_swipt_2021app8, lg1_swipt_2021app15}, sum rate \cite{lg1_swipt_2021app13, lg1_swipt_2021app16, lg1_swipt_2021app28}, throughput \cite{lg1_swipt_2021app22} & \\
 
 \cdashline{2-3}[0.8pt/1pt]
 & mmWave & Secure rate \cite{lg_2022_swipt17} & \\
 
 \cdashline{2-3}[0.8pt/1pt]
 & RSMA & Secure EE \cite{lg_2022_swipt16} & \\
 
 \cdashline{2-3}[0.8pt/1pt]
 & CR & Transmit power/harvested power \cite{ref_lg1_app_15, ref_lg1_app_26, ref_lg1_app_28, ref_lg1_app_39} & \\
 
 \cdashline{2-3}[0.8pt/1pt]
 & Cellular & Mean square error \cite{ref_lg1_app_34}, sum rate \cite{lg1_swipt_2021app10} & \\
 
 \cdashline{2-3}[0.8pt/1pt]
 & HetNet & Secure EE \cite{lg_2021_swipt7}, capacity \cite{lg_2022_swipt15} & \\
 
 \cdashline{2-3}[0.8pt/1pt]
 & IRS-assisted & Transmit power minimization \cite{lg_2021_swipt4,lg_2021_swipt4_1,lg_2022_swipt12}, EE and secure \cite{lg_2021_swipt5}, max-min EE \cite{lg_2021_swipt6}, energy shortage probability and bit error rate \cite{lg_2023app2} & \\
 
 \cdashline{2-3}[0.8pt/1pt]
 & IRS-assisted RSMA & Transmit power minimization \cite{lg_2022_swipt11} & \\
 
 \cdashline{2-3}[0.8pt/1pt]
 & IRS-empowered UAV & Achievable sum rate \cite{lg_2021_swipt9}, energy consumption \cite{lg_2023app8} & \\ 
 
 \cdashline{2-3}[0.8pt/1pt]
 & UAV & Coverage provability-constrained throughput \cite{jrh_uav_1}, coverage performance \cite{jrh_uav_2, jrh_uav_3} & \\ 
 
 \cdashline{2-3}[0.8pt/1pt]
 &Others & Outage \cite{lg1_swipt_2021app18}, chievable rate and region \cite{lg_2022_swipt10}, sum rate \cite{lg_2023app7} & \\

 \hline

 \makecell[c]{\multirow{19}{0.6cm}{WPCN}} & MIMO & Transmit power \cite{ref_lg1_app_23}, EH efficiency \cite{ref_lg1_app_16}, harvested power \cite{ref_lg1_app_47} & \multirow{12}{2.5cm}{Convex optimization theory, probability theory, mathematic} \\
 
 \cdashline{2-3}[0.8pt/1pt]
 & Relay & Outage \cite{ref_lg1_app_13}, transmit power \cite{ref_lg1_app_54}, throughput \cite{lg1_2021app19} & \\
 
 \cdashline{2-3}[0.8pt/1pt]
 & Multi-user & Energy coverage \cite{ref_lg1_app_37}, throughput \cite{ref_lg1_app_56, lg1_2021app5, lg1_2021app12}, sum-rate \cite{lg1_2021app8}, successful transmission probability \cite{lg_2021_wpcn10} & \\
 
 \cdashline{2-3}[0.8pt/1pt]
 & CR & Throughput \cite{ref_lg1_app_18}, secrecy EE \cite{lg1_2021app9, lg1_2021app11}, spectrum efficiency \cite{lg1_2021app15, lg1_2021app16} & \\
 
 \cdashline{2-3}[0.8pt/1pt]
 & NOMA & Rate \cite{lg1_2021app6}, achievable computation rate \cite{lg1_2021app20, lg_2021_wpcn4}, energy consumption \cite{ref_lg1_app_45}, computation efficiency \cite{lg_2021_wpcn2}, EE \cite{lg_2022_wpcn13} & \\
 
 \cdashline{2-3}[0.8pt/1pt]
 & MEC & Energy maximization \cite{lg_2021_wpcn8} & \\
 
 \cdashline{2-3}[0.8pt/1pt]
 & IRS-assisted & Transmit power \cite{lg_2021_wpcn9, lg_2022_wpcn12, lg_2022_wpcn14}, weighted sum throughput \cite{lg_2021_wpcn11}, harvested energy \cite{lg_2023app1}, outage \cite{lg_2023app4} & \\
 
 \cdashline{2-3}[0.8pt/1pt]
 & UAV & Throughput \cite{ref_lg1_app_41, lg1_2021app7}, harvested energy \cite{lg1_2021app13}, AoI \cite{lg1_2021app17}, max-min harvested power \cite{lg_2021_wpcn1}, sum computation bits \cite{lg_2021_wpcn7} & \\
 
 \cdashline{2-3}[0.8pt/1pt]
 & Others & Throughput \cite{lg1_2021app10}, AoI \cite{lg1_swipt_2021app9}, harvested power \cite{lg1_2021app14}, energy consumption \cite{ref_lg1_app_14}, logit Pearson type III distribution of WPT \cite{lg_2021_wpcn0}, transmit power minimization \cite{lg_2021_wpcn5}, EE \cite{lg_2021_wpcn6} & \\

\hline
\hline

\multicolumn{4}{c}{} \\
\multicolumn{4}{c}{\textbf{Hr nonleianr EH model}} \\
\hline
 \textbf{Network model} & \textbf{Scenario} & \makecell[c]{\textbf{Goal}} & \makecell[c]{\textbf{Method}} \\
	 \hline
 \makecell[c]{\multirow{1}{0.6cm}{SWIPT}} & Relay & Achievable rate \cite{ref_hr_lg_app_1} & \multirow{2}{3.0cm}{Convex optimization theory} \\

 \cdashline{1-3}[0.8pt/1pt]
 \makecell[c]{\multirow{1}{0.6cm}{WPCN}} & MIMO & EE \cite{ref_hr_lg_app_2} & \\ 
\hline
\hline

 \end{tabular}
\end{table*}

\subsubsection{\textbf{\textcolor[rgb]{0.00,0.00,0.00}{Lg: Applications}}}


The application of the Lg nonlinear EH model can be summarized into two branches, i.e., mainly focusing on the SWIPT systems and the WPCN systems, respectively, as shown in \ref{tab_lg_app}. In the SWIPT systems, the Lg model has been studied in various wireless communication networks including multiple input single output (MISO) \cite{ref_lg1_app_1, ref_lg1_app_5, ref_lg1_app_2, lg1_swipt_2021app5, lg1_swipt_2021app23, lg1_swipt_2021app30, ref_lg1_app_36, ref_lg1_app_6, ref_lg1_app_11, ref_lg1_app_12, ref_lg1_app_17, ref_lg1_app_42, ref_lg1_app_44, lg1_swipt_2021app3, ref_lg1_app_43, ref_lg1_app_29, ref_lg1_app_51, ref_lg1_app_46, ref_lg1_app_55, lg1_swipt_2021app19, lg1_swipt_2021app29, ref_lg1_app_22}, MIMO \cite{ref_lg1_app_7, ref_lg1_app_9, ref_lg1_app_21, ref_lg1_app_50, lg1_swipt_2021app24, lg1_swipt_2021app26, lg_2021_swipt8}, relay \cite{ref_hr_lg_app_1, lg1_swipt_2021app11, ref_lg1_app_38, lg_2021_swipt3, ref_lg1_app_57, lg1_swipt_2021app1, lg1_swipt_2021app12, lg1_swipt_2021app14, lg1_swipt_2021app20, lg1_swipt_2021app21, lg_2021_swipt2, lg_2021_swipt13}, P2P \cite{ref_lg1_app_8, ref_lg1_app_30, ref_lg1_app_31, ref_lg1_app_32, ref_lg1_app_52, lg1_swipt_2021app17, lg1_swipt_2021app6, lg1_swipt_2021app7}, multi-user \cite{lg1_swipt_2021app4, lg1_swipt_2021app2, ref_lg1_app_35, ref_lg1_app_35_1, lg1_swipt_2021app27, lg_2020_swipt1}, OFDM \cite{lg_2022_swipt14}, NOMA \cite{ref_lg1_app_19, ref_lg1_app_20, ref_lg1_app_53, ref_lg1_app_48, ref_lg1_app_48_0, lg1_swipt_2021app8, lg1_swipt_2021app15, lg1_swipt_2021app13, lg1_swipt_2021app16, lg1_swipt_2021app28, lg1_swipt_2021app22}, mmWave \cite{lg_2022_swipt17}, rate-splitting multiple access (RSMA) \cite{lg_2022_swipt16}, CR \cite{ref_lg1_app_15, ref_lg1_app_26, ref_lg1_app_28, ref_lg1_app_39}, cellular \cite{ref_lg1_app_34, lg1_swipt_2021app10}, HetNet \cite{lg_2021_swipt7, lg_2022_swipt15}, intelligent reflecting surface (IRS) \cite{lg_2021_swipt4,lg_2021_swipt4_1,lg_2022_swipt12, lg_2021_swipt5, lg_2021_swipt6, lg_2023app2, lg_2022_swipt11, lg_2023app8}, UAV \cite{lg_2021_swipt9, jrh_uav_1, jrh_uav_2, jrh_uav_3}, and others \cite{lg1_swipt_2021app18, lg_2022_swipt10, lg_2023app7}. Meanwhile, the WPCN system also has been investigated in many different wireless communication networks, such as MIMO \cite{ref_lg1_app_23, ref_lg1_app_16, ref_lg1_app_47}, relay \cite{ref_lg1_app_13, ref_lg1_app_54, lg1_2021app19}, multi-user \cite{ref_lg1_app_37, ref_lg1_app_56, lg1_2021app5, lg1_2021app12, lg1_2021app8, lg_2021_wpcn10}, CR \cite{ref_lg1_app_18, lg1_2021app9, lg1_2021app11, lg1_2021app15, lg1_2021app16}, NOMA \cite{lg1_2021app6, lg1_2021app20, lg_2021_wpcn4, ref_lg1_app_45, lg_2021_wpcn2, lg_2022_wpcn13}, mobile edge computing (MEC) \cite{lg_2021_wpcn8}, IRS \cite{lg_2021_wpcn9, lg_2022_wpcn12, lg_2022_wpcn14, lg_2021_wpcn11, lg_2023app1, lg_2023app4}, UAV \cite{ref_lg1_app_41, lg1_2021app7, lg1_2021app13, lg1_2021app17, lg_2021_wpcn1, lg_2021_wpcn7}, and others \cite{lg1_2021app10, lg1_swipt_2021app9, lg1_2021app14, ref_lg1_app_14, lg_2021_wpcn0, lg_2021_wpcn5, lg_2021_wpcn6}. The detail shall be introduced as follows.


\textbf{\textit{The SWIPT systems with the Lg EH model:}}
The Lg EH model has been studied in diverse communication network scenarios such as MISO, MIMO, relay, P2P, and multi-user, as well as with various access technologies such as OFDM, NOMA, RSMA, and mmWave. Furthermore, the Lg EH model has been considered in different network topologies like CR, cellular networks, and HetNet, while also being applied and studied with emerging technologies like IRS and UAV. Additionally, the Lg EH model has been examined in other specialized systems.

\begin{itemize}
\item \textbf{MISO SWIPT Systems:}
In \cite{ref_lg1_app_1, ref_lg1_app_2, ref_lg1_app_5, ref_lg1_app_6, ref_lg1_app_11, ref_lg1_app_12, ref_lg1_app_29} and \cite{ref_lg1_app_51}, the multi-user SWIPT system was explored, where the maximization of the total harvested power, the minimization of the total transmit power, the multi-objective optimization, and the outage probability and reliable throughput were involved, respectively. In \cite{ref_lg1_app_17, ref_lg1_app_42, ref_lg1_app_44, ref_lg1_app_46} and \cite{ref_lg1_app_55}, the secure MISO SWIPT networks were investigated, where the minimization of transmit power, the maximization of artificial noise power and the global secrecy EE were involved, respectively. In \cite{lg1_swipt_2021app3}, the heterogeneous multi-user MISO SWIPT networks with coexisting PS and TS users was considered, where the system's required transmit power was minimized by jointly optimizing the transmit beamforming vectors, PS ratios and TS ratios. In \cite{lg1_swipt_2021app5}, the MISO SWIPT system was explored with the PS receiver, where the sum harvested energy and harvested EE were maximized. In \cite{lg1_swipt_2021app19}, the MISO SWIPT networks with multiple passive eavesdroppers eves was studied, where the secrecy EE was involved. In \cite{lg1_swipt_2021app23}, a multiuser downlink MISO SWIPT network with a PS mechanism was considered, where the total harvested energy was maximized. In \cite{lg1_swipt_2021app30}, a multi-user full-duplex MISO SWIPT system with imperfect channel state information (CSI) was investigated, where the minimum average total harvested power was maximized. In \cite{lg1_swipt_2021app29}, a millimeter-wave (MmWave) secure MISO SWIPT system was considered with two RF chain antenna architectures, where the secrecy rate was maximized. In \cite{ref_lg1_app_22}, the MISO cognitive SWIPT network was considered, where the maximization of the system throughput, and the minimization of the energy cost and interference were explored. In \cite{ref_lg1_app_36} and \cite{ref_lg1_app_43}, the multi-cell multi-user MISO SWIPT systems were considered, where the maximization of the EH efficiency and the minimization of the total transmit power were studied, respectively. 

\item \textbf{MIMO SWIPT Systems:}
In \cite{ref_lg1_app_7} and \cite{ref_lg1_app_9}, the MIMO SWIPT system was considered, where the harvested power efficiency and the rate-energy (R-E) region were maximized, respectively. In \cite{ref_lg1_app_21}, a MIMO wiretap SWIPT network was studied, where the outage-constrained secrecy rate was maximized. In \cite{ref_lg1_app_50}, a mMIMO SWIPT system was considered, where the spectral efficiency (SE) and EE were maximized. In \cite{lg1_swipt_2021app24} and \cite{lg_2021_swipt8}, a secure downlink multi-user MIMO SWIPT IoT system was considered with cooperative jamming, where the max-min harvested energy was investigated with norm-bounded channel uncertainties. In \cite{lg1_swipt_2021app26}, a distributed cell-free massive MIMO SWIPT network was considered with the TS receiver, where the harvested energy per-slot and the average achievable downlink rate were studied.

\item \textbf{Relay SWIPT Systems:}
In \cite{ref_hr_lg_app_1}, a MIMO DF relay SWIPT system was considered, where the end-to-end achievable rate was maximized. In \cite{ref_lg1_app_38}, a secure two-hop DF relay SWIPT system was considered, where the secrecy outage probability was minimized. In \cite{ref_lg1_app_57}, a secure two-hop DF relay-assisted CR SWIPT system was considered, where the residual energy was maximized with both Hr and Lg nonlinear EH models. In \cite{lg1_swipt_2021app1}, the two-hop relay-assisted multi-user OFDMA SWIPT network was studied, where the sum rate of the system was maximized. In \cite{lg1_swipt_2021app11}, a multi-relay assisted SWIPT network was considered with imperfect CSI, where the total transmit power of the system was minimized. In \cite{lg1_swipt_2021app12}, a cooperative DF relay network with TS receiver was studied, where the system throughput was evaluated. In \cite{lg1_swipt_2021app14}, a multi-pair DF relay network was considered with a presented PS-based EH architecture, where the fairness of end-to-end rate among user pairs was maximized. In \cite{lg1_swipt_2021app20}, an amplify and forward (AF) multi-antenna relaying system was considered, where the ergodic secrecy capacity was studied with the PS and TS protocols. In \cite{lg1_swipt_2021app21}, a secure SWIPT DF relay network was studied, where the EE of the system was maximized with the PS and TS schemes. In \cite{lg_2021_swipt2}, the cooperative ambient backscatter DF relay SWIPT system was considered, where the outage probability and ergodic capacity were discussed. Similar to \cite{lg_2021_swipt2}, the authors in \cite{lg_2021_swipt3} studied the secrecy outage probability for the cooperative ambient backscatter DF relay SWIPT system with multiple tags and one eavesdropper, where a tag selection scheme was presented. In \cite{lg_2021_swipt13}, the DF relaying cooperative SWIPT system in IoT was considered, where the outage performance was studied over Rayleigh distributed fading channel.


\item \textbf{Point-to-Point SWIPT Systems:}
In \cite{ref_lg1_app_8}, a P2P SWIPT system was studied, where average rate of the system was investigated. In \cite{ref_lg1_app_30, ref_lg1_app_31, ref_lg1_app_32}, the point-to-point SWIPT systems were considered, where the maximizations of the average achievable rate and the average harvested energy, and the R-E tradeoff were explored, respectively. In \cite{ref_lg1_app_52}, a P2P MIMO system was studied, where the R-E region was involved. In \cite{lg1_swipt_2021app6} and \cite{lg1_swipt_2021app7}, the information-energy (I-E) region was studied for SWIPT system in mobility scenarios, where a moving transmitter transmits information and energy to a PS-based receiver. In \cite{lg1_swipt_2021app17}, the R–E region was maximized for a point-to-point SWIPT in an ergodic fading channel, where the dynamic PS scheme was presented.


\item \textbf{Multi-user SWIPT Systems:}
In \cite{lg1_swipt_2021app4}, the R-E region was investigated for SWIPT systems over multi-user interference SISO channels, where both TS and PS schemes were considered. In \cite{ref_lg1_app_35} and \cite{ref_lg1_app_35_1}, a multi-user wireless powered SWIPT system was considered, where the outage probability and the reliable throughput performance were analyzed. In \cite{lg1_swipt_2021app2}, the achievable R-E region was studied for downlink SWIPT system with multiple EH users, where both TS and PS protocols were involved. In \cite{lg1_swipt_2021app27}, a SWIPT-enabled multi-group multicasting system with heterogeneous users was considered, where the maximization of weighted sum-SINR and the maximization of minimum SINR of the intended users were studied. In \cite{lg_2020_swipt1}, the multigroup multicast precoder designs were studied for SWIPT Systems with heterogeneous users, where the total transmit power was minimized and the sum harvested energy was maximized. 

\item \textbf{OFDM SWIPT Systems:} 
In \cite{lg_2022_swipt14}, the achievable sum rate was maximized for the OFDM-based SWIPT networks, where the cooperative transmission strategy was presented. In \cite{lg_2023app6}, the power efficiency problem was studied for a multi-user MIMO-OFDM SWIPT system, where the beamforming design and antenna selection were jointly optimized.

\item \textbf{NOMA SWIPT Systems:}
In \cite{ref_lg1_app_19} and \cite{ref_lg1_app_20}, the downlink NOMA CR SWIPT network was considered, where the harvested energy of each secondary users was maximized. In \cite{ref_lg1_app_53}, a MISO-NOMA SWIPT system was considered, where the system transmit power was minimized. In \cite{lg1_swipt_2021app8}, the downlink NOMA-assisted MISO-SWIPT system serving multiple ID-EH and ID users was studied, where the SE was maximized with the PS protocol. In \cite{lg1_swipt_2021app13}, a NOMA-based cellular massive IoT with SWIPT was considered, where the weighted sum rate was maximized and the total power consumption was minimized. In \cite{lg1_swipt_2021app15} and \cite{lg1_swipt_2021app16}, the multiple users MISO-NOMA downlink SWIPT networks were considered, where the spectral and energy efficiencies (SE and EE), and the sum rate of the system were studied, respectively. In \cite{lg1_swipt_2021app22}, a UAV-aided SWIPT NOMA network with artificial jamming was explored, where the throughput of the system was maximized. In \cite{lg1_swipt_2021app28}, a multi carrier MISO NOMA enabling SWIPT system in HetNets was considered, where the total rate of the system was maximized. In \cite{ref_lg1_app_48} and \cite{ref_lg1_app_48_0}, the MISO-NOMA CR-aided SWIPT under both the bounded and the Gaussian CSI estimation error mode was considered, where the system transmit power was minimized.

\item \textbf{mmWave SWIPT Systems:} 
In \cite{lg_2022_swipt17} a millimeter wave (mmWave) IoT system with SWIPT was considered, where the secrecy rate was maximized in different cases of perfect and imperfect CSI.

\item \textbf{RSMA SWIPT Systems:}
In \cite{lg_2022_swipt16}, a RSMA-based secure SWIPT network under imperfect CSI was considered, where the robust EE beamforming design was studied.

\item \textbf{CR SWIPT Systems:}
In \cite{ref_lg1_app_15}, a downlink secure MISO CR system was considered, where the transmit power of the primary and secondary networks was minimized. In \cite{ref_lg1_app_26} and \cite{ref_lg1_app_28}, the MISO NOMA CR SWIPT networks were considered, where the total transmit power was minimized. In \cite{ref_lg1_app_39}, a downlink MISO CR SWIPT network was studied, where the weighted sum of the transmit power and the total harvested energy was maximized with both Lg and Mlg nonlinear EH models.

\item \textbf{Cellular SWIPT Systems:}
In \cite{ref_lg1_app_34} and \cite{lg1_swipt_2021app10}, the cellular IoT SWIPT networks were considered, where the mean square error was minimized and the weighted sum rate was maximized, respectively.

\item \textbf{HetNet SWIPT Systems:}
In \cite{lg_2021_swipt7}, the downlink two-tier SWIPT-aided heterogeneous networks (HetNets) with multiple eavesdroppers was studied, where the secure EE was maximize over imperfect CSI. In \cite{lg_2022_swipt15}, the downlink and uplink SWIPT system in 5G dense IoT two-tier HetNets was considered, where the total utility for user rates was maximized and discussed.
 
\item \textbf{IRS-assisted RSMA SWIPT Systems:}
In \cite{lg_2022_swipt11}, the authors investigated a IRS-assisted MISO SWIPT system by using rate-splitting multiple access (RSMA), where the transmit power was minimized.
 
\item \textbf{IRS-assisted SWIPT Systems:}
In \cite{lg_2021_swipt4} and \cite{lg_2021_swipt4_1}, the large IRS-assisted SWIPT system was considered, where the transmit power was minimized. In \cite{lg_2021_swipt5}, a secure IRS-assisted SWIPT network was studied, where the EE was maximized. In \cite{lg_2021_swipt6}, a multi-user MISO IRS-aided SWIPT system was considered, where the max-min individual EE was investigated. In \cite{lg_2022_swipt12}, a IRS-aided secure SWIPT terahertz system was investigated, where the total transmit power was minimized with imperfect CSI. In \cite{lg_2023app2}, a frequency-modulated (FM) differential chaos shift keying (DCSK) scheme was presented for the RIS-assisted SWIPT system, where the energy shortage probability and bit error rate were involved over the multipath Rayleigh fading channel.

\item \textbf{IRS-empowered UAV SWIPT Systems:}
In \cite{lg_2021_swipt9}, a IRS-empowered UAV SWIPT system in IoT was considered, where the achievable sum-rate maximization problem was studied. In \cite{lg_2023app8}, the maximum energy consumption of all users was minimized for an IRS-enabled UAV SWIPT system, where the user scheduling and UAV trajectory were jointly designed.

\item \textbf{UAV SWIPT Systems:}
For UAV-assisted SWIPT networks, the information-energy (I-E) coverage provability-constrained throughput was maximized in \cite{jrh_uav_1}, and the coverage performance was investigated in \cite{jrh_uav_2, jrh_uav_3}. Meanwhile, both PS and TS architectures were employed at ground users in \cite{jrh_uav_1, jrh_uav_2, jrh_uav_3}.
 
\item \textbf{Other SWIPT Systems:} 
In \cite{lg1_swipt_2021app18}, an ambient backscatter communication network was studied, where the system outage probability was minimized for any given backscatter link. In \cite{lg_2022_swipt10}, a two-user SWIPT system with multiple access channels was considered, where the achievable rate and maximum departure regions were involved. In \cite{lg_2023app7}, the robust sum-rate of the system was maximized for SWIPT networks with imperfect CSI, where a reconfigurable metasurface-based transceiver was presented.

\end{itemize}

\textbf{\textit{The WPCN systems with the Lg EH model:}}
In WPCN systems, the Lg EH model has been studied in MIMO, relay, multi-user, CR, NOMA, MEC, IRS, UAV and other wireless communication scenarios.

\begin{itemize}
\item \textbf{MIMO WPCN Systems:}
In \cite{ref_lg1_app_23}, the secure multi-user WPCN systems were considered, where the total transmit power was minimized. In \cite{ref_lg1_app_16}, a centralized massive MIMO WPCN system was studied, where the efficiency and fairness of EH users were maximized with the utility functions. In \cite{ref_lg1_app_33}, a MIMO wireless powered sensors network was studied, where the mean square error was minimized. In \cite{ref_lg1_app_47}, a new ER architecture was presented for a P2P MIMO WPT system, where the ER’s harvested DC power was maximized by jointly optimizing the transmit energy beamforming and the adaptive PS ratio.

\item \textbf{Relay WPCN Systems:}
In \cite{ref_lg1_app_13}, a time-slotted single-user WPCN system was explored, where the outage probability and the average throughput were analyzed. In \cite{ref_lg1_app_54}, a wireless powered downlink multi-relay multi-user network was considered with imperfect CSI, where the total transmit power was minimized by joint designing the source and relay beamforming. In \cite{lg1_2021app19}, a relay-based wireless-powered communication network was studied, where the end-to-end sum throughput was maximized by jointly optimizing the power and time fraction for energy and information transmission.

\item \textbf{Multi-user WPCN Systems:}
In \cite{ref_lg1_app_37}, the energy coverage probability, the average harvested energy, and the achievable was analyzed for millimeter wave WPCN networks. In \cite{ref_lg1_app_56}, the buffer-constrained throughput performance was studied for a multi-user wireless-powered communication system, where the finite energy storage capacity, finite data buffer capacity, stochastic channel condition and stochastic traffic arrivals were considered. In \cite{lg1_2021app5}, the secrecy throughput was maximized from a wireless-powered transmitter to a desired receiver with multiple eavesdroppers for WPCN, where a secure two-phase communication protocol was presented. In \cite{lg1_2021app8}, the sum rate was maximized for a WPCN system with considering TDMA and OFDMA, where the ‘‘harvest-then-transmit’’ protocol was involved. In \cite{lg1_2021app12}, the minimum throughput among all the users was maximized for a wireless powered communication network with a remote access point and a group of multiple adjacent users, where a fairness-aware intra-group cooperative transmission protocol was proposed. In \cite{lg_2021_wpcn10}, the wirelessly powered backscatter communication network was considered with large-scale deployment of IoT nodes, where the the successful transmission probability was explored under the imperfect CSI with the stochastic geometry.

\item \textbf{CR WPCN Systems:}
In \cite{ref_lg1_app_18}, a wireless powered CR network was considered, where the sum throughput of the secondary users was maximized. In \cite{lg1_2021app9}, a MISO EH CR network was studied, where the secrecy EE was maximized under given outage probability and transmit power constraints. In \cite{lg1_2021app11}, a downlink EH CR network was considered, where the outage-constrained secrecy EE was maximized. In \cite{lg1_2021app15} and \cite{lg1_2021app16}, a EH-based 5G cooperative CR network was took into account, where both SE and the physical layer security were studied.

\item \textbf{NOMA WPCN Systems:}
In \cite{lg1_2021app6}, the minimum rate of users was maximized for the uplink of wireless powered networks, where by optimizing the allocated time to EH and information transmission were optimized, and NOMA is superior to time-division multiple access (TDMA) for uses. In \cite{lg1_2021app20}, the achievable computation rate of the system was maximized for multi-fog servers assisted NOMA-based wireless powered network, where the time assignment, the power allocation and the computation frequency were jointly optimized. In \cite{ref_lg1_app_45}, the total energy consumption was maximized for an uplink M2M enabled cellular WPCN network, where both NOMA and TDMA were involved. In \cite{lg_2021_wpcn2}, the computation efficiency (i.e., the total computation bits divided by the consumed energy) was maximized for NOMA-assisted wireless powered MEC with user cooperation, where the energy beamforming, time and power allocations were jointly designed. In \cite{lg_2021_wpcn4}, the multi-fog server-assisted NOMA-based wireless powered network was considered, where the achievable computation rate was maximized. In \cite{lg_2022_wpcn13}, the EE maximization problem was studied for massive MIMO NOMA WPCN system, where the joint power, time and antenna selection allocation scheme was presented.

\item \textbf{MEC WPCN Systems:} 
In \cite{lg_2021_wpcn8}, a residual energy maximization problem was studied for wireless powered MEC system with mixed offloading.

\item \textbf{IRS-assisted WPCN Systems:}
In \cite{lg_2021_wpcn9}, the transmit power was minimized for a IRS-assisted WPCN system under imperfect CSI. In \cite{lg_2021_wpcn11}, the weighted sum throughput of system was maximized for a IRS-aided MIMO full duplex (FD)-WPCN system. In \cite{lg_2022_wpcn12} and \cite{lg_2022_wpcn14}, the transmit energy consumption was minimized for a IRS-assisted WPCN system. In \cite{lg_2023app1}, the total harvested energy was maximized of IRS-assisted multiuser wireless energy transfer system, where the static, the dynamic and the low complexity TDMA-based beamforming schemes were studied. In \cite{lg_2023app4}, the outage probability was studied of a double IRS-enabled wireless energy transfer system over different channel fading scenarios, where the power transfer efficiency was involved.

\item \textbf{UAV WPCN Systems:}
In \cite{ref_lg1_app_41}, the weighted throughput was maximized for a UAV empowered WPCN system, where the dirty paper coding scheme and information-theoretic uplink-downlink channel duality were exploited. In \cite{lg1_2021app7}, the minimum throughput of the system was maximized for UAV-aided WPCN system, where the UAVs' trajectories, the uplink power control and the time resource allocation were jointly optimized. In \cite{lg1_2021app13}, the sum energy harvested by ground users was maximized for UAV-enabled WPT networks, where a particle swarm optimization algorithm was presented. In \cite{lg1_2021app17}, the average of AoI of the data was minimized for UAV-assisted wireless powered IoT networks, where the UAV’s trajectory and the transmission time were jointly optimized. In \cite{lg_2021_wpcn1}, the UAV's placement was optimized for UAV-enabled WPT system, where the max-min harvested power was involved. In \cite{lg_2021_wpcn7}, the resource allocation strategy was studied for UAV-assisted MEC system, where the sum of computation bits was maximized.

\item \textbf{Other WPCN Systems:}
In \cite{lg1_2021app10}, a wireless powered multimedia communication system was studied, where the traffic throughput was maximized under statistical data latency requirement. In \cite{lg1_swipt_2021app9}, a WPT enabled network was considered, where the long-term average urgency-aware AoI was minimized. In \cite{lg1_2021app14}, a retrodirective WPT system was considered, where the average harvested power was analyzed over Nakagami-m fading channels. In \cite{ref_lg1_app_14}, a hybrid powered communication system was considered, where the long-term grid energy expenditure was minimized. In \cite{lg_2021_wpcn0}, the logit Pearson type III distribution was utilized in WPT, where the statistical behavior of WPT was investigated. In \cite{lg_2021_wpcn5}, a RF EH assisted quantum battery system was investigated, where the transmit power was minimized. In \cite{lg_2021_wpcn6}, the typical two-layer EH D2D multicast communication was explored, where the EE was maximized with presenting a joint cluster formation scheme. 


\end{itemize}

\subsubsection{\textbf{\textcolor[rgb]{0.00,0.00,0.00}{Hr: Applications}}}


Due to the complexity of the Hr EH model, there may be relatively limited application in the academic. Particularly, in \cite{ref_hr_lg_app_1} and \cite{ref_hr_lg_app_2}, both the Hr and Lg EH models were involved.

\begin{itemize}
\item \textbf{Relay SWIPT Systems:}
In \cite{ref_hr_lg_app_1}, the end-to-end achievable rate was maximized for MIMO decode-and-forward (DF) relay SWIPT system.
\end{itemize}

\begin{itemize}
\item \textbf{MIMO WPCN Systems:}
In \cite{ref_hr_lg_app_2}, the EE was maximized for wireless powered time division duplex (TDD) MIMO WPCN system.
\end{itemize}

\begin{table*}[t!] 
\renewcommand{\arraystretch}{1.40}
\centering 
\caption{Applications of the 2-Pw and 3-Pw nonleianr EH models} \label{tab_2pw_app}
\begin{tabular}{c|c|m{8.0cm}|m{2.5cm}<{\centering}}
\hline
\hline

\multicolumn{4}{c}{} \\
\multicolumn{4}{c}{\textbf{2-Pw nonleianr EH model}} \\
\hline
 \textbf{Network model} & \textbf{Scenario} & \makecell[c]{\textbf{Goal}} & \makecell[c]{\textbf{Method}} \\
	 \hline
 \makecell[c]{\multirow{7}{0.6cm}{SWIPT}} & Relay & Outage \cite{ref_pw1_app_1, ref_pw1_app_2, ref_pw1_app_5, ref_pw1_app_6, ref_pw1_app_7, pw2021app3, pw2021app12, pw2021app4, pw2021app5}, the source-destination mutual information \cite{pw2021app7} security \cite{pw2021app2}, bit error \cite{pw2021app14}, throughput \cite{pw2_2022_swipt5} & \multirow{5}{2.5cm}{Convex optimization theory, probability theory, mathematic} \\
 \cdashline{2-3}[0.8pt/1pt]
 & P2P & Achievable rate \cite{ref_pw1_app_9}, average harvested power \cite{pw2_2020_swipt1} & \\
 \cdashline{2-3}[0.8pt/1pt]
 & Multi-user & Sum rate\cite{pw2_2021_swipt2} & \\
 \cdashline{2-3}[0.8pt/1pt]
 & CR & Outage (physical layer security) \cite{pw2021app1, pw2021app6, pw2021app10, pw2021app13} & \\
 \cdashline{2-3}[0.8pt/1pt]
 & NOMA & Outage \cite{pw2021app8, pw2021app9, pw2021app11, pw2_2021_swipt3, pw2_2021_swipt4} & \\

 \hline

 \makecell[c]{\multirow{7}{0.6cm}{WPCN}} & Relay & The (secrecy) outage performance \cite{ref_pw2_app_3, ref_pw1_app_4} & \multirow{5}{2.5cm}{Convex optimization theory, Probability theory, mathematic} \\
 \cdashline{2-3}[0.8pt/1pt]
 & Multi-user & Sum-rate \cite{ref_pw1_app_8} & \\
 \cdashline{2-3}[0.8pt/1pt]
 & MIMO & Weighted sum of average harvested power \cite{pw2_2022_wpcn1,pw2_2021_wpcn1} & \\
 \cdashline{2-3}[0.8pt/1pt]
 & UAV-assisted & UAV's energy consumption \cite{pw2_2022_wpcn2} & \\
 \cdashline{2-3}[0.8pt/1pt]
 & NOMA & Outage \cite{pw2_2023app2} & \\

\hline
\hline

\multicolumn{4}{c}{} \\
\multicolumn{4}{c}{\textbf{3-Pw nonleianr EH model}} \\
\hline

 \textbf{Network model} & \textbf{Scenario} & \makecell[c]{\textbf{Goal}} & \makecell[c]{\textbf{Method}} \\
	 \hline
\makecell[c]{\multirow{10}{0.6cm}{SWIPT}} & Relay & Capacity \cite{ref_pw2_app_3, pw2_2021app1}, throughput \cite{ref_pw2_app_4}, harvested energy \cite{ref_pw2_app_6}, EE \cite{pw2_2021app3, pw2_2021app8}, outage \cite{pw2_2021app12, pw2_2021app14}, secrecy rate \cite{pw2_2021app13}, system mutual information \cite{pw3_2021_swipt2}, secrecy outage probability \cite{pw3_2021_swipt3}, outage and throughput \cite{pw3_2022_swipt5} & \multirow{7}{2.5cm}{Convex optimization theory, Probability theory, mathematic} \\
 \cdashline{2-3}[0.8pt/1pt]
 & P2P & EH efficiency \cite{pw2_2021app4}, achievable rate \cite{pw2_2021app11}, symbol-error rate \cite{pw2_2021app9}, average achievable rate \cite{pw3_2021_swipt1}, capacity \cite{pw3_2022_swipt4}, average harvested power \cite{pw3_2021_swipt6} & \\
 \cdashline{2-3}[0.8pt/1pt]
 & NOMA & Outage \cite{pw2_2021app5} & \\
 
 \cdashline{2-3}[0.8pt/1pt]
 & RIS-assisted & Block error rate \cite{pw3_2023app2} & \\
 
 \cdashline{2-3}[0.8pt/1pt]
 & URLLC-based IIoT & Outage \cite{pw3_2023app3} & \\
 
 \cdashline{2-3}[0.8pt/1pt]
 & Others & EE \cite{pw2_2021app6}, bit error rate \cite{pw2_2021app7} R-E region \cite{pw2_2021app16}& \\

 \hline
 
\makecell[c]{\multirow{9}{0.6cm}{WPCN}} & Relay & Outage \cite{ref_pw2_app_1}, AoI, end-to-end block error probability \cite{pw3_2022_wpcn1} & \multirow{6}{2.5cm}{Convex optimization theory, Probability theory, mathematic} \\
 \cdashline{2-3}[0.8pt/1pt]
 & MISO & Least-squares estimator \cite{pw2_2021app2} & \\
 \cdashline{2-3}[0.8pt/1pt]
 & NOMA & Outage \cite{pw2_2021app10}, computation EE \cite{pw2_2021app15} & \\
 \cdashline{2-3}[0.8pt/1pt]
 & Distributed WPT & Harvested power maximization \cite{pw3_2022_wpcn2} & \\
 \cdashline{2-3}[0.8pt/1pt]
 & RIS-empowered WPT & Throughput, outage probability, average harvested power \cite{pw3_2022_wpcn3} & \\
 \cdashline{2-3}[0.8pt/1pt]
 & WPT & Outage probability, average harvested energy \cite{pw3_2023app1} & \\
 
 \hline
\hline

 \end{tabular}
\end{table*}

\subsubsection{\textbf{\textcolor[rgb]{0.00,0.00,0.00}{2-Pw: Applications}}} 
The 2-Pw EH model have been studied mainly focusing on the SWIPT and WPCN systems, as shown in \ref{tab_2pw_app}. The SWIPT systems include relay \cite{ref_pw1_app_1, ref_pw1_app_2, ref_pw1_app_5, ref_pw1_app_6, ref_pw1_app_7, pw2021app3, pw2021app12, pw2021app4, pw2021app5, pw2021app7, pw2021app14, pw2_2022_swipt5}, P2P \cite{ref_pw1_app_9, pw2_2020_swipt1}, multi-user \cite{pw2_2021_swipt2}, CR \cite{pw2021app1, pw2021app6, pw2021app10, pw2021app13}, NOMA \cite{pw2021app8, pw2021app9, pw2021app11, pw2_2021_swipt3, pw2_2021_swipt4}, RIS \cite{pw3_2023app2}, and IIoT \cite{pw3_2023app3}. For the WPCN system, the relay \cite{ref_pw2_app_3, ref_pw1_app_4}, multi-user \cite{ref_pw1_app_8}, MIMO \cite{pw2_2022_wpcn1,pw2_2021_wpcn1}, UAV-assisted \cite{pw2_2022_wpcn2}, NOMA \cite{pw2_2023app2}, and WPT \cite{pw3_2023app1} scenarios are studied under the 2-Pw EH model.

\textbf{\textit{The SWIPT systems with the 2-Pw EH model:}}
The 2-Pw EH model has been studied so far in many various wireless communication systems, such as relay, P2P, multi-user, CR, and NOMA networks.

\begin{itemize}
\item \textbf{Relay SWIPT Systems:}
The relaying SWIPT systems were considered in \cite{ref_pw1_app_1, ref_pw1_app_2, ref_pw1_app_5, ref_pw1_app_6, ref_pw1_app_7}, where the system outage probability, the outage capacity, and the bit-error-rate-analysis were investigated correspondingly. In \cite{pw2021app2}, for the cooperative DF multi-relay networks, the outage probability and intercept probability of the system were investigated. In \cite{pw2021app3} and \cite{pw2021app12}, a multiuser overlay spectrum sharing system with using TS receiver was considered, where the outage performance of the system was analysis over the Rayleigh fading channels. In \cite{pw2021app4}, the multi-antenna multi-relay SWIPT-WPCN system was investigated with imperfect CSI, where the outage probability and the reliable throughput were derived. In \cite{pw2021app5}, the cooperative DF relay SWIPT system with spectrum sensing was considered, where the system outage probability was discussed. In \cite{pw2021app7}, a dual-hop AF MIMO relay communication system was studied, where the source-destination mutual information was maximized. In \cite{pw2021app14}, a wireless-powered dual-hop DF relaying system was considered, where the bit error performance was studied over the Nakagami-$m$ fading. In \cite{pw2_2022_swipt5} the two-hop cooperative DF relay SWIPT system with the direct link was investigated, where the expression of throughput was derived.

\item \textbf{P2P SWIPT Systems:}
In \cite{ref_pw1_app_9}, a P2P SWIPT system was involved, where the achievable rate was maximized. In \cite{pw2_2020_swipt1}, a new SWIPT and modulation classification scheme was presented, where the average harvested power was analyzed for different modulation formats over Rayleigh fading channels.

\item \textbf{Multi-user SWIPT Systems:}
In \cite{pw2_2021_swipt2}, a multi-user the combination of filter bank multi-carrier based SWIPT system was studied, where the sum rate was maximized by jointly optimizing time, power and weight allocations.

\item \textbf{CR SWIPT Systems:}
In \cite{pw2021app1}, the outage probability and throughput were studied for CR sensor network with SWIPT under Nakagami-$m$ fading. In \cite{pw2021app6}, the physical layer security was studied for an underlay CR EH network with imperfect CSI, where the secrecy outage probability was discussed. In \cite{pw2021app10}, the outage performance was studied for a hybrid SWIPT CR system over the Nakagami-$m$ fading, where both PS and TS receivers were used. In \cite{pw2021app13}, the outage and throughput performance of the secondary system were studied for three-phase SWIPT-enabled cooperative CR networks under Nakagami-$m$ fading.

\item \textbf{NOMA SWIPT Systems:}
In \cite{pw2021app8}, for a SWIPT-assisted NOMA system, an impartial user cooperation mechanism with the PS architecture was presented, which outperformed the traditional partial cooperation mechanism in terms of outage probability, diversity order and diversity-multiplexing trade-off. In \cite{pw2021app9}, a downlink FD relayed NOMA SWIPT system with the PS architecture was considered, where the outage probability and throughput were investigated. In \cite{pw2021app11}, a wireless powered cooperative NOMA system was studied, where the outage probability and the average throughput were derived with the low-complexity antenna selection and relay selection. In \cite{pw2_2021_swipt3} and \cite{pw2_2021_swipt4}, a FD cooperative NOMA SWIPT relay system was considered, where the outage probability was derived and discussed.

\end{itemize}

\textbf{\textit{The WPCN systems with the 2-Pw EH model:}}
In WPCN system, some works have been carried out in relay, multi-user, MIMO, UAV-assisted scenarios so far with the 2-Pw EH model, see. e.g., \cite{ref_pw2_app_3, ref_pw1_app_8, pw2_2022_wpcn1,pw2_2021_wpcn1, pw2_2022_wpcn2}.

\begin{itemize}
\item \textbf{Relay WPCN systems:}
In \cite{ref_pw2_app_3}, the outage probability was analyzed for a two-hop FD DF relay SWIPT system with the TS protocol. In \cite{ref_pw1_app_4}, the secrecy outage probability performance was studied in multi-antenna relaying WPCN system.

\item \textbf{Multi-User WPCN systems:}
In \cite{ref_pw1_app_8}, the system sum-rate was maximized in the uplink for multi-user TDMA WPCN system.

\item \textbf{MIMO WPCN systems:}
In \cite{pw2_2022_wpcn1} and \cite{pw2_2021_wpcn1}, authors considered multi-user multi-antenna MIMO WPT systems, where the weighted sum of the average harvested powers was maximized.

\item \textbf{UAV-assisted WPCN systems:}
In \cite{pw2_2022_wpcn2}, the rotary-wing UAV-assisted wireless powered fog computing system was studied, where the UAV’s energy consumption was minimized by jointly optimizing the UAV’s trajectory, the task offloading allocation and computing resource allocation.


\item \textbf{NOMA WPCN systems:}
In \cite{pw2_2023app2},the outage performance was studied for a dual-hop free space optical-RF communication system, where the random-maximum and minimum-maximum antenna selection schemes were presented.

\end{itemize}

\subsubsection{\textbf{\textcolor[rgb]{0.00,0.00,0.00}{3-Pw Applications}}}
The 3-Pw EH model have been studied for the SWIPT and WPCN systems, as shown in \ref{tab_2pw_app}. The SWIPT systems include relay \cite{ref_pw2_app_3, pw2_2021app1, ref_pw2_app_4, ref_pw2_app_6, pw2_2021app3, pw2_2021app8, pw2_2021app12, pw2_2021app14, pw2_2021app13, pw3_2021_swipt2, pw3_2021_swipt3, pw3_2022_swipt5}, P2P \cite{pw2_2021app4, pw2_2021app11, pw2_2021app9, pw3_2021_swipt1, pw3_2022_swipt4, pw3_2021_swipt6}, NOMA \cite{pw2_2021app5} and others \cite{pw2_2021app6, pw2_2021app7, pw2_2021app16}. For the WPCN system, the relay \cite{ref_pw2_app_1, pw3_2022_wpcn1}, MISO \cite{pw2_2021app2}, NOMA \cite{pw2_2021app10, pw2_2021app15}, distributed WPT \cite{pw3_2022_wpcn2}, and RIS-Empowered WPT \cite{pw3_2022_wpcn3} scenarios are studied with the 3-Pw EH model.

\textbf{\textit{The SWIPT systems with the 3-Pw EH model:}}
\begin{itemize}
\item \textbf{Relay SWIPT systems:}
The relaying SWIPT networks was considered in \cite{ref_pw2_app_3, ref_pw2_app_4} and \cite{ref_pw2_app_6}, where the maximizations of the system capacity, throughput and EH were corresponding involved, as well as the minimization of the outage probability. In \cite{pw2_2021app1}, the capacity of the system was maximized for SWIPT based three-step two-way DF relay network, where the outage probability with the optimal dynamic PS ratios was discussed. In \cite{pw2_2021app3}, the EE was maximized for SWIPT enabled two-way DF relay network by jointly optimizing the transmit powers, the PS ratios, and the transmission time. In \cite{pw2_2021app8}, the reliability, goodput and EE performance were studied for EH based FD cooperative IoT network for short-packet communications, where the approximate closed-form expressions of the block error rate and outage probability were derived. In \cite{pw2_2021app12}, the outage probability was studied for an FD-AF relay-based system, where the outage throughput and EE of the system were maximized by optimizing the TS parameter. In \cite{pw2_2021app13}, the secrecy rate was maximized for an EH cooperative relay network with multiple eavesdroppers, where the transmission power was optimized with the imperfect self-interference cancellation. In \cite{pw2_2021app14}, the outage probability of the destination node was minimized for SWIPT network with incremental DF relaying protocol, where the system throughput was also derived. In \cite{pw3_2021_swipt2}, a dual-hop DF multicasting MIMO EH-enabled relay system was considered, where the system mutual information was maximized. In \cite{pw3_2021_swipt3}, the secure decode-and-forward (DF) cooperative relay SWIPT network was explored, where the secrecy outage probability was investigated. In \cite{pw3_2022_swipt5}, the relay selection was investigated in dual-hop DF cooperative relay SWIPT network, where the outage and throughput performance were involved.

\item \textbf{P2P SWIPT systems:}
In \cite{pw2_2021app4} and \cite{pw2_2021app11}, to enhance EH efficiency and achievable rate for SWIPT system, a new heterogeneous reconfigurable energy harvester was presented. In \cite{pw2_2021app9}, a novel transmitter-oriented dual-mode SWIPT with deep LSTM RNN-based adaptive mode switching was presented for a P2P SWIPT system over frequency-flat fading channel, where the symbol-error rate performance was discussed for both single-tone and multi-tone SWIPT. In \cite{pw3_2021_swipt1}, the average achievable rate was maximized for a P2P SWIPT system, where a heterogeneously reconfigurable energy harvester was presented. In \cite{pw3_2022_swipt4}, the authors investigated two coherent detection schemes for integrated SWIPT receivers, where the fundamental capacity of the system was studied. In \cite{pw3_2021_swipt6}, a simultaneous wireless power transfer and modulation classification receiver architecture were presented, where the average harvested power was derived for six-class of modulations over AWGN and Rayleigh fading channels.

\item \textbf{NOMA systems:}
In \cite{pw2_2021app5}, the outage probability and the throughput were studied for the SWIPT based NOMA network, where an incremental cooperative NOMA protocol was presented.

\item \textbf{RIS-assisted systems:}
In \cite{pw3_2023app2}, the block error rate was studied in an RIS-assisted multi-user FD communication system, where the short packet transmission was involved.

\item \textbf{URLLC-based IIoT systems:}
In \cite{pw3_2023app3}, the ultra-reliable low latency communication (URLLC)-based cooperative IIoT network was considered, where the outage probability and block error rate were involved. 

\item \textbf{Other SWIPT systems:}
In \cite{pw2_2021app6}, the sum EE of all D2D links was maximized for a D2D underlaid cellular network by optimizing the allocation of spectrum resource and transmit power. In \cite{pw2_2021app7}, the bit error rate performance of the system was investigated for FD overlay CR network with spectrum sharing protocol. In \cite{pw2_2021app16}, the R-E region of the system was investigated for a monostatic wireless powered backscatter communication system with considering PS and TS receivers in a comprehensive way.

\end{itemize}

\textbf{\textit{The WPCN systems with the 3-Pw EH model:}}
\begin{itemize}
\item \textbf{Relay WPCN systems:}
In \cite{ref_pw2_app_1}, considering a relay-assisted FD two-hop WPCN system, the system throughput was maximized and the system outage probability was minimized. In \cite{pw3_2022_wpcn1}, the age of information (AoI) was studied for an AF relay system, where the end-to-end block error probability was involved.

\item \textbf{MISO WPCN systems:}
In \cite{pw2_2021app2}, the practical efficacy of using a single RF chain at a large antenna array power beacon was studied for MISO WPT system over Rician fading channels, where the average harvested power at the user was derived and further discussed.

\item \textbf{NOMA WPCN systems:}
In \cite{pw2_2021app10}, the outage performance of the system was studied for wireless-powered cooperative spectrum sharing system with FD and NOMA transmissions, where the rate of the secondary network was maximized by joint designing beamformer and time-split parameter. In \cite{pw2_2021app15}, the system computation EE was maximized for WPT enabled NOMA-based MEC network, which jointly optimized the computing frequencies and execution time of the MEC server and IoT devices, the offloading and EH time, and the transmit power of each IoT device and beacon.

\item \textbf{Distributed WPT system:}
In \cite{pw3_2022_wpcn2}, the distributed multi-antenna energy beamforming scheme was studied for multi-antenna WPT system over frequency-selective fading channels under joint total and individual power constraints, where the harvested power was maximized.

\item \textbf{RIS-empowered WPT system:}
In \cite{pw3_2022_wpcn3}, a RIS-empowered WPT system was considered, where the closed-formed expressions of throughput, outage probability and average harvested power were derived.

\item \textbf{WPT system:}
In \cite{pw3_2023app1}, the average harvested energy and outage performance were investigated for wireless powered communication systems, where Beaulieu-Xie fading model was presented to characterize the manifold of channels.


\end{itemize}

\begin{table*}[t!] 
\renewcommand{\arraystretch}{1.40}
\centering 
\caption{Applications of the Mlg, 2-ord and Fr nonleianr EH models} \label{tab_multimd_app}
\begin{tabular}{c|c|m{8.0cm}|m{2.5cm}<{\centering}}
\hline
\hline

%
%

\multicolumn{4}{c}{} \\
\multicolumn{4}{c}{\textbf{Mlg nonleianr EH model}} \\
\hline
 \textbf{Network model} & \textbf{Scenario} & \textbf{\makecell[c]{Goal}} & \textbf{\makecell[c]{Method}} \\
	 \hline
	 \makecell[c]{\multirow{7}{0.6cm}{SWIPT}} & MISO & Transmit power minimization \cite{ref_lg2_app_1, ref_lg2_app_2} & \multirow{5}{2.5cm}{Convex optimization theory, Lagrangian, mathematic} \\
 \cdashline{2-3}[0.8pt/1pt]
 & Relay & EE fairness \cite{ref_lg2_app_4} & \\
 \cdashline{2-3}[0.8pt/1pt]
 & CR & Transmit power/harvested power \cite{ref_lg1_app_39} & \\
 \cdashline{2-3}[0.8pt/1pt]
 & IRS-assisted & Achievable rate and harvested energy \cite{mlg_2021_swipt1}, achievable data rate \cite{mlg_2022_swipt2} & \\
 \cdashline{2-3}[0.8pt/1pt]
 & Cell-free & Achievable rate and harvested energy \cite{mlg_2022_swipt3} & \\ 
 \hline
 \makecell[c]{\multirow{5.5}{0.6cm}{WPCN}} & Cellular & Transmit energy minimization \cite{lg2_2021app1} & \multirow{4}{2.5cm}{Convex optimization theory} \\
 \cdashline{2-3}[0.8pt/1pt]
 & NOMA & Computation efficiency \cite{lg2_2021app2, lg2_2021app3} & \\
 \cdashline{2-3}[0.8pt/1pt]
 & Secure & Secure EE \cite{mlg_2021_wpcn1} & \\
 \cdashline{2-3}[0.8pt/1pt]
 & Ambient & Outage \cite{lg_2023app3}, secrecy outage and EE \cite{lg_2023app5} & \\

\hline
\hline
\multicolumn{4}{c}{} \\
\multicolumn{4}{c}{\textbf{2-ord nonleianr EH model}} \\
\hline

 \textbf{Network model} & \textbf{Scenario} & \makecell[c]{\textbf{Goal}} & \makecell[c]{\textbf{Method}} \\
	 \hline
\makecell[c]{\multirow{1}{0.6cm}{WPCN}} & WSN & Signal reconstruction error \cite{ref_2rd_app_1, 2rd_2021app1} & \multirow{1}{2.5cm}{Probability theory} \\

\hline
\hline

\multicolumn{4}{c}{} \\
\multicolumn{4}{c}{\textbf{Fr nonleianr EH model}} \\
\hline

\textbf{Network model} & \textbf{Scenario} & \makecell[c]{\textbf{Goal}} & \makecell[c]{\textbf{Method}} \\
	 \hline
 \makecell[c]{\multirow{5}{0.6cm}{SWIPT}} & Relay & Outage \cite{ref_frac_app_1}, capacity \cite{ref_frac_app_1}, target rate \cite{fr_2021app3}, instantaneous rate \cite{fr_2021app2}, data rate \cite{frc_2022_swipt1} & \multirow{3}{2.5cm}{Convex optimization theory} \\
 \cdashline{2-3}[0.8pt/1pt]
 & MISO & SINR and harvested energy \cite{fr_2021app4} & \\
 \cdashline{2-3}[0.8pt/1pt]
 & Secure & Secrecy throughput \cite{fr_2021app5} & \\

 \hline

 \makecell[c]{\multirow{7}{0.6cm}{WPCN}} & Relay & Outage \cite{fr_2021app6}, throughput \cite{fr_2021app7, fr_2021app10}, outage \cite{frc_2021_wpcn2}, transmission time minimization \cite{frc_2021_wpcn3} & \multirow{5}{2.5cm}{Convex optimization theory, probability theory} \\
 \cdashline{2-3}[0.8pt/1pt]
 & MIMO & LMMSE \cite{fr_2021app9} & \\
 \cdashline{2-3}[0.8pt/1pt]
 & MEC & Weighted sum computation bits \cite{fr_2021app8}, computational EE fairness \cite{frc_2021_wpcn1} & \\
 \cdashline{2-3}[0.8pt/1pt]
 & IRS-assisted & Sum throughput \cite{frc_2022_wpcn1} & \\ 
 \cdashline{2-3}[0.8pt/1pt]
 & NOMA & Outage \cite{fr_2023app1} & \\ 

\hline
\hline

\multicolumn{4}{c}{} \\
\multicolumn{4}{c}{\textbf{Log nonleianr EH model}} \\
\hline

\textbf{Network model} & \textbf{Scenario} & \makecell[c]{\textbf{Goal}} & \makecell[c]{\textbf{Method}} \\
	 \hline
 \makecell[c]{\multirow{1}{0.6cm}{WPCN}} & MEC & max-min energy fairness and sum power \cite{han2021joint_log_aap_1} & Convex optimization theory \\ 

\hline
\hline

 \end{tabular}
\end{table*}

\subsubsection{\textbf{\textcolor[rgb]{0.00,0.00,0.00}{Mlg Applications}}}
The Mlg EH model have been studied for the SWIPT and WPCN systems, as shown in \ref{tab_multimd_app}. The SWIPT systems include cellular \cite{ref_lg2_app_1, ref_lg2_app_2}, relay \cite{ref_lg2_app_4}, CR \cite{ref_lg1_app_39}, IRS-assisted \cite{mlg_2021_swipt1, mlg_2022_swipt2}, and cell-free massive MIMO \cite{mlg_2022_swipt3} scenarios. For the WPCN system, the MISO \cite{lg2_2021app1}, NOMA \cite{lg2_2021app2, lg2_2021app3}, secure \cite{mlg_2021_wpcn1}, and ambient backscatter \cite{lg_2023app3, lg_2023app5} scenarios are studied with the Mlg EH model.

\textbf{\textit{The SWIPT systems with the Mlg EH model:}}
\begin{itemize}
\item \textbf{MISO SWIPT systems:}
In \cite{ref_lg2_app_1} and \cite{ref_lg2_app_2}, the multi-user MISO SWIPT systems were considered, where the transmit power was minimized.

\item \textbf{Relay SWIPT systems:}
In \cite{ref_lg2_app_4}, the multi-pair AF relay SWIPT system was studied, where the max-min EE fairness was involved.

\item \textbf{IRS-assisted SWIPT systems:}
In \cite{mlg_2021_swipt1}, an IRS-assisted SWIPT system was considered, where the achievable rate and harvested energy were studied with considering the TS and PS protocols. In \cite{mlg_2022_swipt2}, a P2P IRS-assisted MIMO-OFDM SWIPT system was investigated, where the achievable data rate was maximized.

\item \textbf{Cell-free massive MIMO SWIPT systems:}
In \cite{mlg_2022_swipt3}, a cell-free massive MIMO SWIPT system was considered, where the downlink harvested energy and downlink/uplink achievable rates were studied.


\end{itemize}

\textbf{\textit{The WPCN systems with the Mlg EH model:}}
\begin{itemize}
\item \textbf{Cellular WPCN systems:}
In \cite{lg2_2021app1}, the total transmitted energy was minimized for wireless-powered M2M multicasting in cellular network, where the routing and the scheduling of multicast messages and the scheduling of EH were considered.

\item \textbf{NOMA WPCN systems:}
In \cite{lg2_2021app2} and \cite{lg2_2021app3}, the computation efficiency was maximized for wireless-powered NOMA MEC networks under both partial and binary computation offloading modes, where the energy harvesting time, the local computing frequency, the operation mode selection, the offloading time and power were jointly optimized.

\item \textbf{Secure WPCN systems:}
In \cite{mlg_2021_wpcn1}, authors considered a multi-tag wireless-powered backscatter communication network with an eavesdropper, where the max-min EE problem was studied with imperfect CSI.

\item \textbf{Ambient systems:}
In \cite{lg_2023app3}, the outage probability was investigated for the multi-tag ambient backscatter communication system over time-selective fading channel, where the effect of dynamic reflection coefficient was involved. In \cite{lg_2023app5}, the wireless powered ambient backscatter cooperative DF relay communication system was considered, where the secrecy outage probability and EE were studied.

\end{itemize}

\subsubsection{\textbf{\textcolor[rgb]{0.00,0.00,0.00}{2-ord Applications}}} \label{2ord_app}
So far, the 2-ord EH model has been studied in WPCN systems, see e.g., \cite{ref_2rd_app_1, 2rd_2021app1}.

\begin{itemize}
\item \textbf{WPT-enabled WSN systems:}
In \cite{ref_2rd_app_1} and \cite{2rd_2021app1}, the signal reconstruction error was minimized for a wireless sensor network by using a deep reinforcement learning (DRL) based approach, where multi-layer neural networks were utilized.

\end{itemize}

\subsubsection{\textbf{\textcolor[rgb]{0.00,0.00,0.00}{Fr: Applications}}} 
The Fr EH model have been studied for the SWIPT and WPCN systems, as shown in \ref{tab_multimd_app}. The SWIPT systems include relay \cite{ref_frac_app_1, ref_frac_app_1, fr_2021app3, fr_2021app2, frc_2022_swipt1}, MISO \cite{fr_2021app4}, and secure \cite{fr_2021app5} scenarios. For the WPCN system, the relay \cite{fr_2021app6, fr_2021app7, fr_2021app10, frc_2021_wpcn2, frc_2021_wpcn3}, MIMO \cite{fr_2021app9}, MEC \cite{fr_2021app8, frc_2021_wpcn1}, IRS \cite{frc_2022_wpcn1}, and NOMA \cite{fr_2023app1} scenarios are studied with the Fr EH model.

\textbf{\textit{The SWIPT systems with the Fr EH model:}}
\begin{itemize}
\item \textbf{Relay SWIPT systems:}
In \cite{ref_frac_app_1}, the system outage probability and throughput performance was studied in AF relay SWIPT networks. In \cite{ref_frac_app_2}, the system capacity was maximized in DF relay SWIPT networks. In \cite{fr_2021app1}, the system capacity was maximized for a DF relay SWIPT system with the “harvest-then-forward” scheme. In \cite{fr_2021app4}, the target rate was maximized for disaster-recovery communications utilizing SWIPT enabled DF relay D2D network by jointly optimizing the transmit power, the PS ratio and the location of the relay. In \cite{fr_2021app3}, the instantaneous rate was maximized for wireless powered cooperative communications system by considering multiple wireless powered AF self-energy recycling relays and relay selection. In \cite{frc_2022_swipt1}, a two-way relaying device-to-device (D2D) model sharing the same resources with the underlying cellular network was investigated, where the data rate was maximized.

\item \textbf{MISO SWIPT systems:}
In \cite{fr_2021app2}, the optimal transmit beamforming was designed to maximize the received SINR and the harvested energy for MISO downlink SWIPT system with specific absorption rate constraints.

\item \textbf{Secure SWIPT systems:}
In \cite{fr_2021app5}, the secrecy throughput of the system was maximized for SWIPT-enabled network, where a PS-based FD jamming scheme was presented.

\end{itemize}

\textbf{\textit{The WPCN systems with the Fr EH model:}}
\begin{itemize}
\item \textbf{Relay WPCN systems:}
In \cite{fr_2021app6}, the outage probability performance was studied for a WPCN network to explore the problem of the $k$-th best relay (device) selection. In \cite{fr_2021app7} and \cite{fr_2021app10}, the max-min throughput was maximized for a wireless powered IoT (relay) network to ensure the fairness among multiple sensor nodes. In \cite{frc_2021_wpcn2}, the outage performance was studied for a WPCN system with batteryless devices, where several selection/scheduling schemes were involved corresponding to different implementation complexities and CSI requirements. In \cite{frc_2021_wpcn3}, a hybrid transmission scheme of wireless powered active communications and passive backscatter communications was presented, where the total transmission time was minimized.

\item \textbf{MIMO WPCN systems:}
In \cite{fr_2021app9}, the uplink SE was investigated for wireless uplink information and downlink power transfer in cell-free massive MIMO systems with considering Rician fading and maximum ratio processing based on either linear minimum mean-squared error (LMMSE) or least-squares channel estimation.

\item \textbf{MEC WPCN systems:}
In \cite{fr_2021app8} and \cite{frc_2021_wpcn1}, the weighted sum computation bits and the computational EE fairness were respectively maximized for a backscatter assisted wirelessly powered MEC network, where the partial offloading scheme was considered at each users. 

\item \textbf{IRS-assisted WPCN systems:}
In \cite{frc_2022_wpcn1}, an IRS-assisted wireless powered IoT network was considered, where the sum throughput was maximized.

\item \textbf{NOMA WPCN systems:}
In \cite{fr_2023app1}, the system outage performance was studied for an uplink WPT-NOMA in IoT network, where a hybrid hybrid successive interference cancellation (SIC) scheme was presented.

\end{itemize}

\subsubsection{\textbf{\textcolor[rgb]{0.00,0.00,0.00}{Log Applications}}}
So far, the Log EH model has been studied in WPCN systems, see e.g., \cite{han2021joint_log_aap_1}.

\begin{itemize}
\item \textbf{MEC WPCN systems:}
In \cite{han2021joint_log_aap_1}, two resource allocation problems were studied for OFDMA-based WPT-MEC systems, i.e., a max–min energy fairness problem and a power sum maximization problem. The Log EH model is used, which lead to nonconvex mixed-integer nonlinear programming (MINLP) problems.
\end{itemize}

\subsubsection{\textbf{\textcolor[rgb]{0.00,0.00,0.00}{Multiple EH Models Together: Applications}}}
For comparison, several EH models are explored in a SWIPT system at the same time, see e.g., \cite{2022_all_mod1, 2022_all_mod2, 2022_all_mod3, 2022_all_mod4}.

\begin{itemize}
\item \textbf{SWIPT systems:}
In \cite{2022_all_mod1}, this work compared different linear and nonlinear EH models (i.e., the Lg, 3-Pw, Mlg, 2-ord EH models) to study the sensitivity and the nonlinearity of the harvester for OFDM SWIPT system, where the probability of successful SWIPT reception was discussed. In \cite{2022_all_mod2}, this paper studied the optimization of multi-tone signals for co-existing wireless power and information transfer system with the different nonlinear EH models (i.e., the 2-ord, Lg, Hr, 2-Pw and 3-Pw EH models). In \cite{2022_all_mod3}, this paper considered a two-hop SWIPT AF MIMO relay communication system with imperfect CSI and the nonlinear EH models (i.e., the Lg, 3-Pw and Hr EH models), where the mutual information between the source and destination nodes was maximized. In \cite{2022_all_mod4}, a multi-hop clustering routing protocol was presented for EH-enabled CR sensor networks, where the various curvilinear EH models (i.e., the Lg, Hr, 2-ord, Mlg, Fr EH models) were fit with the original data \cite[9]{2022_all_mod4}, and found that the Hr nonlinear EH model was best and used among the fitting results.

\end{itemize}

\section{\textcolor[rgb]{0.00,0.00,0.00}{Challenges of Utilizing Nonlinear EH Models}} \label{section_challenge_apps_ehmodels}

\begin{figure*}[htp!]
\centering
\includegraphics[width=0.98\textwidth]{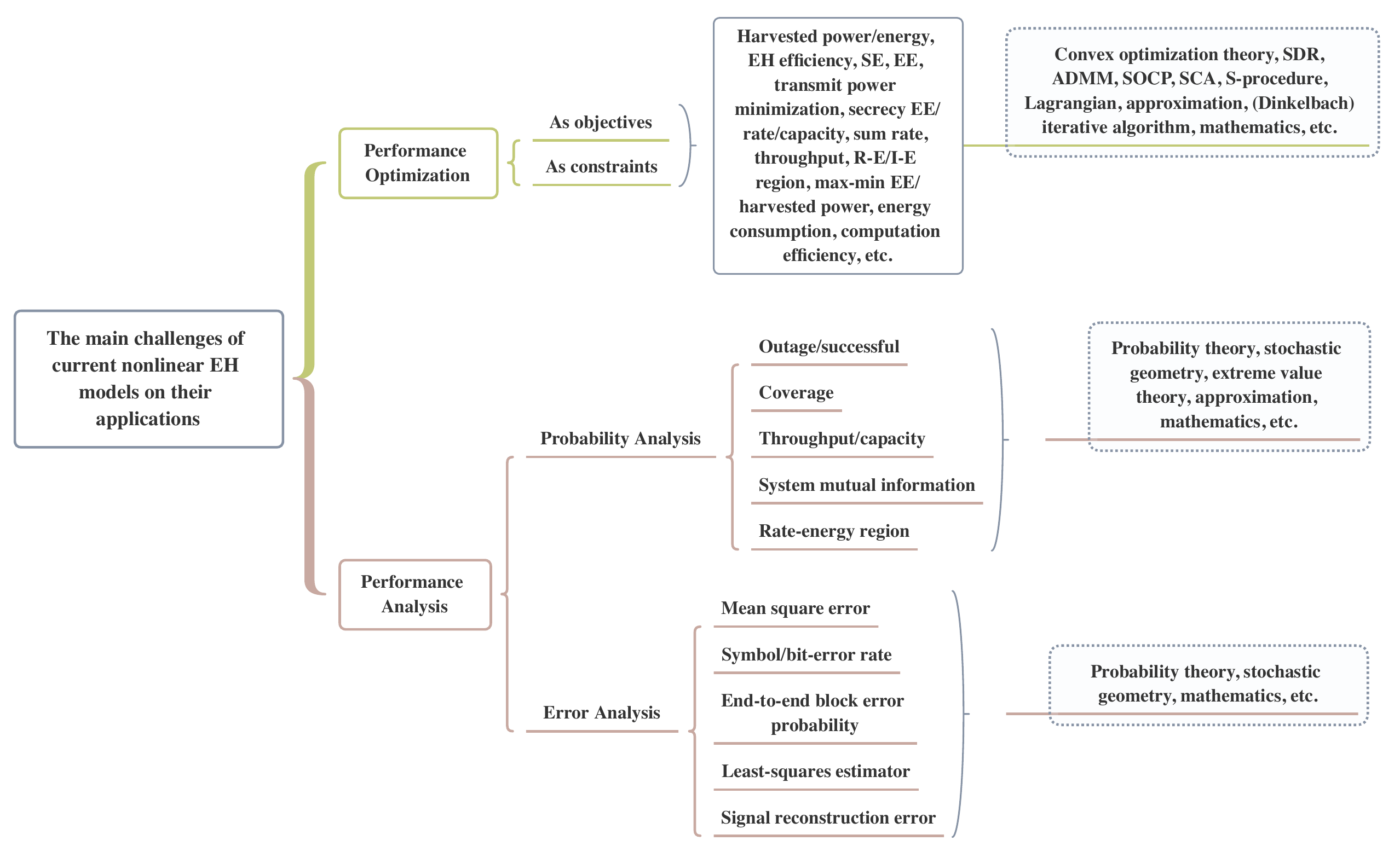}
\caption{\textcolor[rgb]{0.00,0.00,0.00}{The main application challenges of the nonlinear EH models}}
\label{challenges_models}
\end{figure*}

Based on the previous section, the application challenges of each nonlinear EH model can be mainly summarized into two aspects: the optimization of the system performance and the analysis of the system performance, as shown in Fig. \ref{challenges_models}.

On the optimization of the system performance, the study of RF-based EH with the nonlinear model is generally used as an objective function or constraint in an optimization problem, which brings different difficulties and challenges. For example, the optimization problems, such as the harvested power/energy, EH efficiency, SE, EE, transmit power minimization, secrecy EE/rate/capacity, sum rate, throughput, R-E/I-E region, max-min EE/harvested power, energy consumption, computation efficiency, etc., are normally nonconvex and cannot be solved directly. Correspondingly, the methods, based on the convex optimization theory, semidefinite relaxation (SDR), alternating direction method of multipliers (ADMM), second-order cone program (SOCP), successive convex approximation (SCA), S-procedure, Lagrangian, approximate, iterative algorithm, mathematics, etc., are presented and used to solve these optimization problems.

On the analysis of the system performance, we divide it into two parts: the system probability analysis and the system error analysis. For the system probability analysis, the outage/successful transmission, coverage, throughput/capacity, mutual information, R-E region, etc. of the system are often to be studied in academia. For the system error analysis, the related works mainly focus on the mean square error, symbol/bit-error rate, end-to-end block error probability, least-squares estimator signal reconstruction error, etc. Unsurprisingly, the presence of certain nonlinear EH models, such as Lg, Mlg and 3-Pw, poses significant challenges to the theoretical derivation and analysis of the system. To tackle these challenges, various well-established theories are commonly employed, including probability theory, stochastic geometry, extreme value theory, approximation, and game theory etc.




To provide a clearer description, we first classify the nonlinear EH models into four distinct groups based on their forms: $\{$Lg, Mlg, Ef$\}$, $\{$2-Pw, 3-Pw$\}$, $\{$Log$\}$, and $\{$Jm$\}$. Subsequently, we shall introduce each group respectively.

\subsection{\textcolor[rgb]{0.00,0.00,0.00}{Application Challenges: Lg, Mlg and Ef Models}}

\subsubsection{Performance optimization}
Due to the nonconvex and nonconcave properties of the Lg, Mlg and Ef models, the studied optimization problems are not convex, which results in being solved and obtaining some analytical solutions not easily. Considering the similar form of the Lg, Mlg and Ef models, we will take the Lg nonlinear EH model as an example in the following introduction. In particular, since the Ef model contains the error function, i.e., $\textrm{erf}(x) = \frac{2}{\sqrt{\pi}} \int_0^x e^{-t^2} \textit{dt}$, there exist more challenges in practical application.

\begin{itemize}
\item \textbf{Used in objective functions}
\end{itemize}

Many works take the (sum/total) harvested energy as the studied goal with the Lg model in \eqref{mod_Logi}, which is given by
\begin{flalign} \label{lg_pb1}
\max_{P_{\textrm{in},i}, ...} & \,\,\,\, \sum\limits_{i=1}^{N} \frac{\frac{{\rm Q}_{\max}}{1+e^{-a(P_{\textrm{in},i} - b)}} - \frac{{\rm Q}_{\max}}{1+e^{a b}}} {1-\frac{1}{1+e^{a b}}} \\
\textrm{s.t.} & \,\,\,\,\,\, ... \notag
\end{flalign}
where $P_{\textrm{in}, n}$ is an optimal variable and $N$ is the number of EH receivers. To address this kind of optimization problem, the following approaches can be adopted.

\begin{itemize}
\item[(a)] \textit{Function simplification:} Since the term of $\frac{{\rm Q}_{\max}}{1+e^{a b}}$ and $\left( 1-\frac{1}{1+e^{a b}} \right)$ are constant in \eqref{lg_pb1}, which have no effect on the system design. Therefore, the problem in \eqref{lg_pb1} can be rewritten as
\begin{flalign}
\max_{P_{\textrm{in}, i}, ...} & \,\,\,\, \sum\limits_{i=1}^{N} \frac{{\rm Q}_{\max}}{1+e^{-a(P_{\textrm{in},i} - b)}} \\
\textrm{s.t.} & \,\,\,\,\,\, ... \notag
\end{flalign}
Then, fractional programming can be adopted to solve the related problems. Besides, when the objective is the sum of the harvested energy, it becomes a nonlinear sum-of-ratios problem, which can be solved by using the iterative algorithm presented in \cite{frac_opt_1, ref_mod_lg1_0}.

\item[(b)] \textit{First-order approximation:} The objective function can be approximated by the first-order Taylor expansion function of ${\rm Q}_{\rm Lg} (P_{\textrm{in},i})$, which is expressed as
\begin{flalign}
{\rm Q}_{\rm Lg} (P_{\textrm{in},i}) \le {\rm Q}_{\rm Lg}^{\prime} (P_{\textrm{in},i}),
\label{approx_eh_lg}
\end{flalign}
where ${\rm Q}_{\rm Lg}^{\prime} (P_{\textrm{in},i}) = {\rm Q}_{\rm Lg} (\bar{P}_{\textrm{in},i}) + \partial_{P_{\textrm{in},i}} {\rm Q}_{\rm Lg} (\bar{P}_{\textrm{in},i}) (P_{\textrm{in},i} - \bar{P}_{\textrm{in},i})$, with $\partial_{P_{\textrm{in},i}} {\rm Q}_{\rm Lg} ( P_{\textrm{in},i})$ being the derivative of ${\rm Q}_{\rm Lg} (P_{\textrm{in},i})$ w.r.t. $P_{\textrm{in},i}$, and $\bar{P}_{\textrm{in},i}$ being a feasible point of the problem\footnote{The first-order Taylor expansion function of ${\rm Q}_{\rm Lg} (P_{\textrm{in}})$ can also be expressed in other forms of its transformation \cite{ref_lg1_app_55}.}. Thus, the problem in \eqref{lg_pb1} can be transformed as
\begin{flalign}
\max_{P_{\textrm{in}, i}, ...} & \,\,\,\, \sum\nolimits_{i=1}^{N} {\rm Q}_{\rm Lg}^{\prime} (P_{\textrm{in},i}) \\
\textrm{s.t.} & \,\,\,\,\,\, ... \notag
\end{flalign}
And then, the problem can be solved by SCA-based algorithm \cite{2004Convex, 2017Convex}.
\end{itemize}

\begin{itemize}
\item \textbf{Used in constraint}
\end{itemize}

As a constraint condition of optimization problems, the harvested energy is usually assumed to be less than or equal to a certain threshold. That is,
\begin{flalign}
\max_{P_{\textrm{in}, i}, ...} & \,\,\,\, ... \\
\textrm{s.t.} & \,\,\,\,\,\, {\rm Q}_{\rm Lg} (P_{\textrm{in},i}) \le q_0 \notag, \forall i = 1, 2, ..., N
\label{lg_pb2}
\end{flalign}
To deal with this type of optimization problem, the following ways can be used.

\begin{itemize}
\item[(a)] \textit{First-order approximation:} Through variable substitution and first-order approximation in \eqref{approx_eh_lg}, the corresponding constraints can be converted into convex. So, the the problem in \eqref{lg_pb2} is transformed as
\begin{flalign}
\max_{P_{\textrm{in}, i}, ...} & \,\,\,\, ... \\
\textrm{s.t.} & \,\,\,\,\,\, {\rm Q}_{\rm Lg}^{\prime} (P_{\textrm{in},i}) \le q_0 \notag, \forall i = 1, 2, ..., N
\end{flalign}
Then, the problem can be solved by the SCA method.

\item[(b)] \textit{Inverse function:} Let $\mathcal{P}_{\rm in}(\cdot)$ be the inverse function of ${\rm Q}_{\rm Lg} (P_{\rm in})$,\footnote{The inverse function of ${\rm Q}_{\rm Lg} (P_{\rm in})$ can be given by $\mathcal{P}_{\rm in} ({\rm Q}_{\rm Lg}) = b- \frac{1}{a} \ln \left( \frac{e^{a b} ({\rm Q}_{\max} - {\rm Q}_{\rm Lg}) }{e^{a b} {\rm Q}_{\rm Lg} + {\rm Q}_{\max}} \right)$.} the related constraints can be transformed to be convex. Thus, the problem in \eqref{lg_pb2} is reformulated as
\begin{flalign}
\max_{P_{\textrm{in}, i}, ...} & \,\,\,\, ... \\
\textrm{s.t.} & \,\,\,\,\,\, P_{\textrm{in},i} \le \mathcal{P}_{\rm in}(q_0) \notag, \forall i = 1, 2, ..., N
\end{flalign}
which can be efficiently solved via some optimization tools such as CVX \cite{2008CVX}.
\end{itemize}

\subsubsection{Performance analysis}
For the analysis of system performance, Rayleigh, Rice and Nakagami-$m$ fading channel are generally considered \cite{ref_lg1_app_35, ref_lg1_app_51, lg1_swipt_2021app18,lg_2021_swipt3,lg1_2021app11}. Due to the complexity of different channel models, the difficulty of analyzing system performance is gradually increasing. Meanwhile, the complexity of the Lg model also can bring even more challenges to system performance analysis, which makes it harder to derive closed form results. Additionally, in the different systems studied in existing works, the used methods for theoretical derivation are not the same, including approximation, variable substitution, some theories in special cases, etc. For example, to address this issue, in \cite{ref_lg1_app_13}, the Lg model can be rewritten as 
\begin{flalign}
{\rm Q}_{\rm Lg} (P_{\rm in}) &= \frac{ {\rm Q}_{\max} (1-e^{-a P_{\rm in}})}{1+e^{-a(P_{\rm in}-b)}}.
\end{flalign}
Then, with the help of some auxiliary variables and formula transformation, these operations can contribute to some theoretical derivation and analysis.

In \cite{lg1_2021app11}, Bernstein-type inequality approach was adopted to approximate the outage probabilistic constraints conservatively. That is, suppose the chance constraint
\begin{flalign}
\Pr \{ \bm{x}^H \bm{A} \bm{x} + 2\mathcal{R}\{ \bm{x}^H \bm{r} \} + \theta \ge 0 \} \ge 1 - \alpha,
\end{flalign}
where $(\bm{A}, \bm{r}, \theta) \in \mathbb{R}^N \times \mathbb{C}^N \times \mathbb{R}, \bm{x} \sim \mathcal{CN}(0,1)$ and $\alpha \in (0, 1]$. Introducing two slack variables $\phi$ and $\omega$, the following implication always holds
\begin{flalign}
\begin{cases}
 \bm{Tr}(\bm{A}) - \sqrt{-2 \ln(\alpha)} \phi + \ln(\alpha) \omega + \theta, \\
\left\| \begin{bmatrix} \rm{rec} (\bm{A}) \\ \sqrt{2} \bm{r} \end{bmatrix} \right\| \le \phi, \\
\omega \bm{I} + \bm{A} \succeq \bm{0}, \omega \ge 0. \\
\end{cases}
\end{flalign}
In this way, the corresponding probabilistic constraint can be transformed into a set of inequalities to solve the studied problems.

\subsection{\textcolor[rgb]{0.00,0.00,0.00}{Application Challenges: 2-Pw and 3-Pw Models}}

\subsubsection{Performance optimization}
In fact, the 2-Pw and 3-Pw nonlinear EH models are piecewise linear function in terms of the input power, i.e., $P_{\rm in}$. Therefore, compared with the Lg, Mlg and Ef nonlinear EH models, the difficulty on exploring the optimization problems of the system performance is relatively small. In the optimization problems, the 2-Pw and 3-Pw models can be respectively rewritten as $\min (\eta P_{\rm in}, {\rm Q}_{\max})$ and $\min (0, \eta P_{\rm in}, {\rm Q}_{\max})$ function, which are convex and relatively easy to solve by using some known optimization methods such as linear programming \cite{2004Convex}.

\subsubsection{Performance analysis}
Since the 2-Pw and 3-Pw nonlieanr EH models are piecewise, the system performance analysis can also be processed in piecewise, which can effectively perform theoretical analysis. Further, by using some mathematical theories, some closed-form analysis results can be obtained \cite{ref_pw1_app_2, ref_pw1_app_7, pw2021app4, pw2021app5, ref_pw2_app_3, pw2_2021app5, pw3_2021_swipt3}.

\subsection{\textcolor[rgb]{0.00,0.00,0.00}{Application Challenges: Hr, 2-ord and Fr Models}}
As for the Hr model, its form (higher order polynomial) is too complex, which brings great challenges for the system optimization and analysis, which may be one of the reasons why it is not widely used.

For the 2-ord model, it is a concave function, which can avoid some non-convex troubles in optimization problems \cite{ref_mod_2or_0}. However, it is also a second-order function, which often leads to the optimization variables of the quadratic term, non-convex and difficult to handle. Similarly, these bring the same difficulty in the theoretical derivation and discussion of the system performance analysis.

For the Fr model, it is a simplified form of the Hr one (first-order fractional form). In optimization problems, if it is used as the objective function, the methods of fractional programming, Lagrangian and Dinkelbach can be adopted to solve \cite{frc_2021_wpcn1, 2004Convex}; If it is used as a constraint, it can be solved through mathematical transformation. Compared with the Hr nonlinear EH model, the difficulty of exploring problems is relatively small. Meanwhile, in the aspect of the system performance analysis, due to its relatively simple form, some good results can be obtained through some mathematical means.

\subsection{\textcolor[rgb]{0.00,0.00,0.00}{Application Challenges: Log Model}}
For the Log model, modeled by a logarithmic function, its form seems to be similar to the Shannon formula for channel capacity, which generally is concave in terms of the optimal variable. 

\subsubsection{Performance optimization}
Therefore, in an optimization problem, as an objective function, the methods of projection function and epigraph reformulation can be used to deal with it, which can make the problem convex \cite{han2021joint_log_aap_1}. As a constraint, if it is a constraint greater than or equal to a certain threshold, it is directly convex and does not need to be processed; If it is a constraint less than or equal to a certain threshold, it needs some formula transformation or variable substitution to convert it to convex \cite{2017Convex}.

\subsubsection{Performance analysis}
Similar to the Lg, Mlg, Jm models, the system performance analysis under the Log model needs to be explored in combination with the considered channels, where the variable substitution, approximation and some special theories can be adopted to deal with.

\subsection{\textcolor[rgb]{0.00,0.00,0.00}{Application Challenges: Jm Model}}
In the Jm EH model, the energy conversion efficiency, i.e., $\eta$, is modeled as a function of the power and frequency. Therefore, no matter the research on the optimization or analysis of the system, there will be both the influence of the power and frequency models. The impact of the power model can be addressed by the corresponding strategies of different EH models mentioned above. For the frequency model, it also needs to be analyzed according to its specific form and the explored problems.


\section{\textcolor[rgb]{0.00,0.00,0.00}{AI Powered RF-based Nonlinear EH Systems}} \label{section_AI_ehmodels}
\subsection{\textcolor[rgb]{0.00,0.00,0.00}{AI-based Approaches for EH Communications}} 


Facing complex and dynamic network environments, new wireless communication technology is required to equip with the ability to sense the surroundings and the ability of real-time decisions. Traditional optimization-based methods become unmanageable for sophisticated network optimizations \cite{du2021millimeter, du2022performance, du2022_2}. Recently, AI, the core force of the next industrial revolution, has been referred as a notable innovative technique for future wireless communication systems \cite{ai_0, du2023exploring}. Particularly, machine learning (ML) algorithms such as deep learning (DL), reinforcement learning (RL), deep RL (DRL), deep deterministic policy gradient (DDPG), etc., have become an emerging and promising technology and has attracted extensive attention \cite{2022_ai_0, ai_2033app1, du2023generative}.



\begin{figure*}[htp!]
\centering
\includegraphics[width=0.90\textwidth]{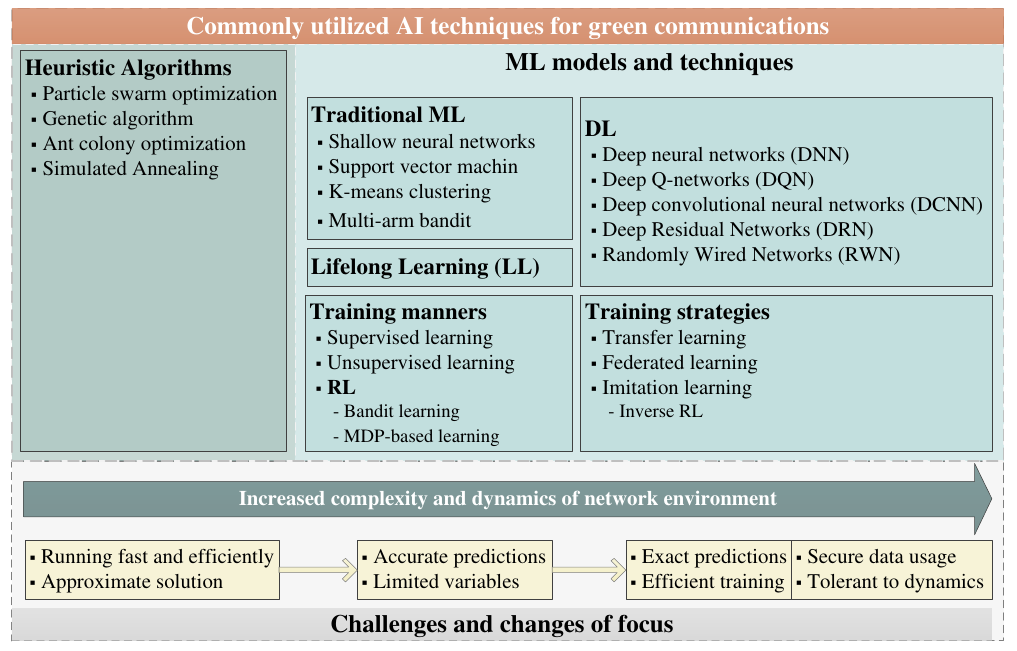}
\caption{Commonly utilized AI techniques for green communications}
\label{ai_method}
\end{figure*}

Up to this point, various AI techniques and models have been developed with different performances and complexity of communication systems. Moreover, the methods, applications and challenges of AI have been also surveyed well. For a better description, we draw some commonly utilized AI techniques in Fig. \ref{ai_method} according to the existing overviews of AI \cite[Fig. 3]{eh_2022_2}. Conventional AI techniques include heuristic algorithms and ML-based methods. At present, ML-based methods are adopted more in wireless communication optimization, particularly, DL, RL, and the combination of the two.


Since some commonly used AI algorithms have been fully introduced and applied in the field of wireless communication in the existing research and reviews, we shall briefly introduce two relatively new learning algorithms applied in the field of wireless communications, namely the inverse RL (IRL) and the lifelong learning (LL), as follows:
\begin{itemize}
\item[-] \textit{IRL:} In the field of Imitation learning, there is debate about whether the IRL method should be considered part of imitation learning or treated as a separate category. IRL is an algorithm that aims to determine the reward function of MDPs by working backward from a given strategy (whether optimal or not) or demonstration data. By doing so, agents can learn to make decisions on complex problems by following expert trajectories. IRL is commonly applied in domains where accurately quantifying the reward function is challenging \cite{ai_2023aap2}.

\item[-] \textit{LL:} It is also known as continuous learning, incremental learning or never-ending learning. In typical ML, a single model is designed to solve one or a few specific tasks. When faced with new tasks, we usually retrain a new model. While, LL first uses a model on $\rm{task}_1$, and then still uses the model on $\rm{task}_2$ until $\rm{task}_n$, which aims to explore whether a lifelong model can perform well on many tasks. As this continues, the model capability becomes stronger and stronger, similar to that human beings keep learning new knowledge and thus master many different kinds of knowledge. LL is equivalent to an adaptive algorithm that can learn from continuous information flows. As time progresses, the information becomes available. Hence, to truly achieve intelligence, the LL-based algorithm is indispensable \cite{rostami2020using}.


\end{itemize}


However, training an advanced AI model takes time, money and high-quality labelled data, it also takes energy. AI-based algorithms increase the energy consumption of wireless nodes in communication networks. Between storing data in large-scale data centers and then using the data to train an ML or DL model, the energy consumption of AI-based algorithms is high. Therefore, it is important to apply energy preservation techniques. Two known energy preservation methods are energy saving and EH. Most of the energy saving techniques are implemented through the estimate and control of the uptime of end devices. In addition, ML-based EH techniques also have attracted extensive attention and research so far \cite{ref_2rd_app_1, 2rd_2021app1, lg1_swipt_2021app25, ai_eh_0,ai_eh_1, ai_eh_2, ai_eh_3, ai_eh_4, lg_2023app9, pw2_2023app1}, summarized in Table \ref{tab_ai_app}.

\begin{table*}[t!] 
\renewcommand{\arraystretch}{1.40}
\centering 
\caption{Applications utilizing AI-based algorithms under nonlinear EH models} \label{tab_ai_app}
\begin{tabular}{c|c|c|cl}
\hline
\hline

 \textbf{Scenario} & \textbf{Nonlinear EH model} & \makecell[c]{\textbf{Goal}} & \makecell[c]{\textbf{AI-based approache}} \\
	 \hline

 P2P SWITP \cite{lg1_swipt_2021app25} & Lg & Error rate & DL \\
 \cdashline{1-4}[0.8pt/1pt]
 Wireless powered HetNets \cite{ai_eh_0} & Lg & Sum rate & DUL \\
 \cdashline{1-4}[0.8pt/1pt]
 Wireless powered HetNets \cite{ai_eh_2} & Lg & Throughput & DQL \\
 \cdashline{1-4}[0.8pt/1pt]
 UAV-assisted IoT \cite{ai_eh_3} & Lg & Multi-objective & DDPG \\
 \cdashline{1-4}[0.8pt/1pt]
 Wireless powered CR \cite{ai_eh_4} & Lg & Transmit rate & DRL \\
 \cdashline{1-4}[0.8pt/1pt]
 UAV-assisted D2D \cite{lg_2023app9} & Lg & Throughput, EE & Multi-agent DRL \\
 \cdashline{1-4}[0.8pt/1pt]
 D2D-assisted MEC \cite{pw2_2023app1} & 2-Pw & Task delay & Multi-agent RL \\
 \cdashline{1-4}[0.8pt/1pt]
 Wireless powered WSN \cite{ref_2rd_app_1, 2rd_2021app1} & 2-ord & Signal reconstruction error & DRL \\

\hline
\hline

 \end{tabular}
\end{table*}

Particularly, in \cite{lg1_swipt_2021app25}, the system symbol error rate was investigated for P2P SWIPT system with the Lg EH model, where a DL-based approach was involved. In \cite{ai_eh_0}, the sum-rate of the system was maximized for renewable energy source-assisted HetNets under the Lg EH model, where a deep unsupervised learning (DUL)-based user association scheme and a DRL-based power control scheme were presented, respectively. In \cite{ai_eh_1}, the capacity of femtocells was maximized for wireless powered HetNets with the Lg EH model, where a Q-learning (QL) based algorithm was presented. In \cite{ai_eh_2}, the longterm average throughput was maximized for wireless powered communication system with the Lg EH model, where the deep Q-learning (DQL) and actor-critic approaches were involved. In \cite{ai_eh_3}, a multi-objective optimization problem (e.i., maximization of sum rate, maximization of total harvested energy and minimization of UAV’s energy consumption) was studied for UAV-assisted wireless powered IoT network with the Lg EH model, where a DDPG algorithm was designed. In \cite{ai_eh_4}, the average transmit rate was maximized for a wideband wireless powered CR network with the Lg EH model, where a DRL method was presented. In \cite{lg_2023app9}, the system throughput and EE were jointly maximized for a UAV-assisted D2D communication network with the Lg EH model, where a multi-agent DRL algorithm was proposed. In \cite{pw2_2023app1}, a multi-agent RL algorithm was presented to minimize the long-term average task delay for the D2D-assisted MEC system with the 2-Pw EH model. Moreover, the DRL-based approaches were also investigated with the 2-ord EH model for WSN in \cite{ref_2rd_app_1, 2rd_2021app1}, as mentioned in subsection \ref{2ord_app}. 


\subsection{\textcolor[rgb]{0.00,0.00,0.00}{Effect of Nonlinear EH Models on AI-based Methods}} 
EH model is a description of EH behavior, and the complexity of its mathematical modeling has a great impact on the optimization-based algorithms, especially in the aspect of mathematical theoretical analysis as mentioned above. However, AI-based algorithms can effectively avoid the difficulty brought by the EH model without requiring some mathematical theoretical analysis of it. That is, the EH model has little influence on AI-based algorithms. In AI-based algorithm design, the computational complexity introduced by the EH model is generally one. This is an advantage of AI-based algorithms, but they also have disadvantages. The results obtained by AI-based algorithms cannot be analyzed theoretically and are approximate. Therefore, based on some theoretical results of convex optimization, the design of corresponding AI-based algorithms has also been explored in many existing works \cite{ai_convex_1,ai_convex_2,ai_convex_3}.

On the other hand, there are also many challenges in designing AI-based algorithms, such as exact predictions, efficient training, secure data usage and tolerant to dynamics, as shown in Fig. \ref{ai_method}. It can be summarized as follows:
\begin{itemize}
\item[-] \textit{Learning framework design:} Difficulties are often not caused by mathematics, and AI-based algorithms do not need strong mathematical ability.

\item[-] \textit{Exponential debugging:} What's unique about machine learning is that when an algorithm doesn't work as expected, it's hard to figure out why it's wrong. Few algorithms work the first time, so most of the time is spent designing algorithms.

\item[-] \textit{Computing capacity:} As the complexity of the model increases, the computational complexity also increases generally, even exponentially.
\end{itemize}

\subsection{\textcolor[rgb]{0.00,0.00,0.00}{RF-based EH in AI-driven Applications}}
The increasing global demand for wireless devices such as mobile phones and computers underscores the importance of wireless applications. However, these devices require a continuous power supply or long battery life. To address safety concerns associated with battery usage, RF-based EH systems that provide wireless power can greatly benefit the application market, which is projected to grow by 22\% between 2020 and 2025 \cite{eh_ai_apps}.

Several factors contribute to this growth, with AI being the primary driver \cite{du2023spear}. While AI-based algorithms have achieved significant advancements, they increasingly rely on large volumes of data to improve their effectiveness. Meanwhile, RF-based EH has demonstrated reliability and efficiency, offering the potential to enhance decision-making capabilities of AI algorithms. The implementation and adoption of RF-based EH can have substantial benefits in various AI-related domains, such as IoT, intelligent medical applications and smart cities \cite{eh_2022_2, zhang2021joint, zhang2023energy}. Nevertheless, it is important to give careful attention to key components of RF-based EH, including the receiver antenna and power conditioning circuits. These elements play a crucial role in ensuring the successful implementation of this approach across different applications.

\section{\textcolor[rgb]{0.00,0.00,0.00}{Lessons Learned and Design Guidelines} \label{section_lessons_eh}}
\textcolor[rgb]{0.00,0.00,0.00}{By adhering to these lessons learned and design guidelines, engineers can create efficient and reliable RF-based EH systems with accurate nonlinear models. These systems have the potential to harvest energy from ambient RF signals and power various wireless devices, sensors, and IoT applications in an environmentally friendly and sustainable manner.}

\subsection{Lessons Learned}

\begin{itemize}
\item \textcolor[rgb]{0.00,0.00,0.00}{\textit{Importance of Nonlinearity:} One of the key lessons learned is the significance of considering nonlinearities in RF-based EH systems. The energy harvesting process often exhibits nonlinear characteristics due to the behavior of diodes and other nonlinear elements. Neglecting these nonlinearities can lead to inaccurate modeling and suboptimal system performance.}

\item \textcolor[rgb]{0.00,0.00,0.00}{\textit{Nonlinear Model Accuracy:} Accurate representation of nonlinear behavior in EH systems is essential for reliable predictions of energy harvesting performance. The nonlinearity introduced by diodes and other components should be carefully modeled to reflect real-world characteristics.}

\item \textcolor[rgb]{0.00,0.00,0.00}{\textit{Importance of Efficiency:} One of the key lessons learned is the significance of optimizing the efficiency of RF-based EH systems. Efficient energy capture and conversion are crucial to maximize the amount of harvested energy and ensure the sustainability of the system.}

\item \textcolor[rgb]{0.00,0.00,0.00}{\textit{Frequency Dependence:} Another important lesson is the frequency dependence of RF-based EH systems. The efficiency of energy harvesting can vary significantly at different RF frequencies. Understanding and accounting for this frequency dependency is crucial for designing efficient and reliable RF-based EH systems.}

\item \textcolor[rgb]{0.00,0.00,0.00}{\textit{Real-World Environment:} Real-world environmental factors, such as variations in RF signal strength and interference, can impact the performance of RF-based EH systems. Considering these factors during the design phase is essential to ensure the system's robustness and adaptability to different environments.}

\item \textcolor[rgb]{0.00,0.00,0.00}{\textit{Validation with Experiments:} Validating the performance of RF-based EH systems and nonlinear models using experimental data is essential for verifying their accuracy and effectiveness. Real-world testing helps identify and address potential issues that might not be evident in simulations alone.}
\end{itemize}

\subsection{Design Guidelines}

\begin{itemize}
\item \textcolor[rgb]{0.00,0.00,0.00}{\textit{Nonlinear Model Integration:} Incorporate nonlinear models of diodes and other nonlinear components into the design of RF-based EH systems. These models should accurately represent the voltage-current characteristics and other nonlinear behaviors to achieve accurate predictions of energy harvesting performance.}

\item \textcolor[rgb]{0.00,0.00,0.00}{\textit{Frequency Spectrum Analysis:} Perform a comprehensive analysis of the frequency spectrum in the target environment to determine the optimal frequency range for energy harvesting. This analysis will aid in selecting suitable antennas and optimizing the system's frequency-dependent performance.}

\item \textcolor[rgb]{0.00,0.00,0.00}{\textit{Efficient Antenna Design:} Select antennas with high gain and appropriate resonant frequency to maximize energy harvesting efficiency. Proper antenna design is crucial for capturing and converting RF energy effectively.}

\item \textcolor[rgb]{0.00,0.00,0.00}{\textit{Realistic Environmental Considerations:} Consider real-world environmental factors, such as temperature, interference, and RF signal fluctuations, during system design. This will help create a more robust and reliable RF-based EH system capable of adapting to various operating conditions.}

\item \textcolor[rgb]{0.00,0.00,0.00}{\textit{Energy Storage and Management:} Design efficient energy storage and management circuits to store harvested energy and ensure a stable power supply for the target application. Balancing energy collection and consumption is essential for optimal system performance.}

\item \textcolor[rgb]{0.00,0.00,0.00}{\textit{Validation and Calibration:} Validate the RF-based EH system and nonlinear model using experimental data. Calibration based on real-world measurements can enhance the accuracy of the model and ensure reliable performance.}

\item \textcolor[rgb]{0.00,0.00,0.00}{\textit{Energy Harvesting Efficiency Metrics:} Define appropriate metrics to evaluate the energy harvesting efficiency of the RF-based EH system. These metrics will assist in assessing the system's performance and identifying areas for improvement.}

\item \textcolor[rgb]{0.00,0.00,0.00}{\textit{Integration with Target Applications:} Design the RF-based EH system with integration in mind, ensuring compatibility with the target application and other electronic components. Scalability and adaptability are crucial for broad applicability.}

\end{itemize}

\section{\textcolor[rgb]{0.00,0.00,0.00}{Emerging Research Challenges}} \label{section_direction_ehmodels}

\subsection{\textcolor[rgb]{0.00,0.00,0.00}{Limitations of RF-based EH Applications}}
In the field of RF-based EH, there exist several challenges and limitations that impact its applications and ongoing progress. Here, we present some common challenges and limitations, including technical barriers, system performance optimization, and the development of practical solutions. Overcoming these issues is crucial to fully harness the benefits of RF-based EH across various applications.

\subsubsection{EH Efficiency}
RF-based EH systems often face the challenge of low EH efficiency. RF energy in the environment is relatively scarce and weak, requiring EH devices to have efficient energy capture and conversion capabilities. Improving EH efficiency is an important challenge that requires optimization in antenna design, rectifier performance, and energy conversion circuits. For zero-power communications, OPPO is working with partners across the industry to make zero-power communications devices a core part of next-generation communications technologies for greater convenience, inclusivity, and sustainability. And, Texas Instruments has developed RF-based EH solutions for industrial applications, such as wireless sensors in industrial automation systems, to improve energy efficiency and reduce battery maintenance.

\subsubsection{Environmental Dependency}
The effectiveness of RF-based EH is influenced by environmental factors. RF energy intensity and spectral distribution vary across different regions and environments, which can lead to stability and reliability issues in EH. Changes in environmental factors can result in interruptions or reductions in EH, requiring strategies to address this environmental dependency challenge. For instant, ABB, a leading industrial technology company, is researching RF energy harvesting for wireless sensor networks in smart factories, where environmental variations can be effectively managed to power various IoT devices.

\subsubsection{Energy Management and Storage}
Effective energy management and storage are crucial elements of RF-based EH systems. Due to the variability and instability of RF energy, energy management circuits need to be designed and optimized to achieve efficient energy storage and supply. Additionally, certain applications require a balance between energy harvesting and energy consumption to ensure system stability. For instant, Schneider Electric, a multinational corporation specializing in energy management and automation, is exploring RF energy harvesting for self-powered wireless sensors in building automation systems, enabling efficient energy utilization and reducing the reliance on batteries.

\subsubsection{Miniaturization and Integration}
Many RF-based EH applications require small-sized and highly integrated EH systems. This necessitates minimizing the size, weight, and power consumption of energy harvesting devices, circuits, and interfaces, while enabling close integration with other electronic components. Achieving small size and high integration is a technical challenge. For instant, Siemens, a global technology company, is researching miniaturized RF energy harvesting solutions for wireless sensors in industrial equipment, enabling self-powered devices with reduced maintenance needs.

\subsubsection{Security and Privacy}
RF-based EH involves the reception and processing of wireless signals, raising concerns about security and privacy. For example, security measures need to be implemented to protect energy harvesting systems and transmission processes from unauthorized access or attacks. Additionally, for devices and sensors powered by RF energy, ensuring data security and privacy protection is important. For instant, Honeywell, a multinational conglomerate, is working on secure RF energy harvesting solutions for industrial control systems, ensuring data integrity and preventing unauthorized access to critical infrastructure.


 \subsection{\textcolor[rgb]{0.00,0.00,0.00}{Future Directions}}
\subsubsection{RF-based EH and IRS}
IRS technology is considered as one of the key technologies of 6G. RIS can gather the power of RF signals in the environment, provide more accurately focused energy beams for wireless devices, and improve EH efficiency. Besides, RIS can realize controllable backscattering by controlling the phase, amplitude, polarization and other parameters of the signals, and forwarding it to the receiver, thus improving the receiving performance. The combination of RIS and EH in the 6G system will contribute to providing an approach for ultra-low power IoT. How to use RIS technology to improve the efficiency of EH and the performance of wireless communications is an important opportunity worth further exploration.

\subsubsection{RF-based EH and Integrated Sensing and Communication}
The combination of RF-based EH and integrated sensing and communication (ISAC) can significantly improve the EE of the system and meet the requirement of green communications. Obtaining energy through RF-based EH can fundamentally eliminate the dependence on batteries. At the same time, RF-based EH also provides an effective means for ISAC. For example, the destination is configured with an EH unit, which can trigger EH when sending a sense signal to it. Then, the information of the destination can be reported to the source through the RF-based EH communication systems, so as to achieve accurate sensing function. Meanwhile, integrated sensing, computing and communication (ISCAC) also attracts significant interest with the demands of huge-volume data processing. Cloud computing was commonly adopted by exploring the powerful computing capability of cloud servers. The study with the combination of EH and ISCAC is also a major trend.

\subsubsection{RF-based EH and Semantic Communication}
Semantic communication (SemCom) techniques enable wireless edge devices to extract and transmit the meaning of original data (called semantic information) to reduce the heavy congestion of current wireless networks thus improving communication efficiency. With the growing number of IoTs with EH capabilities, it is important to determine the importance of information with the help of semantics. In an RF-based EH communication system, users can consume the harvested energy for semantic information transmission. At present, the application of SemCom techniques in RF-based EH systems is still in the early phases. Many SemCom networks with EH-enabled devices are not explored, e.g., IRS/UAV-assisted networks with SWIPT. Therefore, the combination of RF-based EH and SemCom still has many open problems to be studied.

\subsubsection{RF-based EH and THz Communication}
RF-based EH is employed to capture and convert RF energy into usable electrical power, which can be then utilized to support THz communication devices and systems. By integrating RF-based EH capabilities, THz communication devices can harvest energy from the surrounding RF signals, reducing or eliminating the need for external power sources such as batteries or wired connections. The RF-based EH process involves capturing RF energy using antennas, rectifying and converting it into DC power using rectifiers, and regulating and storing the harvested energy for subsequent use in THz communication systems. This enables self-sustaining and potentially battery-free operation of THz communication devices, leading to increased mobility and flexibility in deployment.

\subsubsection{RF-based EH and AI}
The performance of AI systems depends largely on sufficient data sources. RF-based EH provides a relatively low-cost data acquisition solution, which can promote the development of AI technology and improve the performance of AI systems. For example, in some intelligent plant scenarios, the RF-based EH system is used to collect environmental information such as temperature, humidity, dynamic frequency, etc., and then AI is used to predict environmental changes in advance, predict the working state changes of the system, trigger early warnings, and provide other intelligent devices with environmental regulation indication information.

\section{\textcolor[rgb]{0.00,0.00,0.00}{Conclusion}} \label{section_conclusion_ehmodels}

\textcolor[rgb]{0.00,0.00,0.00}{This article provided a comprehensive overview of the existing mathematical models that accurately characterize the nonlinear characteristics of practical EH circuits. It serves as a handbook of mathematical nonlinear EH models. Furthermore, we summarized the application of each nonlinear EH model, highlighting the associated challenges and precautions. Additionally, we analyzed the influence and progress of each nonlinear EH model on wireless communication systems utilizing RF-based EH, leveraging the power of AI. Lastly, we emphasized emerging research directions in nonlinear RF-based EH. This article contributes to the future application of RF-based EH in cutting-edge communication research domains to some extent.}



\bibliographystyle{IEEEtran}
\bibliography{IEEEabrv,Reference}

\begin{thebibliography}{100}
\providecommand{\url}[1]{#1}
\csname url@samestyle\endcsname
\providecommand{\newblock}{\relax}
\providecommand{\bibinfo}[2]{#2}
\providecommand{\BIBentrySTDinterwordspacing}{\spaceskip=0pt\relax}
\providecommand{\BIBentryALTinterwordstretchfactor}{4}
\providecommand{\BIBentryALTinterwordspacing}{\spaceskip=\fontdimen2\font plus
\BIBentryALTinterwordstretchfactor\fontdimen3\font minus
  \fontdimen4\font\relax}
\providecommand{\BIBforeignlanguage}[2]{{%
\expandafter\ifx\csname l@#1\endcsname\relax
\typeout{** WARNING: IEEEtran.bst: No hyphenation pattern has been}%
\typeout{** loaded for the language `#1'. Using the pattern for}%
\typeout{** the default language instead.}%
\else
\language=\csname l@#1\endcsname
\fi
#2}}
\providecommand{\BIBdecl}{\relax}
\BIBdecl

\bibitem{6g_1}
M.~Alsabah, M.~A. Naser, B.~M. Mahmmod, S.~H. Abdulhussain, M.~R. Eissa,
  A.~Al-Baidhani, N.~K. Noordin, S.~M. Sait, K.~A. Al-Utaibi, and F.~Hashim,
  ``{6G wireless communications networks: A comprehensive survey},'' \emph{IEEE
  Access}, vol.~9, pp. 148\,191--148\,243, 2021.

\bibitem{du2022semantic}
H.~Du, J.~Wang, D.~Niyato, J.~Kang, Z.~Xiong, J.~Zhang \emph{et~al.},
  ``{Semantic communications for wireless sensing: RIS-aided encoding and
  self-supervised decoding},'' \emph{IEEE J. Sel. Areas Commun.}, to appear,
  2023.

\bibitem{du2023attention}
H.~Du, J.~Liu, D.~Niyato, J.~Kang, Z.~Xiong, J.~Zhang, and D.~I. Kim,
  ``{Attention-aware resource allocation and QoE analysis for metaverse xURLLC
  services},'' \emph{IEEE J. Sel. Areas Commun.}, to appear, 2023.

\bibitem{zhang2023generative}
R.~Zhang, K.~Xiong, H.~Du, D.~Niyato, J.~Kang, X.~Shen, and H.~V. Poor,
  ``{Generative AI-enabled vehicular networks: Fundamentals, framework, and
  case study},'' \emph{arXiv preprint arXiv:2304.11098}, 2023.

\bibitem{du2023ai}
H.~Du, J.~Wang, D.~Niyato, J.~Kang, Z.~Xiong, and D.~I. Kim, ``{AI-generated
  incentive mechanism and full-duplex semantic communications for information
  sharing},'' \emph{IEEE J. Sel. Areas Commun.}, to appear, 2023.

\bibitem{du2022exploring}
H.~Du, J.~Wang, D.~Niyato, J.~Kang, Z.~Xiong, X.~S. Shen, and D.~I. Kim,
  ``{Exploring attention-aware network resource allocation for customized
  metaverse services},'' \emph{IEEE Netw.}, to appear, 2023.

\bibitem{eh_2023app_2}
V.~Fernandes, N.~Cravo, H.~L.~M.~Monteiro, D.~N. K.~Jayakody, H.~V. Poor, and
  M.~V.~Ribeiro, ``{Energy harvesting in the UNB-PLC spectrum: Hidden
  opportunities for IoT devices},'' \emph{IEEE Internet Things J.}, vol.~10,
  no.~2, pp. 1236--1247, 2023.

\bibitem{mao_2023_1}
W.~Mao, K.~Xiong, Y.~Lu, P.~Fan, and Z.~Ding, ``{Energy consumption
  minimization in secure multi-antenna UAV-assisted MEC networks with channel
  uncertainty},'' \emph{IEEE Transactions on Wireless Communications}, pp.
  1--1, to appear, 2023.

\bibitem{eh_2023app_1}
Y.~Kim, B.~C. Jung, and Y.~Song, ``{Online learning for joint energy harvesting
  and information decoding optimization in IoT-enabled smart city},''
  \emph{IEEE Internet Things J.}, vol.~10, no.~12, pp. 10\,675--10\,686, 2023.

\bibitem{jiang2015outage}
R.~Jiang, K.~Xiong, P.~Fan, and Z.~Zhong, ``{Outage performance of
  SWIPT-enabled two-way relay networks},'' in \emph{IEEE HMWC}, 2015, pp.
  106--110.

\bibitem{jiang2017outage}
R.~Jiang, K.~Xiong, Y.~Zhang, Z.~Zhong \emph{et~al.}, ``{Outage analysis and
  optimization of SWIPT in network-coded two-way relay networks},''
  \emph{Mobile Inf. Syst.}, vol. 2017, 2017.

\bibitem{se_eh}
\emph{{Energy harvesting starting to gain traction}},
  https://semiengineering.com/energy-harvesting-starting-to-gain-traction/.

\bibitem{product_eh}
\emph{{Top key players – STMicroelectronics, Texas Instruments, EnOcean GmbH,
  etc. – designer women}},
  https://mitchspoolservice.net/top-key-players-stmicroelectronics-texas-instruments-enocean-gmbh-etc-designer-women/.

\bibitem{Sharma21}
A.~Sharma and P.~Sharma, ``{Energy harvesting technology for IoT edge
  applications},'' in \emph{Smart Manufacturing}, T.~Y. Kheng, Ed.\hskip 1em
  plus 0.5em minus 0.4em\relax Rijeka: IntechOpen, 2021, ch.~6.

\bibitem{cast2019power}
\emph{{RF energy harvest \& wireless power}}, https://www.powercastco.com/.

\bibitem{clex2019nonlinear}
B.~{Clerckx}, R.~{Zhang}, R.~{Schober}, D.~W.~K. {Ng}, D.~I. {Kim}, and H.~V.
  {Poor}, ``{Fundamentals of wireless information and power transfer: From RF
  energy harvester models to signal and system designs},'' \emph{IEEE J. Sel.
  Areas Commun.}, vol.~37, no.~1, pp. 4--33, 2019.

\bibitem{Brono_2016}
B.~Clerckx and E.~Bayguzina, ``{Waveform design for wireless power transfer},''
  \emph{IEEE Trans. Signal Process.}, vol.~64, no.~23, pp. 6313--6328, 2016.

\bibitem{eh_2011}
S.~Sudevalayam and P.~Kulkarni, ``{Energy harvesting sensor nodes: Survey and
  implications},'' \emph{IEEE Commun. Surveys Tuts.}, vol.~13, no.~3, pp.
  443--461, 2011.

\bibitem{eh_2014}
R.~V. Prasad, S.~Devasenapathy, V.~S. Rao, and J.~Vazifehdan, ``{Reincarnation
  in the ambiance: Devices and networks with energy harvesting},'' \emph{IEEE
  Commun. Surveys Tuts.}, vol.~16, no.~1, pp. 195--213, 2014.

\bibitem{eh_2015}
X.~Huang, T.~Han, and N.~Ansari, ``{On green-energy-powered cognitive radio
  networks},'' \emph{IEEE Commun. Surveys Tuts.}, vol.~17, no.~2, pp. 827--842,
  2015.

\bibitem{ref_eh_2}
X.~{Lu}, P.~{Wang}, D.~{Niyato}, D.~I. {Kim}, and Z.~{Han}, ``{Wireless
  networks with RF energy harvesting: A contemporary survey},'' \emph{IEEE
  Commun. Surveys Tuts.}, vol.~17, no.~2, pp. 757--789, 2015.

\bibitem{ref_eh_1}
M.~{Ku}, W.~{Li}, Y.~{Chen}, and K.~J. {Ray Liu}, ``{Advances in energy
  harvesting communications: Past, present, and future challenges},''
  \emph{IEEE Commun. Surveys Tuts.}, vol.~18, no.~2, pp. 1384--1412, 2016.

\bibitem{ref_eh_5}
X.~{Lu}, P.~{Wang}, D.~{Niyato}, D.~I. {Kim}, and Z.~{Han}, ``{Wireless
  charging technologies: Fundamentals, standards, and network applications},''
  \emph{IEEE Commun. Surveys Tuts.}, vol.~18, no.~2, pp. 1413--1452, 2016.

\bibitem{eh_2018_1}
N.~Van~Huynh, D.~T. Hoang, X.~Lu, D.~Niyato, P.~Wang, and D.~I. Kim, ``{Ambient
  backscatter communications: A contemporary survey},'' \emph{IEEE Commun.
  Surveys Tuts.}, vol.~20, no.~4, pp. 2889--2922, 2018.

\bibitem{alsaba2018beamforming}
Y.~Alsaba, S.~K.~A. Rahim, and C.~Y. Leow, ``{Beamforming in wireless energy
  harvesting communications systems: A survey},'' \emph{IEEE Commun. Surveys
  Tuts.}, vol.~20, no.~2, pp. 1329--1360, 2018.

\bibitem{ref_eh_3}
T.~D. {Ponnimbaduge Perera}, D.~N.~K. {Jayakody}, S.~K. {Sharma},
  S.~{Chatzinotas}, and J.~{Li}, ``{Simultaneous wireless information and power
  transfer (SWIPT): Recent advances and future challenges},'' \emph{IEEE
  Commun. Surveys Tuts.}, vol.~20, no.~1, pp. 264--302, 2018.

\bibitem{eh_2019}
V.~L. Quintero, C.~Estevez, M.~E. Orchard, and A.~Pérez, ``{Improvements of
  energy-efficient techniques in WSNs: A MAC-protocol approach},'' \emph{IEEE
  Commun. Surveys Tuts.}, vol.~21, no.~2, pp. 1188--1208, 2019.

\bibitem{eh_2020_1}
P.~Tedeschi, S.~Sciancalepore, and R.~Di~Pietro, ``{Security in energy
  harvesting networks: A survey of current solutions and research
  challenges},'' \emph{IEEE Commun. Surveys Tuts.}, vol.~22, no.~4, pp.
  2658--2693, 2020.

\bibitem{eh_2020_2}
D.~Ma, G.~Lan, M.~Hassan, W.~Hu, and S.~K. Das, ``{Sensing, computing, and
  communications for energy harvesting IoTs: A survey},'' \emph{IEEE Commun.
  Surveys Tuts.}, vol.~22, no.~2, pp. 1222--1250, 2020.

\bibitem{eh_2022_1}
A.~Kaswan, P.~K. Jana, and S.~K. Das, ``{A survey on mobile charging techniques
  in wireless rechargeable sensor networks},'' \emph{IEEE Commun. Surveys
  Tuts.}, vol.~24, no.~3, pp. 1750--1779, 2022.

\bibitem{eh_2022_2}
B.~Mao, F.~Tang, Y.~Kawamoto, and N.~Kato, ``{AI models for green
  communications towards 6G},'' \emph{IEEE Commun. Surveys Tuts.}, vol.~24,
  no.~1, pp. 210--247, 2022.

\bibitem{iot_2021}
M.~M. Sandhu, S.~Khalifa, R.~Jurdak, and M.~Portmann, ``{Task scheduling for
  energy-harvesting-based IoT: A survey and critical analysis},'' \emph{IEEE
  Internet Things J.}, vol.~8, no.~18, pp. 13\,825--13\,848, 2021.

\bibitem{access_2017}
N.~Zhao, S.~Zhang, F.~R. Yu, Y.~Chen, A.~Nallanathan, and V.~C.~M. Leung,
  ``{Exploiting interference for energy harvesting: A survey, research issues,
  and challenges},'' \emph{IEEE Access}, vol.~5, pp. 10\,403--10\,421, 2017.

\bibitem{access_2019}
M.~A. Hossain, R.~M. Noor, K.-L.~A. Yau, I.~Ahmedy, and S.~S. Anjum, ``{A
  survey on simultaneous wireless information and power transfer with
  cooperative relay and future challenges},'' \emph{IEEE Access}, vol.~7, pp.
  19\,166--19\,198, 2019.

\bibitem{access_2021_1}
A.~J. Williams, M.~F. Torquato, I.~M. Cameron, A.~A. Fahmy, and J.~Sienz,
  ``{Survey of energy harvesting technologies for wireless sensor networks},''
  \emph{IEEE Access}, vol.~9, pp. 77\,493--77\,510, 2021.

\bibitem{access_2021_2}
N.~Ashraf, S.~A. Sheikh, S.~A. Khan, I.~Shayea, and M.~Jalal, ``{Simultaneous
  wireless information and power transfer with cooperative relaying for
  next-generation wireless networks: A review},'' \emph{IEEE Access}, vol.~9,
  pp. 71\,482--71\,504, 2021.

\bibitem{access_2021_3}
A.~Padhy, S.~Joshi, S.~Bitragunta, V.~Chamola, and B.~Sikdar, ``{A survey of
  energy and spectrum harvesting technologies and protocols for next generation
  wireless networks},'' \emph{IEEE Access}, vol.~9, pp. 1737--1769, 2021.

\bibitem{access_2022}
H.~Ojukwu, B.-C. Seet, and S.~U. Rehman, ``{Metasurface-aided wireless power
  transfer and energy harvesting for future wireless networks},'' \emph{IEEE
  Access}, vol.~10, pp. 52\,431--52\,450, 2022.

\bibitem{eh_2023_1}
Y.~C. Lee, H.~Ramiah, A.~Choo, K.~K.~P. Churchill, N.~S. Lai, C.~C. Lim,
  Y.~Chen, P.-I. Mak, and R.~P. Martins, ``{High-performance multiband ambient
  RF energy harvesting front-end system for sustainable IoT applications—a
  review},'' \emph{IEEE Access}, vol.~11, pp. 11\,143--11\,164, 2023.

\bibitem{eh_2023_2}
Z.~Cai, Q.~Chen, T.~Shi, T.~Zhu, K.~Chen, and Y.~Li, ``{Battery-free wireless
  sensor networks: A comprehensive survey},'' \emph{IEEE Internet Things J.},
  vol.~10, no.~6, pp. 5543--5570, 2023.

\bibitem{eh_2023_3}
T.~Jiang, Y.~Zhang, W.~Ma, M.~Peng, Y.~Peng, M.~Feng, and G.~Liu,
  ``{Backscatter communication meets practical battery-free Internet of Things:
  A survey and outlook},'' \emph{IEEE Commun. Surveys Tuts.}, pp. 1--1, to
  appear, 2023.

\bibitem{eh_2023_4}
M.~A. Halimi, T.~Khan, D.~Surender, N.~Nasimuddin, and S.~R. Rengarajan,
  ``{Dielectric resonator antennas for RF energy-harvesting/wireless power
  transmission applications: A state-of-the-art review},'' \emph{IEEE Antennas
  Propag. Mag.}, pp. 2--13, 2023.

\bibitem{hujie_eh}
J.~Hu, K.~Yang, G.~Wen, and L.~Hanzo, ``{Integrated data and energy
  communication network: A comprehensive survey},'' \emph{IEEE Commun. Surveys
  Tuts.}, vol.~20, no.~4, pp. 3169--3219, 2018.

\bibitem{hertz1990h}
H.~Hertz, ``{Dictionary of scientific biography},'' \emph{Scribner}, vol.~6,
  pp. 340--349, 1990.

\bibitem{steer2006ieee}
M.~Cheney, \emph{{Tesla: Man out of time}}.\hskip 1em plus 0.5em minus
  0.4em\relax Simon and Schuster, 2011.

\bibitem{Gla1986ieee}
G.~P. E, ``{Power from the sun: Its future},'' \emph{Science}, vol. 162, no.
  3856, pp. 857--861, 1968.

\bibitem{mou2015wireless}
G.~W. Jull, A.~Lillemark, and R.~M. Turner., ``{SHARP (stationary high altitude
  relay platform) telecommunications missions and systems},'' in \emph{IEEE
  GLOBECOM}, New Orleans, USA, Dec 1989, pp. 955--959.

\bibitem{shi2013beam}
N.~Shinohara, ``{Beam control technologies with a high-efficiency phased array
  for microwave power transmission in Japan},'' \emph{Proceedings of the IEEE},
  vol. 101, no.~6, pp. 1448--1463, 2013.

\bibitem{wei1994power}
W.~Lin, Y.~Zhao, and G.~Wen, ``{Power transmission by microwave-a propulsion
  for modernization construction},'' \emph{Science and Technology Review},
  vol.~3, pp. 31--34, 1994.

\bibitem{luk2013ptop}
L.~S. Luk, ``{Point-to-point wireless power transportation in reunion
  island},'' in \emph{International Astronautical Congress}, Turin, Italy, Oct
  1997.

\bibitem{kurs1994wireless}
e.~a. Kurs, Andre, ``{Wireless power transfer via strongly coupled magnetic
  resonances},'' \emph{Science}, vol. 317, no. 5834, pp. 83--86, 2007.

\bibitem{whi2008wired}
L.~H. Whitesides, ``{Researchers beam ‘space' solar power in Hawaii},''
  \emph{Wired}, 2008.

\bibitem{zeng2017communications}
Y.~Zeng, B.~Clerckx, and R.~Zhang, ``{Communications and signals design for
  wireless power transmission},'' \emph{IEEE Trans. Commun.}, vol.~65, no.~5,
  pp. 2264--2290, 2017.

\bibitem{ref_eh_4}
J.~{Huang}, C.~{Xing}, and C.~{Wang}, ``{Simultaneous wireless information and
  power transfer: Technologies, applications, and research challenges},''
  \emph{IEEE Commun. Mag.}, vol.~55, no.~11, pp. 26--32, 2017.

\bibitem{dia2019eng}
\emph{{Dialog semiconductor and Energous}},
  https://www.dialog-semiconductor.com/node/3708.

\bibitem{zhang2019wifi}
e.~a. Zhang, Xu, ``{Two-dimensional MoS2-enabled flexible rectenna for
  Wi-Fi-band wireless energy harvesting},'' \emph{Nature}, vol. 566, no. 7744,
  pp. 368--372, 2019.

\bibitem{mop2020charge}
\emph{{Mophie introduces multi-device 4-in-1 wireless charging mat to combat
  cable-clutter}}, https://stockhouse.com/news/press-releases/2020/10/20/.

\bibitem{xiaomi2020charge}
\emph{{Xiaomi introduces pioneering 80W Mi wireless charging technology}},
  https://blog.mi.com/en/2020/10/19/xiaomi-introduces-pioneering-80w-mi-wireless-charging-technology/.

\bibitem{eh_ai_apps}
\emph{{RF energy harvesting finds growing role in AI-driven applications}},
  https://www.embedded.com/rf-energy-harvesting-finds-growing-role-in-ai-driven-applications/.

\bibitem{whatup_app}
\emph{{RF-based wireless charging from 2 meters? new team-up combines RF energy
  harvesting and RF WPT}},
  https://www.allaboutcircuits.com/news/atmosic-energous-rf-wireless-power-transfer-2-meters-away/.

\bibitem{Belgian_1}
\emph{{IoT energy harvesting challenges: E-peas tackles it head-on with new
  PMICs}},
  https://www.allaboutcircuits.com/news/iot-energy-harvesting-challenges-e-peas-tackles-it-head-on-with-new-pmics/.

\bibitem{Atmosic_1}
\emph{{Atmosic brings wireless energy harvesting for IoT devices directly onto
  the SoC}},
  https://www.allaboutcircuits.com/news/atmosic-brings-wireless-energy-harvesting-for-internet-of-things-devices-directly-onto-the-system-on-a-chip/.

\bibitem{Florida_1}
\emph{{Researchers create “perfect EM absorption” rectenna for RF energy
  harvesting}},
  https://www.allaboutcircuits.com/news/researchers-create-perfect-electromagnetic-absorption-rectenna-rf-energy-harvesting.

\bibitem{Florida_2}
\emph{{Energy harvesting finds new source: RF waves}},
  https://www.designnews.com/electronics/energy-harvesting-finds-new-source-rf-waves.

\bibitem{Ossia_1}
\emph{Three way deal for wireless power to RF ePaper tags},
  https://blog.ossia.com/news/three-way-deal-for-wireless-power-to-rf-epaper-tags.

\bibitem{ref_swipt_0}
L.~R. Varshney, ``{Transporting information and energy simultaneously},'' in
  \emph{IEEE ISIT}, Toronto, Ontario Canada, Jul. 2008, pp. 1612--1616.

\bibitem{ref_wpcn_0}
H.~Ju and R.~Zhang, ``{Throughput maximization in wireless powered
  communication networks},'' \emph{IEEE Trans. Wireless Commun.}, vol.~13,
  no.~1, pp. 418--428, 2013.

\bibitem{ref_pst}
R.~Zhang and C.~K. Ho, ``{MIMO broadcasting for simultaneous wireless
  information and power transfer},'' \emph{IEEE Trans. Wireless Commun.},
  vol.~12, no.~5, pp. 1989--2001, 2013.

\bibitem{ref_sw_wsn_0}
S.~{Guo}, F.~{Wang}, Y.~{Yang}, and B.~{Xiao}, ``{Energy-efficient cooperative
  for simultaneous wireless information and power transfer in clustered
  wireless sensor networks},'' \emph{IEEE Trans. Commun.}, vol.~63, no.~11, pp.
  4405--4417, 2015.

\bibitem{ref_sw_wsn_1}
G.~Pan, H.~Lei, Y.~Yuan, and Z.~Ding, ``{Performance analysis and optimization
  for SWIPT wireless sensor networks},'' \emph{IEEE Trans. Commun.}, vol.~65,
  no.~5, pp. 2291--2302, 2017.

\bibitem{ref_sw_relay_1}
Y.~{Feng}, Z.~{Yang}, W.~{Zhu}, Q.~{Li}, and B.~{Lv}, ``{Robust cooperative
  secure beamforming for simultaneous wireless information and power transfer
  in amplify-and-forward relay networks},'' \emph{IEEE Trans. Veh. Technol.},
  vol.~66, no.~3, pp. 2354--2366, 2017.

\bibitem{ref_sw_relay_2}
K.~Xiong, C.~Chen, G.~Qu, P.~Fan, and K.~B. Letaief, ``{Group cooperation with
  optimal resource allocation in wireless powered communication networks},''
  \emph{IEEE Trans. Wireless Commun.}, vol.~16, no.~6, pp. 3840--3853, 2017.

\bibitem{ref_sw_mimo_1}
G.~{Amarasuriya}, E.~G. {Larsson}, and H.~V. {Poor}, ``{Wireless information
  and power transfer in multiway massive MIMO relay networks},'' \emph{IEEE
  Trans. Wireless Commun.}, vol.~15, no.~6, pp. 3837--3855, 2016.

\bibitem{ref_sw_mimo_2}
Q.~{Li} and J.~{Qin}, ``{Joint source and relay secure beamforming for
  nonregenerative MIMO relay systems with wireless information and power
  transfer},'' \emph{IEEE Trans. Veh. Technol.}, vol.~66, no.~7, pp.
  5853--5865, 2017.

\bibitem{ref_sw_ofdm_1}
K.~Xiong, P.~Fan, C.~Zhang, and K.~B. Letaief, ``{Wireless information and
  energy transfer for two-hop non-regenerative MIMO-OFDM relay networks},''
  \emph{IEEE J. Sel. Areas Commun.}, vol.~33, no.~8, pp. 1595--1611, 2015.

\bibitem{ref_sw_noma_1}
Y.~{Liu}, Z.~{Ding}, M.~{Elkashlan}, and H.~V. {Poor}, ``{Cooperative
  non-orthogonal multiple access with simultaneous wireless information and
  power transfer},'' \emph{IEEE J. Sel. Areas Commun.}, vol.~34, no.~4, pp.
  938--953, 2016.

\bibitem{ref_sw_cr_1}
Z.~{Yang}, Z.~{Ding}, P.~{Fan}, and G.~K. {Karagiannidis}, ``{Outage
  performance of cognitive relay networks with wireless information and power
  transfer},'' \emph{IEEE Trans. Veh. Technol.}, vol.~65, no.~5, pp.
  3828--3833, 2016.

\bibitem{ref_sw_cr_2}
D.~W.~K. {Ng}, E.~S. {Lo}, and R.~{Schober}, ``{Multiobjective resource
  allocation for secure communication in cognitive radio networks with wireless
  information and power transfer},'' \emph{IEEE Trans. Veh. Technol.}, vol.~65,
  no.~5, pp. 3166--3184, 2016.

\bibitem{ref_sw_d2d_1}
K.~{Ali}, H.~X. {Nguyen}, Q.~{Vien}, P.~{Shah}, and Z.~{Chu}, ``{Disaster
  management using D2D communication with power transfer and clustering
  techniques},'' \emph{IEEE Access}, vol.~6, pp. 14\,643--14\,654, 2018.

\bibitem{ref_sw_d2d_2}
D.~{Lim}, J.~{Kang}, C.~{Chun}, and H.~{Kim}, ``{Joint transmit power and
  time-switching control for device-to-device communications in SWIPT cellular
  networks},'' \emph{IEEE Commun. Lett.}, vol.~23, no.~2, pp. 322--325, 2019.

\bibitem{ref_sw_uav_1}
X.~{Hong}, P.~{Liu}, F.~{Zhou}, S.~{Guo}, and Z.~{Chu}, ``{Resource allocation
  for secure UAV-assisted SWIPT systems},'' \emph{IEEE Access}, vol.~7, pp.
  24\,248--24\,257, 2019.

\bibitem{ref_sw_uav_2}
X.~{Sun}, W.~{Yang}, Y.~{Cai}, Z.~{Xiang}, and X.~{Tang}, ``{Secure
  transmissions in millimeter wave SWIPT UAV-based relay networks},''
  \emph{IEEE Wireless Commun. Lett.}, vol.~8, no.~3, pp. 785--788, 2019.

\bibitem{phdthesis}
B.~E, ``{Practical non-linear energy harvesting model and resource allocation
  in SWIPT systems},'' Ph.D. dissertation, Universit\"at Erlangen-N\"urnberg,
  Germany, Sept. 2015.

\bibitem{ref_mod_lg1_0}
E.~Boshkovska, D.~W.~K. Ng, N.~Zlatanov, and R.~Schober, ``{Practical
  non-linear energy harvesting model and resource allocation for SWIPT
  systems},'' \emph{IEEE Commun. Lett.}, vol.~19, no.~12, pp. 2082--2085, 2015.

\bibitem{ref_lg1_app_1}
E.~Boshkovska, A.~Koelpin, D.~W.~K. Ng, N.~Zlatanov, and R.~Schober, ``{Robust
  beamforming for SWIPT systems with non-linear energy harvesting model},'' in
  \emph{IEEE SPAWC}, Edinburgh, UK, Jul. 2016, pp. 1--5.

\bibitem{ref_lg1_app_5}
E.~Boshkovska, N.~Zlatanov, L.~Dai, D.~W.~K. Ng, and R.~Schober, ``{Secure
  SWIPT networks based on a non-linear energy harvesting model},'' in
  \emph{IEEE WCNCW}, San Francisco, CA, Mar. 2017, pp. 1--6.

\bibitem{ref_lg1_app_2}
E.~Boshkovska, R.~Morsi, D.~W.~K. Ng, and R.~Schober, ``{Power allocation and
  scheduling for SWIPT systems with non-linear energy harvesting model},'' in
  \emph{IEEE ICC}, Kuala Lumpur, Malaysia, May 2016, pp. 1--6.

\bibitem{ref_lg1_app_23}
E.~Boshkovska, D.~W.~K. Ng, L.~Dai, and R.~Schober, ``{Power-efficient and
  secure WPCNs with hardware impairments and non-linear EH circuit},''
  \emph{IEEE Trans. Commun.}, vol.~66, no.~6, pp. 2642--2657, 2017.

\bibitem{ref_lg1_app_4}
E.~Boshkovska, D.~W.~K. Ng, N.~Zlatanov, A.~Koelpin, and R.~Schober, ``{Robust
  resource allocation for MIMO wireless powered communication networks based on
  a non-linear EH model},'' \emph{IEEE Trans. Commun.}, vol.~65, no.~5, pp.
  1984--1999, 2017.

\bibitem{ref_lg1_app_6}
E.~Boshkovska, X.~Chen, L.~Dai, D.~W.~K. Ng, and R.~Schober, ``{Max-min fair
  beamforming for SWIPT systems with non-linear EH model},'' in \emph{IEEE VTC
  Fall}, Toronto, Canada, Sep. 2017, pp. 1--6.

\bibitem{ref_lg1_app_13}
R.~Morsi, E.~Boshkovska, E.~Ramadan, D.~W.~K. Ng, and R.~Schober, ``{On the
  performance of wireless powered communication with non-linear energy
  harvesting},'' in \emph{IEEE SPAWC}, Sapporo, Japan, Jul. 2017, pp. 1--5.

\bibitem{ref_mod_hr_0}
Y.~Chen, K.~T. Sabnis, and R.~A. Abd-Alhameed, ``{New formula for conversion
  efficiency of RF EH and its wireless applications},'' \emph{IEEE Trans. Veh.
  Technol.}, vol.~65, no.~11, pp. 9410--9414, 2016.

\bibitem{ref_mod_pw1_0}
Y.~{Dong}, M.~J. {Hossain}, and J.~{Cheng}, ``{Performance of wireless powered
  amplify and forward relaying over Nakagami-$m$ fading channels with nonlinear
  energy harvester},'' \emph{IEEE Commun. Lett.}, vol.~20, no.~4, pp. 672--675,
  2016.

\bibitem{ref_mod_pw2_0}
P.~N. Alevizos and A.~Bletsas, ``{Sensitive and nonlinear far-field RF energy
  harvesting in wireless communications},'' \emph{IEEE Trans. Wireless
  Commun.}, vol.~17, no.~6, pp. 3670--3685, 2018.

\bibitem{ref_mod_pw2_1}
L.~{Shi}, L.~{Zhao}, K.~{Liang}, X.~{Chu}, G.~{Wu}, and H.~{Chen}, ``{Profit
  maximization in wireless powered communications with improved non-linear
  energy conversion and storage efficiencies},'' in \emph{IEEE ICC}, 2017, pp.
  1--6.

\bibitem{ref_pw2_app_1}
D.~{Mishra}, S.~{De}, and D.~{Krishnaswamy}, ``{Dilemma at RF energy harvesting
  relay: downlink energy relaying or uplink information transfer?}'' \emph{IEEE
  Trans. Wireless Commun.}, vol.~16, no.~8, pp. 4939--4955, 2017.

\bibitem{ref_mod_lg2_0}
S.~Wang, M.~Xia, K.~Huang, and Y.-C. Wu, ``{Wirelessly powered two-way
  communication with nonlinear energy harvesting model: Rate regions under
  fixed and mobile relay},'' \emph{IEEE Trans. Wireless Commun.}, vol.~16,
  no.~12, pp. 8190--8204, 2017.

\bibitem{ref_lg2_app_1}
S.~Wang, M.~Xia, and Y.-C. Wu, ``{Multicast wirelessly powered network with
  large number of antennas via first-order method},'' \emph{IEEE Trans.
  Wireless Commun.}, vol.~17, no.~6, pp. 3781--3793, 2018.

\bibitem{ref_mod_2or_0}
X.~Xu, A.~{\"O}z{\c{c}}elikkale, T.~McKelvey, and M.~Viberg, ``{Simultaneous
  information and power transfer under a non-linear RF energy harvesting
  model},'' in \emph{IEEE ICC Wkshps}, Paris, France, May 2017, pp. 179--184.

\bibitem{ref_mod_frac_0}
Y.~Chen, N.~Zhao, and M.-S. Alouini, ``{Wireless energy harvesting using
  signals from multiple fading channels},'' \emph{IEEE Trans. Commun.},
  vol.~65, no.~11, pp. 5027--5039, 2017.

\bibitem{erf_mod_2020_0}
D.~Wang, F.~Rezaei, and C.~Tellambura, ``{Performance analysis and resource
  allocations for a WPCN with a new nonlinear energy harvester model},''
  \emph{IEEE Open J. Commun. Soc.}, vol.~1, pp. 1403--1424, 2020.

\bibitem{han2020joint}
J.~Han, G.~H. Lee, S.~Park, and J.~K. Choi, ``{Joint orthogonal band and power
  allocation for energy fairness in WPT system with nonlinear logarithmic
  energy harvesting model},'' \emph{arXiv preprint arXiv:2003.13255}, 2020.

\bibitem{Joint_mod_0}
D.~Alqahtani, Y.~Chen, W.~Feng, and M.-S. Alouini, ``{A new non-linear joint
  model for RF energy harvesters in wireless networks},'' \emph{IEEE Trans.
  Green Commun. Netw.}, vol.~5, no.~2, pp. 895--907, 2021.

\bibitem{fit_data}
T.~Le, K.~Mayaram, and T.~Fiez, ``{Efficient far-field radio frequency energy
  harvesting for passively powered sensor networks},'' \emph{IEEE Journal of
  Solid-State Circuits}, vol.~43, no.~5, pp. 1287--1302, 2008.

\bibitem{lg1_swipt_2021app5}
P.~V. {Tuan} and I.~{Koo}, ``{Optimizing efficient energy transmission on a
  SWIPT interference channel under linear/nonlinear EH models},'' \emph{IEEE
  Syst. J.}, vol.~14, no.~1, pp. 457--468, 2020.

\bibitem{lg1_swipt_2021app23}
T.~X. {Vu}, S.~{Chatzinotas}, S.~{Gautam}, E.~{Lagunas}, and B.~{Ottersten},
  ``{Joint optimization for PS-based SWIPT multiuser systems with non-linear
  energy harvesting},'' in \emph{IEEE WCNC}, 2020, pp. 1--6.

\bibitem{lg1_swipt_2021app30}
L.~{Zheng}, D.~{Liu}, Z.~{Wen}, and J.~{Zou}, ``{Robust beamforming for
  multi-user MISO full-duplex swipt system under non-linear energy harvesting
  model},'' \emph{IEEE Access}, vol.~9, pp. 14\,387--14\,397, 2021.

\bibitem{ref_lg1_app_36}
S.~Jang, H.~Lee, S.~Kang, T.~Oh, and I.~Lee, ``{Energy efficient SWIPT systems
  in multi-cell MISO networks},'' \emph{IEEE Trans. Wireless Commun.}, vol.~17,
  no.~12, pp. 8180--8194, 2018.

\bibitem{ref_lg1_app_11}
R.~{Jiang}, K.~{Xiong}, P.~{Fan}, S.~{Zhong}, and Z.~{Zhong}, ``{Optimal
  beamforming and power splitting design for SWIPT under non-linear energy
  harvesting model},'' in \emph{IEEE GLOBECOM}, 2017, pp. 1--6.

\bibitem{ref_lg1_app_12}
R.~{Jiang}, K.~{Xiong}, P.~{Fan}, Y.~{Zhang}, and Z.~{Zhong}, ``{Optimal design
  of SWIPT systems with multiple heterogeneous users under non-linear energy
  harvesting model},'' \emph{IEEE Access}, vol.~5, pp. 11\,479--11\,489, 2017.

\bibitem{ref_lg1_app_17}
Y.~Lu, K.~Xiong, P.~Fan, T.~Liu, and Z.~Zhong, ``{SWIPT for MISO wiretap
  networks: Channel uncertainties and nonlinear energy harvesting features},''
  in \emph{IEEE GLOBECOM}, Singapore, Singapore, Dec. 2017, pp. 1--7.

\bibitem{ref_lg1_app_42}
Y.~Lu, K.~Xiong, J.~Liu, D.~Wang, P.~Fan, and Z.~Zhong, ``{Secrecy energy
  efficiency in SWIPT networks with two-layer power-splitting receiver},'' in
  \emph{IEEE GLOBECOM Wkshps}, Abu Dhabi, United Arab Emirates, Dec. 2018, pp.
  1--7.

\bibitem{ref_lg1_app_44}
Y.~Lu, K.~Xiong, P.~Fan, Z.~Zhong, and K.~B. Letaief, ``{Robust transmit
  beamforming with artificial redundant signals for secure SWIPT system under
  non-linear EH model},'' \emph{IEEE Trans. Wireless Commun.}, vol.~17, no.~4,
  pp. 2218--2232, 2018.

\bibitem{lg1_swipt_2021app3}
R.~{Jiang}, K.~{Xiong}, P.~{Fan}, Y.~{Zhang}, and Z.~{Zhong}, ``{Power
  minimization in SWIPT networks with coexisting power-splitting and
  time-switching users under nonlinear eh model},'' \emph{IEEE Internet Things
  J.}, vol.~6, no.~5, pp. 8853--8869, 2019.

\bibitem{ref_lg1_app_43}
Y.~Lu, K.~Xiong, P.~Fan, Z.~Zhong, and K.~B. Letaief, ``{Coordinated
  beamforming with artificial noise for secure SWIPT under non-linear EH model:
  Centralized and distributed designs},'' \emph{IEEE J. Sel. Areas Commun.},
  vol.~36, no.~7, pp. 1544--1563, 2018.

\bibitem{ref_lg1_app_29}
H.-V. Tran, G.~Kaddoum, and K.~T. Truong, ``{Resource allocation in SWIPT
  networks under a nonlinear energy harvesting model: Power efficiency, user
  fairness, and channel nonreciprocity},'' \emph{IEEE Trans. Veh. Technol.},
  vol.~67, no.~9, pp. 8466--8480, 2018.

\bibitem{ref_lg1_app_51}
R.~Jiang, K.~Xiong, Y.~Zhang, L.~Zhou, T.~Liu, and Z.~Zhong, ``{Outage and
  throughput of WPCN-SWIPT networks with nonlinear EH model in Nakagami-$m$
  fading},'' \emph{Electronics}, vol.~8, no.~2, p. 138, 2019.

\bibitem{ref_lg1_app_46}
M.~Zhang, K.~Cumanan, L.~Ni, H.~Hu, A.~G. Burr, and Z.~Ding, ``{Robust
  beamforming for AN aided MISO SWIPT system with unknown eavesdroppers and
  non-linear EH model},'' in \emph{IEEE GLOBECOM Wkshps}, Abu Dhabi, United
  Arab Emirates, Dec. 2018, pp. 1--7.

\bibitem{ref_lg1_app_55}
Y.~Lu, K.~Xiong, P.~Fan, Z.~Ding, Z.~Zhong, and K.~B. Letaief, ``{Global energy
  efficiency in secure MISO SWIPT systems with non-linear power-splitting EH
  model},'' \emph{IEEE J. Sel. Areas Commun.}, vol.~37, no.~1, pp. 216--232,
  2019.

\bibitem{lg1_swipt_2021app19}
Y.~{Lu}, K.~{Xiong}, P.~{Fan}, Z.~{Ding}, Z.~{Zhong}, and K.~B. {Letaief},
  ``{Secrecy energy efficiency in multi-antenna SWIPT networks with dual-layer
  PS Receivers},'' \emph{IEEE Trans. Wireless Commun.}, vol.~19, no.~6, pp.
  4290--4306, 2020.

\bibitem{lg1_swipt_2021app29}
G.~{Sun}, M.~{Ma}, Z.~{Zhu}, J.~{Xu}, and W.~{Hao}, ``{Secrecy rate
  maximization in millimeter wave SWIPT systems based on non-linear energy
  harvesting},'' in \emph{IEEE ICCC}, 2020, pp. 1110--1115.

\bibitem{ref_lg1_app_22}
B.~Liu, Y.~Bai, G.~Lu, J.~Wang, and H.~Huang, ``{Optimal spectrum sensing
  interval in MISO cognitive small cell networks},'' \emph{IEEE Access},
  vol.~6, pp. 3479--3490, 2018.

\bibitem{ref_lg1_app_7}
H.~Niu, D.~Guo, Y.~Huang, and B.~Zhang, ``{Robust energy efficiency
  optimization for secure MIMO SWIPT systems with non-linear EH model},''
  \emph{IEEE Commun. Lett.}, vol.~21, no.~12, pp. 2610--2613, 2017.

\bibitem{ref_lg1_app_9}
K.~Xiong, B.~Wang, and K.~R. Liu, ``{Rate-energy region of SWIPT for MIMO
  broadcasting under nonlinear energy harvesting model},'' \emph{IEEE Trans.
  Wireless Commun.}, vol.~16, no.~8, pp. 5147--5161, 2017.

\bibitem{ref_lg1_app_21}
Y.~{Yuan} and Z.~{Ding}, ``{Secrecy outage design in MIMO-SWIPT systems based
  on a non-linear EH model},'' in \emph{IEEE GLOBECOM Wkshps}, 2017, pp. 1--6.

\bibitem{ref_lg1_app_50}
K.~Xu, Z.~Shen, M.~Zhang, Y.~Wang, X.~Xia, W.~Xie, and D.~Zhang, ``{Beam-domain
  SWIPT for mMIMO system with nonlinear energy harvesting legitimate terminals
  and a non-cooperative terminal},'' \emph{IEEE Trans. Green Commun. Netw.},
  vol.~3, no.~3, pp. 703--720, 2019.

\bibitem{lg1_swipt_2021app24}
Z.~{Zhu}, N.~{Wang}, W.~{Hao}, Z.~{Wang}, and I.~{Lee}, ``{Robust beamforming
  designs in secure MIMO SWIPT IoT networks with a nonlinear channel model},''
  \emph{IEEE Internet Things J.}, vol.~8, no.~3, pp. 1702--1715, 2021.

\bibitem{lg_2021_swipt8}
Z.~Zhu, Z.~Wang, Y.~Lin, P.~Liu, W.~Hao, Z.~Wang, and I.~Lee, ``{Robust max-min
  fair beamforming of secrecy SWIPT IoT systems under a non-linear EH model},''
  in \emph{IEEE ICC Wkshps}, 2021, pp. 1--6.

\bibitem{lg1_swipt_2021app26}
S.~{Kusaladharma}, W.~P. {Zhu}, W.~{Ajib}, and G.~A.~A. {Baduge}, ``{Stochastic
  geometry based performance characterization of SWIPT in cell-free massive
  MIMO},'' \emph{IEEE Trans. Veh. Technol.}, vol.~69, no.~11, pp.
  13\,357--13\,370, 2020.

\bibitem{ref_hr_lg_app_1}
F.~Benkhelifa and M.-S. Alouini, ``{Practical nonlinear energy harvesting model
  in MIMO DF relay system with channel uncertainty},'' in \emph{IEEE GLOBECOM},
  Abu Dhabi, United Arab Emirates, Dec. 2018, pp. 1--7.

\bibitem{lg1_swipt_2021app11}
G.~{Xu}, W.~{Tan}, and E.~{de Carvalho}, ``{Decentralized robust beamforming
  and power splitting for multi-relay assisted SWIPT systems},'' in \emph{IEEE
  WCNC}, 2019, pp. 1--7.

\bibitem{ref_lg1_app_38}
X.~Liu, Z.~Li, and C.~Wang, ``{Secure decode-and-forward relay SWIPT systems
  with power splitting schemes},'' \emph{IEEE Trans. Veh. Technol.}, vol.~67,
  no.~8, pp. 7341--7354, 2018.

\bibitem{lg_2021_swipt3}
Y.~Liu, Y.~Ye, and R.~Q. Hu, ``{Secrecy outage probability in backscatter
  communication systems with tag selection},'' \emph{IEEE Wireless Commun.
  Lett.}, vol.~10, no.~10, pp. 2190--2194, 2021.

\bibitem{ref_lg1_app_57}
A.~Banerjee and S.~P. Maity, ``{On residual energy maximization in cognitive
  relay networks with eavesdropping},'' \emph{IEEE Syst. J.}, vol.~13, no.~4,
  pp. 3836--3846, 2018.

\bibitem{lg1_swipt_2021app1}
S.~{Gautam}, E.~{Lagunas}, S.~{Chatzinotas}, and B.~{Ottersten}, ``{Relay
  selection and resource allocation for SWIPT in multi-user OFDMA systems},''
  \emph{IEEE Trans. Wireless Commun.}, vol.~18, no.~5, pp. 2493--2508, 2019.

\bibitem{lg1_swipt_2021app12}
B.~{Prasad}, R.~{Bhattacharjee}, and S.~K. {Bose}, ``{Cooperative communication
  under nonlinear energy harvesting model and interference signal},'' in
  \emph{IEEE TENCON}, 2019, pp. 1975--1980.

\bibitem{lg1_swipt_2021app14}
V.~{Bui}, V.~{Nguyen}, H.~V. {Nguyen}, O.~A. {Dobre}, and O.~{Shin},
  ``{Optimization of rate fairness in multi-pair wireless-powered relaying
  systems},'' \emph{IEEE Commun. Lett.}, vol.~24, no.~3, pp. 603--607, 2020.

\bibitem{lg1_swipt_2021app20}
Y.~{Zhang}, X.~{Jiang}, H.~{Hai}, J.~{Hau}, and K.~{Peng}, ``{Generalized
  non-linear energy harvesting protocol for enhancing security of AF
  multi-antenna relaying systems},'' in \emph{IEEE ICECE}, 2019, pp. 195--201.

\bibitem{lg1_swipt_2021app21}
W.~{Luo} and Y.~{Shen}, ``{Secure energy efficiency in SWIPT multicarrier relay
  transmission with direct links},'' in \emph{IEEE ICCC}, 2019, pp. 877--883.

\bibitem{lg_2021_swipt2}
Y.~Liu, Y.~Ye, J.~Ma, and Y.~Zhao, ``{Performance analysis for cooperative
  communication systems with a pure backscatter relay},'' in \emph{IEEE ICCC},
  2021, pp. 324--328.

\bibitem{lg_2021_swipt13}
S.~Kurma, P.~K. Sharma, V.~Panse, and K.~Singh, ``{Cooperative user selection
  with non-linear energy harvesting in IoT wnvironment},'' in \emph{IEEE VTC
  Fall}, 2021, pp. 1--6.

\bibitem{ref_lg1_app_8}
J.~{Kang}, I.~{Kim}, and D.~I. {Kim}, ``{Mode switching for SWIPT over fading
  channel with nonlinear energy harvesting},'' \emph{IEEE Wireless Commun.
  Lett.}, vol.~6, no.~5, pp. 678--681, 2017.

\bibitem{ref_lg1_app_30}
J.~H. Moon, J.~J. Park, and D.~I. Kim, ``{New reconfigurable nonlinear energy
  harvester: Boosting rate-energy tradeoff},'' in \emph{IEEE VTC Spring},
  Porto, Portugal, Jun. 2018, pp. 1--5.

\bibitem{ref_lg1_app_31}
J.~{Kang}, I.~{Kim}, and D.~I. {Kim}, ``{Joint optimal mode switching and power
  adaptation for nonlinear energy harvesting SWIPT system over fading
  channel},'' \emph{IEEE Trans. Commun.}, vol.~66, no.~4, pp. 1817--1832, 2018.

\bibitem{ref_lg1_app_32}
J.-M. Kang, I.-M. Kim, and D.~I. Kim, ``{Wireless information and power
  transfer: Rate-energy tradeoff for nonlinear energy harvesting},'' \emph{IEEE
  Trans. Wireless Commun.}, vol.~17, no.~3, pp. 1966--1981, 2017.

\bibitem{ref_lg1_app_52}
S.~{Kang}, H.~{Lee}, S.~{Jang}, H.~{Kim}, and I.~{Lee}, ``{Dynamic time
  switching for MIMO wireless information and power transfer},'' \emph{IEEE
  Trans. Commun.}, vol.~67, no.~6, pp. 3978--3990, 2019.

\bibitem{lg1_swipt_2021app17}
J.~{Kang}, C.~{Chun}, I.~{Kim}, and D.~I. {Kim}, ``{Dynamic power splitting for
  SWIPT with nonlinear energy harvesting in ergodic fading channel},''
  \emph{IEEE Internet Things J.}, vol.~7, no.~6, pp. 5648--5665, 2020.

\bibitem{lg1_swipt_2021app6}
R.~{Jiang}, K.~{Xiong}, P.~{Fan}, Z.~{Zhong}, and K.~B. {Letaief},
  ``{Information-energy region for SWIPT networks in mobility scenarios},''
  \emph{IEEE Trans. Veh. Technol.}, vol.~69, no.~7, pp. 7264--7280, 2020.

\bibitem{lg1_swipt_2021app7}
R.~{Jiang}, K.~{Xiong}, P.~{Fan}, D.~{Wang}, and Z.~{Zhong},
  ``{Information-energy region of mobile SWIPT networks with nonlinear EH
  model},'' in \emph{IEEE ICC}, 2019, pp. 1--6.

\bibitem{lg1_swipt_2021app4}
L.~{Li}, R.~{Cai}, H.~{Jiang}, and X.~{Su}, ``{Rate-energy tradeoff for SWIPT
  systems with multi-user interference channels under non-linear energy
  harvesting model},'' in \emph{IEEE VTC Spring}, 2019, pp. 1--6.

\bibitem{lg1_swipt_2021app2}
S.~A. {Tegos}, P.~D. {Diamantoulakis}, K.~N. {Pappi}, P.~C. {Sofotasios},
  S.~{Muhaidat}, and G.~K. {Karagiannidis}, ``{Toward efficient integration of
  information and energy reception},'' \emph{IEEE Trans. Commun.}, vol.~67,
  no.~9, pp. 6572--6585, 2019.

\bibitem{ref_lg1_app_35}
S.~{Gao}, K.~{Xiong}, R.~{Jiang}, L.~{Zhou}, and H.~{Tang}, ``{Outage
  performance of wireless-powered SWIPT networks with non-linear EH model in
  Nakagami-$m$ fading},'' in \emph{IEEE ICSP}, 2018, pp. 668--671.

\bibitem{ref_lg1_app_35_1}
Y.~Zhang, R.~Jiang, S.~Gao, K.~Xiong, L.~Zhou, and T.~Liu, ``{Outage
  Performance of Two-hop Networks with Multiple SWIPT Relays under Nonlinear EH
  Model},'' in \emph{IEEE ICSP}, 2018, pp. 752--757.

\bibitem{lg1_swipt_2021app27}
S.~{Gautam}, E.~{Lagunas}, S.~K. {Sharma}, S.~{Chatzinotas}, B.~{Ottersten},
  and L.~{Vandendorpe}, ``{Weighted sum-SINR and fairness optimization for
  SWIPT-multigroup multicasting systems with heterogeneous users},'' \emph{IEEE
  Open J. Commun. Soc.}, vol.~1, pp. 1470--1484, 2020.

\bibitem{lg_2020_swipt1}
S.~Gautam, E.~Lagunas, A.~Bandi, S.~Chatzinotas, S.~K. Sharma, T.~X. Vu,
  S.~Kisseleff, and B.~Ottersten, ``{Multigroup multicast precoding for energy
  optimization in swipt systems with heterogeneous users},'' \emph{IEEE Open J.
  Commun. Soc.}, vol.~1, pp. 92--108, 2020.

\bibitem{lg_2022_swipt14}
H.~An and H.~Park, ``{Energy-balancing resource allocation for wireless
  cooperative IoT networks with SWIPT},'' \emph{IEEE Internet Things J.},
  vol.~9, no.~14, pp. 12\,258--12\,271, 2022.

\bibitem{lg_2023app6}
J.~Jalali, A.~Khalili, A.~Rezaei, J.~Famaey, and W.~Saad, ``{Power-efficient
  antenna switching and beamforming design for multi-user SWIPT with non-linear
  energy harvesting},'' in \emph{IEEE CCNC}, 2023, pp. 746--751.

\bibitem{ref_lg1_app_19}
Y.~{Wang}, Y.~{Wu}, F.~{Zhou}, Y.~{Wu}, Z.~{Chu}, and Y.~{Wang},
  ``{Multi-objective resource allocation in NOMA cognitive radios based on a
  practical non-linear energy harvesting model},'' in \emph{WCSP}, 2017, pp.
  1--6.

\bibitem{ref_lg1_app_20}
Y.~{Wang}, Y.~{Wu}, F.~{Zhou}, Z.~{Chu}, Y.~{Wu}, and F.~{Yuan},
  ``{Multi-objective resource allocation in a NOMA cognitive radio network with
  a practical non-linear energy harvesting model},'' \emph{IEEE Access},
  vol.~6, pp. 12\,973--12\,982, 2018.

\bibitem{ref_lg1_app_53}
S.~{Mao}, S.~{Leng}, J.~{Hu}, and K.~{Yang}, ``{Power minimization resource
  allocation for underlay MISO-NOMA SWIPT systems},'' \emph{IEEE Access},
  vol.~7, pp. 17\,247--17\,255, 2019.

\bibitem{ref_lg1_app_48}
H.~{Sun}, F.~{Zhou}, R.~Q. {Hu}, and L.~{Hanzo}, ``{Robust Beamforming Design
  in a NOMA Cognitive Radio Network Relying on SWIPT},'' \emph{IEEE J. Sel.
  Areas Commun.}, vol.~37, no.~1, pp. 142--155, 2019.

\bibitem{ref_lg1_app_48_0}
H.~{Sun}, F.~{Zhou}, and Z.~{Zhang}, ``{Robust Beamforming Design in a NOMA
  Cognitive Radio Network Relying on SWIPT},'' in \emph{IEEE ICC}, 2018, pp.
  1--6.

\bibitem{lg1_swipt_2021app8}
V.~{Nguyen} and O.~{Shin}, ``{An efficient design for NOMA-assisted MISO-SWIPT
  systems with AC computing},'' \emph{IEEE Access}, vol.~7, pp.
  97\,094--97\,105, 2019.

\bibitem{lg1_swipt_2021app15}
T.~V. {Nguyen}, V.~D. {Nguyen}, D.~B. {da Costa}, and B.~{An}, ``{Hybrid user
  pairing for spectral and energy efficiencies in multiuser MISO-NOMA networks
  with SWIPT},'' \emph{IEEE Trans. Commun.}, vol.~68, no.~8, pp. 4874--4890,
  2020.

\bibitem{lg1_swipt_2021app13}
Q.~{Qi}, X.~{Chen}, and D.~W.~K. {Ng}, ``{Robust beamforming for NOMA-based
  cellular massive IoT with SWIPT},'' \emph{IEEE Trans. Signal Process.},
  vol.~68, pp. 211--224, 2020.

\bibitem{lg1_swipt_2021app16}
T.~{Nguyen}, V.~{Nguyen}, T.~{Do}, D.~B. {da Costa}, and B.~{An}, ``{Spectral
  efficiency maximization for multiuser MISO-NOMA downlink systems with
  SWIPT},'' in \emph{IEEE GLOBECOM}, 2019, pp. 1--6.

\bibitem{lg1_swipt_2021app28}
S.~{Bayat}, A.~{Khalili}, and Z.~{Han}, ``{Resource allocation for MC MISO-NOMA
  SWIPT-enabled HetNets with non-linear energy harvesting},'' \emph{IEEE
  Access}, vol.~8, pp. 192\,270--192\,281, 2020.

\bibitem{lg1_swipt_2021app22}
W.~{Wang}, J.~{Tang}, N.~{Zhao}, X.~{Liu}, X.~Y. {Zhang}, Y.~{Chen}, and
  Y.~{Qian}, ``{Joint precoding optimization for secure SWIPT in UAV-aided NOMA
  networks},'' \emph{IEEE Trans. Commun.}, vol.~68, no.~8, pp. 5028--5040,
  2020.

\bibitem{lg_2022_swipt17}
Z.~Zhu, M.~Ma, G.~Sun, W.~Hao, P.~Liu, Z.~Chu, and I.~Lee, ``{Secrecy rate
  optimization in nonlinear energy harvesting model-based mmWave IoT systems
  with SWIPT},'' \emph{IEEE Syst. J.}, pp. 1--11, 2022.

\bibitem{lg_2022_swipt16}
Y.~Lu, K.~Xiong, P.~Fan, Z.~Zhong, B.~Ai, and K.~B. Letaief, ``{Worst-case
  energy efficiency in secure SWIPT networks with rate-splitting ID and
  power-splitting eh receivers},'' \emph{IEEE Trans. Wireless Commun.},
  vol.~21, no.~3, pp. 1870--1885, 2022.

\bibitem{ref_lg1_app_15}
Y.~Huang, Z.~Li, F.~Zhou, and R.~Zhu, ``{Robust AN-aided beamforming design for
  secure MISO cognitive radio based on a practical nonlinear EH model},''
  \emph{IEEE Access}, vol.~5, pp. 14\,011--14\,019, 2017.

\bibitem{ref_lg1_app_26}
F.~Zhou, Z.~Chu, H.~Sun, R.~Q. Hu, and L.~Hanzo, ``{Artificial noise aided
  secure cognitive beamforming for cooperative MISO-NOMA using SWIPT},''
  \emph{IEEE J. Sel. Areas Commun.}, vol.~36, no.~4, pp. 918--931, 2018.

\bibitem{ref_lg1_app_28}
H.~Sun, F.~Zhou, R.~Q. Hu, and L.~Hanzo, ``{Robust beamforming design in a NOMA
  cognitive radio network relying on SWIPT},'' \emph{IEEE J. Sel. Areas
  Commun.}, vol.~37, no.~1, pp. 142--155, 2018.

\bibitem{ref_lg1_app_39}
X.~Zhang, Y.~Wang, F.~Zhou, N.~Al-Dhahir, and X.~Deng, ``{Robust resource
  allocation for MISO cognitive radio networks under two practical non-linear
  energy harvesting models},'' \emph{IEEE Commun. Lett.}, vol.~22, no.~9, pp.
  1874--1877, 2018.

\bibitem{ref_lg1_app_34}
Q.~{Qi} and X.~{Chen}, ``{Wireless powered massive access for cellular internet
  of things with imperfect SIC and nonlinear EH},'' \emph{IEEE Internet Things
  J.}, vol.~6, no.~2, pp. 3110--3120, 2019.

\bibitem{lg1_swipt_2021app10}
Q.~{Qi}, X.~{Chen}, D.~W.~K. {Ng}, C.~{Zhong}, and Z.~{Zhang}, ``{Robust
  beamforming design for SWIPT in cellular internet of things},'' in \emph{IEEE
  ICCC}, 2019, pp. 523--528.

\bibitem{lg_2021_swipt7}
Y.~Xu, H.~Xie, C.~Liang, and F.~R. Yu, ``{Robust secure energy-efficiency
  optimization in SWIPT-aided heterogeneous networks with a nonlinear
  energy-harvesting model},'' \emph{IEEE Internet Things J.}, vol.~8, no.~19,
  pp. 14\,908--14\,919, 2021.

\bibitem{lg_2022_swipt15}
J.~Zheng, L.~Gao, H.~Zhang, D.~Niyato, J.~Ren, H.~Wang, H.~Guo, and Z.~Wang,
  ``{eICIC configuration of downlink and uplink decoupling with SWIPT in 5G
  dense IoT HetNets},'' \emph{IEEE Trans. Wireless Commun.}, vol.~20, no.~12,
  pp. 8274--8287, 2021.

\bibitem{lg_2021_swipt4}
D.~Xu, V.~Jamali, X.~Yu, D.~W.~K. Ng, and R.~Schober, ``{Optimal resource
  allocation design for large IRS-assisted SWIPT systems: A scalable
  optimization framework},'' \emph{IEEE Trans. Commun.}, vol.~70, no.~2, pp.
  1423--1441, 2022.

\bibitem{lg_2021_swipt4_1}
D.~Xu, X.~Yu, V.~Jamali, D.~W.~K. Ng, and R.~Schober, ``{Resource allocation
  for large IRS-assisted SWIPT systems with non-linear energy harvesting
  model},'' in \emph{IEEE WCNC}, 2021, pp. 1--7.

\bibitem{lg_2022_swipt12}
Z.~Zhu, J.~Xu, G.~Sun, W.~Hao, Z.~Chu, C.~Pan, and I.~Lee, ``{Robust
  beamforming design for IRS-aided secure SWIPT Terahertz systems with
  non-linear EH model},'' \emph{IEEE Wireless Commun. Lett.}, vol.~11, no.~4,
  pp. 746--750, 2022.

\bibitem{lg_2021_swipt5}
J.~Liu, K.~Xiong, Y.~Lu, D.~W.~K. Ng, Z.~Zhong, and Z.~Han, ``{Energy
  efficiency in secure IRS-sided SWIPT},'' \emph{IEEE Wireless Commun. Lett.},
  vol.~9, no.~11, pp. 1884--1888, 2020.

\bibitem{lg_2021_swipt6}
S.~Zargari, A.~Khalili, Q.~Wu, M.~Robat~Mili, and D.~W.~K. Ng, ``{Max-min fair
  energy-efficient beamforming design for intelligent reflecting surface-aided
  SWIPT systems with non-linear energy harvesting model},'' \emph{IEEE Trans.
  Veh. Technol.}, vol.~70, no.~6, pp. 5848--5864, 2021.

\bibitem{lg_2023app2}
Y.~Fang, Y.~Tao, H.~Ma, Y.~Li, and M.~Guizani, ``{Design of a reconfigurable
  intelligent surface- assisted FM-DCSK-SWIPT scheme with non-linear energy
  harvesting model},'' \emph{IEEE Trans. Commun.}, vol.~71, no.~4, pp.
  1863--1877, 2023.

\bibitem{lg_2022_swipt11}
M.~R. Camana, C.~E. Garcia, and I.~Koo, ``{Rate-splitting multiple access in a
  MISO SWIPT system assisted by an intelligent reflecting surface},''
  \emph{IEEE Trans. Green Commun. Netw.}, vol.~6, no.~4, pp. 2084--2099, 2022.

\bibitem{lg_2021_swipt9}
Z.~Li, W.~Chen, H.~Cao, H.~Tang, K.~Wang, and J.~Li, ``{Joint communication and
  trajectory design for intelligent reflecting surface empowered UAV SWIPT
  networks},'' \emph{IEEE Trans. Veh. Technol.}, pp. 1--16, 2022.

\bibitem{lg_2023app8}
S.~Zargari, A.~Hakimi, C.~Tellambura, and S.~Herath, ``{User scheduling and
  trajectory optimization for energy-efficient IRS-UAV networks with SWIPT},''
  \emph{IEEE Trans. Veh. Technol.}, vol.~72, no.~2, pp. 1815--1830, 2023.

\bibitem{jrh_uav_1}
R.~Jiang, K.~Xiong, T.~Liu, D.~Wang, and Z.~Zhong, ``{Coverage
  probability-constrained maximum throughput in UAV-aided SWIPT networks},'' in
  \emph{IEEE ICC Workshops}, 2020, pp. 1--6.

\bibitem{jrh_uav_2}
R.~Jiang, K.~Xiong, H.-C. Yang, J.~Cao, Z.~Zhong, and B.~Ai, ``{Coverage
  performance of UAV-assisted SWIPT networks with directional antennas},''
  \emph{IEEE Internet Things J.}, vol.~9, no.~13, pp. 10\,600--10\,609, 2022.

\bibitem{jrh_uav_3}
R.~Jiang, K.~Xiong, H.-C. Yang, P.~Fan, Z.~Zhong, and K.~B. Letaief, ``{On the
  coverage of UAV-assisted SWIPT networks with Nonlinear EH model},''
  \emph{IEEE Trans. Wireless Commun.}, vol.~21, no.~6, pp. 4464--4481, 2022.

\bibitem{lg1_swipt_2021app18}
Y.~{Ye}, L.~{Shi}, X.~{Chu}, and G.~{Lu}, ``{On the outage performance of
  ambient backscatter communications},'' \emph{IEEE Internet Things J.},
  vol.~7, no.~8, pp. 7265--7278, 2020.

\bibitem{lg_2022_swipt10}
P.~Nezhadmohammad, M.~Abedi, M.~J. Emadi, and R.~Wichman, ``{SWIPT-enabled
  multiple access channel: Effects of decoding cost and non-linear EH model},''
  \emph{IEEE Trans. Commun.}, vol.~70, no.~1, pp. 306--316, 2022.

\bibitem{lg_2023app7}
Z.~Li, W.~Chen, Z.~Zhang, Q.~Wu, H.~Cao, and J.~Li, ``{Robust sum-rate
  maximization in transmissive RMS transceiver-enabled SWIPT networks},''
  \emph{IEEE Internet Things J.}, vol.~10, no.~8, pp. 7259--7271, 2023.

\bibitem{ref_lg1_app_16}
Y.~{Kuang}, L.~{Zhao}, and H.~{Zhao}, ``{Efficiency and fairness of energy
  broadcasting systems with centralized massive MIMO},'' in \emph{IEEE ICCC},
  2017, pp. 1--5.

\bibitem{ref_lg1_app_47}
G.~{Ma}, J.~{Xu}, Y.~{Zeng}, and M.~R.~V. {Moghadam}, ``{A generic receiver
  architecture for MIMO wireless power transfer with nonlinear energy
  harvesting},'' \emph{IEEE Signal Process. Lett.}, vol.~26, no.~2, pp.
  312--316, 2019.

\bibitem{ref_lg1_app_54}
X.~{Jia}, C.~{Zhang}, and I.~{Kim}, ``{Worst-case robust beamforming design for
  wireless powered multirelay multiuser network with a nonlinear EH model},''
  \emph{IEEE Trans. Veh. Technol.}, vol.~68, no.~3, pp. 3038--3042, 2019.

\bibitem{lg1_2021app19}
T.~T. {Nguyen}, V.~D. {Nguyen}, Q.~V. {Pham}, J.~H. {Lee}, and Y.~H. {Kim},
  ``{Resource allocation for AF relaying wireless-powered networks with
  nonlinear energy harvester},'' \emph{IEEE Commun. Lett.}, vol.~25, no.~1, pp.
  229--233, 2021.

\bibitem{ref_lg1_app_37}
T.~X. {Tran}, W.~{Wang}, S.~{Luo}, and K.~C. {Teh}, ``{Nonlinear energy
  harvesting for millimeter wave networks with large-scale antennas},''
  \emph{IEEE Trans. Veh. Technol.}, vol.~67, no.~10, pp. 9488--9498, 2018.

\bibitem{ref_lg1_app_56}
Z.~{Li}, Y.~{Jiang}, Y.~{Gao}, L.~{Sang}, and D.~{Yang}, ``{On
  buffer-constrained throughput of a wireless-powered communication system},''
  \emph{IEEE J. Sel. Areas Commun.}, vol.~37, no.~2, pp. 283--297, 2019.

\bibitem{lg1_2021app5}
X.~{Liu}, Y.~{Gao}, M.~{Guo}, and N.~{Sha}, ``{Secrecy throughput optimization
  for the WPCNs with non-linear EH model},'' \emph{IEEE Access}, vol.~7, pp.
  59\,477--59\,490, 2019.

\bibitem{lg1_2021app12}
Z.~{Mao}, F.~{Hu}, D.~{Sun}, S.~{Ma}, and X.~{Liu}, ``{Fairness-aware
  intragroup cooperative transmission in wireless powered communication
  networks},'' \emph{IEEE Trans. Veh. Technol.}, vol.~69, no.~6, pp.
  6463--6472, 2020.

\bibitem{lg1_2021app8}
T.~{Nguyen}, V.~{Nguyen}, J.~{Lee}, and Y.~{Kim}, ``{Sum rate maximization for
  multi-user wireless powered IoT network with non-linear energy harvester:
  Time and power allocation},'' \emph{IEEE Access}, vol.~7, pp.
  149\,698--149\,710, 2019.

\bibitem{lg_2021_wpcn10}
J.~Zan, Y.~Ye, R.~Q. Hu, and G.~Lu, ``{Stochastic geometry based performance
  study for wireless-powered backscatter communications},'' \emph{IEEE Trans.
  Veh. Technol.}, pp. 1--14, 2022.

\bibitem{ref_lg1_app_18}
Y.~{Wang}, Y.~{Wang}, F.~{Zhou}, Y.~{Wu}, and H.~{Zhou}, ``{Resource allocation
  in wireless powered cognitive radio networks based on a practical non-linear
  energy harvesting model},'' \emph{IEEE Access}, vol.~5, pp. 17\,618--17\,626,
  2017.

\bibitem{lg1_2021app9}
L.~{Ni}, X.~{Da}, H.~{Hu}, M.~{Zhang}, and K.~{Cumanan}, ``{Outage constrained
  robust secrecy energy efficiency maximization for EH cognitive radio
  networks},'' \emph{IEEE Wireless Commun. Lett.}, vol.~9, no.~3, pp. 363--366,
  2020.

\bibitem{lg1_2021app11}
L.~{Ni}, X.~{Da}, H.~{Hu}, Y.~{Yuan}, Z.~{Zhu}, and Y.~{Pan},
  ``{Outage-constrained secrecy energy efficiency optimization for CRNs with
  non-linear energy harvesting},'' \emph{IEEE Access}, vol.~7, pp.
  175\,213--175\,221, 2019.

\bibitem{lg1_2021app15}
F.~{Wang} and X.~{Zhang}, ``{Secure resource allocation for polarization-based
  non-linear energy harvesting over 5G cooperative CRNs},'' \emph{IEEE Wireless
  Commun. Lett.}, pp. 1--1, 2020.

\bibitem{lg1_2021app16}
------, ``{Secure resource allocation for polarization-based non-linear energy
  harvesting over 5G cooperative cognitive radio networks},'' in \emph{IEEE
  ICC}, 2020, pp. 1--6.

\bibitem{lg1_2021app6}
C.~K. {Vranas}, P.~S. {Bouzinis}, V.~K. {Papanikolaou}, P.~D. {Diamantoulakis},
  and G.~K. {Karagiannidis}, ``{On the gain of NOMA in wireless powered
  networks with circuit power consumption},'' \emph{IEEE Commun. Lett.},
  vol.~23, no.~9, pp. 1657--1660, 2019.

\bibitem{lg1_2021app20}
H.~Zheng, K.~Xiong, P.~Fan, Z.~Zhong, Z.~Ding, and K.~B. Letaief, ``{Achievable
  computation rate in NOMA-based wireless-powered networks assisted by multiple
  fog servers},'' \emph{IEEE Internet Things J.}, vol.~8, no.~6, pp.
  4802--4815, 2021.

\bibitem{lg_2021_wpcn4}
------, ``{Achievable computation rate in NOMA-based wireless-powered networks
  assisted by multiple fog servers},'' \emph{IEEE Internet Things J.}, vol.~8,
  no.~6, pp. 4802--4815, 2021.

\bibitem{ref_lg1_app_45}
Z.~{Yang}, W.~{Xu}, Y.~{Pan}, C.~{Pan}, and M.~{Chen}, ``{Energy efficient
  resource allocation in machine-to-machine communications with multiple access
  and energy harvesting for IoT},'' \emph{IEEE Internet Things J.}, vol.~5,
  no.~1, pp. 229--245, 2018.

\bibitem{lg_2021_wpcn2}
B.~Su, Q.~Ni, W.~Yu, and H.~Pervaiz, ``{Optimizing computation efficiency for
  NOMA-assisted mobile edge computing with user cooperation},'' \emph{IEEE
  Trans. Green Commun. Netw.}, vol.~5, no.~2, pp. 858--867, 2021.

\bibitem{lg_2022_wpcn13}
Z.~Wang, T.~Lv, and W.~Li, ``{Energy efficiency maximization in massive
  MIMO-NOMA networks with non-linear energy harvesting},'' in \emph{IEEE WCNC},
  2021, pp. 1--6.

\bibitem{lg_2021_wpcn8}
M.~Wu, W.~Qi, J.~Park, P.~Lin, L.~Guo, and I.~Lee, ``{Residual energy
  maximization for wireless powered mobile edge computing systems with
  mixed-offloading},'' \emph{IEEE Trans. Veh. Technol.}, vol.~71, no.~4, pp.
  4523--4528, 2022.

\bibitem{lg_2021_wpcn9}
Z.~Li, W.~Chen, Q.~Wu, H.~Cao, K.~Wang, and J.~Li, ``{Robust beamforming design
  and time allocation for IRS-assisted wireless powered communication
  networks},'' \emph{IEEE Trans. Commun.}, vol.~70, no.~4, pp. 2838--2852,
  2022.

\bibitem{lg_2022_wpcn12}
M.~Hua, Q.~Wu, and H.~V. Poor, ``{Power-efficient passive beamforming and
  resource allocation for IRS-Aaided WPCNs},'' \emph{IEEE Trans. Commun.},
  vol.~70, no.~5, pp. 3250--3265, 2022.

\bibitem{lg_2022_wpcn14}
M.~Hua and Q.~Wu, ``{Energy minimization for IRS-aided WPCNs with non-linear
  power-splitting EH model},'' in \emph{IEEE WCNC}, 2022, pp. 1057--1062.

\bibitem{lg_2021_wpcn11}
------, ``{Throughput maximization for IRS-aided MIMO FD-WPCN with non-linear
  EH model},'' \emph{IEEE J. Sel. Topics Signal Process.}, vol.~16, no.~5, pp.
  918--932, 2022.

\bibitem{lg_2023app1}
C.~Qiu, Q.~Wu, M.~Hua, X.~Guan, and Y.~Wu, ``{Achieving multi-beam gain in
  intelligent reflecting surface assisted wireless energy transfer},''
  \emph{IEEE Trans. Veh. Technol.}, vol.~72, no.~3, pp. 4052--4057, 2023.

\bibitem{lg_2023app4}
C.~Kumar and S.~Kashyap, ``{On the power transfer efficiency and feasibility of
  wireless energy transfer using double IRS},'' \emph{IEEE Trans. Veh.
  Technol.}, vol.~72, no.~5, pp. 6165--6180, 2023.

\bibitem{ref_lg1_app_41}
Y.~Dong, J.~Cheng, M.~J. Hossain, and V.~C. Leung, ``{Extracting the most
  weighted throughput in UAV empowered wireless systems with nonlinear energy
  harvester},'' in \emph{Biennial Symposium on Communications (BSC)}, Toronto,
  ON, Canada, Jun. 2018, pp. 1--5.

\bibitem{lg1_2021app7}
J.~{Park}, H.~{Lee}, S.~{Eom}, and I.~{Lee}, ``{UAV-aided wireless powered
  communication networks: Trajectory optimization and resource allocation for
  minimum throughput maximization},'' \emph{IEEE Access}, vol.~7, pp.
  134\,978--134\,991, 2019.

\bibitem{lg1_2021app13}
Q.~{Zhang}, Z.~{Wang}, P.~{Zhang}, H.~{Zhang}, X.~{Wan}, and Z.~{Fan}, ``{Sum
  energy maximization for UAV-enabled wireless power transfer networks with
  nonlinear energy harvesting model},'' in \emph{IEEE ITNEC}, vol.~1, 2020, pp.
  1417--1420.

\bibitem{lg1_2021app17}
H.~Hu, K.~Xiong, G.~Qu, Q.~Ni, P.~Fan, and K.~B. Letaief, ``{AoI-minimal
  trajectory planning and data collection in UAV-assisted wireless powered IoT
  networks},'' \emph{IEEE Internet Things J.}, vol.~8, no.~2, pp. 1211--1223,
  2021.

\bibitem{lg_2021_wpcn1}
P.~Zhang, Z.~Wang, Q.~Zhang, Y.~Liu, X.~Wan, and Z.~Fan, ``{Max-min placement
  optimization for UAV enabled wireless powered networks with non-linear energy
  harvesting model},'' in \emph{IEEE ICCCBDA}, 2020, pp. 436--439.

\bibitem{lg_2021_wpcn7}
X.~He, Y.~Zhao, Z.~Xu, and Y.~Chen, ``{Resource allocation strategy for
  UAV-assisted non-linear energy harvesting MEC system},'' in \emph{IEEE VTC
  Spring}, 2022, pp. 1--7.

\bibitem{lg1_2021app10}
Y.~{Zou} and Z.~{Yang}, ``{Throughput maximization for wireless powered
  multimedia communication systems under statistical latency constraint},''
  \emph{IEEE Access}, vol.~7, pp. 175\,816--175\,826, 2019.

\bibitem{lg1_swipt_2021app9}
Y.~{Lu}, K.~{Xiong}, P.~{Fan}, Z.~{Zhong}, and K.~B. {Letaief}, ``{Online
  transmission policy in wireless powered networks with urgency-aware age of
  information},'' in \emph{IWCMC}, 2019, pp. 1096--1101.

\bibitem{lg1_2021app14}
S.~{Idrees}, X.~{Zhou}, S.~{Durrani}, and D.~{Niyato}, ``{Design of ambient
  backscatter training for wireless power transfer},'' \emph{IEEE Trans.
  Wireless Commun.}, vol.~19, no.~10, pp. 6316--6330, 2020.

\bibitem{ref_lg1_app_14}
Y.~{Dong}, M.~J. {Hossain}, J.~{Cheng}, and V.~C.~M. {Leung}, ``{Dynamic
  cross-layer beamforming in hybrid powered communication systems with
  harvest-use-trade strategy},'' \emph{IEEE Trans. Wireless Commun.}, vol.~16,
  no.~12, pp. 8011--8025, 2017.

\bibitem{lg_2021_wpcn0}
S.~A. Tegos, G.~K. Karagiannidis, P.~D. Diamantoulakis, and N.~D.
  Chatzidiamantis, ``{Nonlinear energy harvesting evaluation through the Logit
  pearson distribution},'' in \emph{IEEE SPAWC}, 2021, pp. 611--615.

\bibitem{lg_2021_wpcn5}
S.~Gautam, S.~K. Sharma, S.~Chatzinotas, and B.~Ottersten, ``{Modeling and
  optimization of RF-energy harvesting-assisted quantum battery system},'' in
  \emph{IEEE VTC Spring}, 2021, pp. 1--6.

\bibitem{lg_2021_wpcn6}
M.~Zeng, Y.~Luo, H.~Jiang, and Y.~Wang, ``{A joint cluster formation scheme
  with multi-layer awareness for energy-harvesting supported D2D multicast
  communication},'' \emph{IEEE Trans. Wireless Commun.}, vol.~21, no.~9, pp.
  7595--7608, 2022.

\bibitem{ref_hr_lg_app_2}
L.~Shi, L.~Zhao, and K.~Liang, ``{Power allocation for wireless powered MIMO
  transmissions with non-linear RF energy conversion models},'' \emph{China
  Commun.}, vol.~14, no.~2, pp. 57--64, 2017.

\bibitem{ref_lg1_app_33}
N.~K.~D. {Venkategowda}, H.~{Lee}, and I.~{Lee}, ``{Joint transceiver designs
  for MSE minimization in MIMO wireless powered sensor networks},'' \emph{IEEE
  Trans. Wireless Commun.}, vol.~17, no.~8, pp. 5120--5131, 2018.

\bibitem{ref_pw1_app_1}
T.~M. Hoang, T.~T. Duy, and V.~N.~Q. Bao, ``{On the performance of non-linear
  wirelessly powered partial relay selection networks over Rayleigh fading
  channels},'' in \emph{NICS}, Danang, Vietnam, Sept. 2016, pp. 6--11.

\bibitem{ref_pw1_app_2}
J.~Zhang and G.~Pan, ``{Outage analysis of wireless-powered relaying MIMO
  systems with non-linear energy harvesters and imperfect CSI},'' \emph{IEEE
  Access}, vol.~4, pp. 7046--7053, 2016.

\bibitem{ref_pw1_app_5}
K.~Wang, Y.~Li, Y.~Ye, and H.~Zhang, ``{Dynamic power splitting schemes for
  non-linear EH relaying networks: Perfect and imperfect CSI},'' in \emph{IEEE
  VTC Fall}, Toronto, ON, Canada, Sept. 2017, pp. 1--5.

\bibitem{ref_pw1_app_6}
Y.~Feng, M.~Wen, F.~Ji, and V.~C. Leung, ``{Performance analysis for BDPSK
  modulated SWIPT cooperative systems with nonlinear energy harvesting
  model},'' \emph{IEEE Access}, vol.~6, pp. 42\,373--42\,383, 2018.

\bibitem{ref_pw1_app_7}
Y.~{Zhang}, R.~{Jiang}, S.~{Gao}, K.~{Xiong}, L.~{Zhou}, and T.~{Liu},
  ``{Outage performance of two-hop networks with multiple SWIPT relays under
  nonlinear EH model},'' in \emph{IEEE ICSP}, 2018, pp. 752--757.

\bibitem{pw2021app3}
S.~{Solanki}, P.~K. {Upadhyay}, D.~B. {da Costa}, H.~{Ding}, and J.~M.
  {Moualeu}, ``{Non-linear energy harvesting based cooperative spectrum sharing
  networks},'' in \emph{ISWCS}, 2019, pp. 566--570.

\bibitem{pw2021app12}
S.~{Solanki}, P.~K. {Upadhyay}, D.~B.~D. {Costa}, H.~{Ding}, and J.~M.
  {Moualeu}, ``{Performance analysis of piece-wise linear model of energy
  harvesting-based multiuser overlay spectrum sharing networks},'' \emph{IEEE
  Open J. Commun. Soc.}, vol.~1, pp. 1820--1836, 2020.

\bibitem{pw2021app4}
R.~{Jiang}, K.~{Xiong}, P.~{Fan}, L.~{Zhou}, and Z.~{Zhong}, ``{Outage
  probability and throughput of multirelay SWIPT-WPCN networks with nonlinear
  EH model and imperfect CSI},'' \emph{IEEE Syst. J.}, vol.~14, no.~1, pp.
  1206--1217, 2020.

\bibitem{pw2021app5}
T.~L.~N. {Nguyen} and Y.~{Shin}, ``{Outage probability analysis for SWIPT
  systems with nonlinear energy harvesting model},'' in \emph{ICTC}, 2019, pp.
  196--199.

\bibitem{pw2021app7}
B.~{Li}, M.~{Zhang}, H.~{Cao}, Y.~{Rong}, and Z.~{Han}, ``{Transceiver design
  for AF MIMO relay systems with a power splitting based energy harvesting
  relay node},'' \emph{IEEE Trans. Veh. Technol.}, vol.~69, no.~3, pp.
  2376--2388, 2020.

\bibitem{pw2021app2}
X.~{Li}, M.~{Huang}, C.~{Zhang}, D.~{Deng}, K.~M. {Rabie}, Y.~{Ding}, and
  J.~{Du}, ``{Security and reliability performance analysis of cooperative
  multi-relay systems with nonlinear energy harvesters and hardware
  impairments},'' \emph{IEEE Access}, vol.~7, pp. 102\,644--102\,661, 2019.

\bibitem{pw2021app14}
M.~{Babaei}, L.~{Durak-Ata}, and U.~{Ayg\"ol\"u}, ``{Performance analysis of
  linear/nonlinear energy harvesting with dual-hop DF relaying},'' in
  \emph{SIU}, 2020, pp. 1--4.

\bibitem{pw2_2022_swipt5}
K.~Agrawal, A.~Jee, and S.~Prakriya, ``{Performance of SWIPT in cooperative
  networks with direct link and nonlinear energy harvesting at the
  battery-assisted relay},'' \emph{IEEE Trans. Green Commun. Netw.}, vol.~6,
  no.~2, pp. 1198--1215, 2022.

\bibitem{ref_pw1_app_9}
J.-M. Kang, I.-M. Kim, and D.~I. Kim, ``{Joint Tx power allocation and Rx power
  splitting for SWIPT system with multiple nonlinear energy harvesting
  circuits},'' \emph{IEEE Wireless Commun. Lett.}, vol.~8, no.~1, pp. 53--56,
  2018.

\bibitem{pw2_2020_swipt1}
R.~Gupta and I.~Krikidis, ``{A new receiver design: Simultaneous wireless power
  transfer with modulation classification},'' in \emph{IEEE WPTC}, 2020, pp.
  331--333.

\bibitem{pw2_2021_swipt2}
S.~Mahama, D.~K.~P. Asiedu, Y.~J. Harbi, K.-J. Lee, D.~Grace, and A.~G. Burr,
  ``{Multi-user wireless information and power transfer in FBMC-based IoT
  networks},'' \emph{IEEE Open J. Commun. Soc.}, vol.~2, pp. 545--563, 2021.

\bibitem{pw2021app1}
D.~S. {Gurjar}, H.~H. {Nguyen}, and P.~{Pattanayak}, ``{Performance of wireless
  powered cognitive radio sensor networks with nonlinear energy harvester},''
  \emph{IEEE Sensors Lett.}, vol.~3, no.~8, pp. 1--4, 2019.

\bibitem{pw2021app6}
P.~{Maji}, S.~D. {Roy}, and S.~{Kundu}, ``{Physical layer security with
  non-linear energy harvesting relay},'' in \emph{ICCCNT}, 2019, pp. 1--6.

\bibitem{pw2021app10}
A.~{Prathima}, D.~S. {Gurjar}, H.~H. {Nguyen}, and A.~{Bhardwaj},
  ``{Performance analysis and optimization of bidirectional overlay cognitive
  radio networks with hybrid-SWIPT},'' \emph{IEEE Trans. Veh. Technol.},
  vol.~69, no.~11, pp. 13\,467--13\,481, 2020.

\bibitem{pw2021app13}
A.~{Prathima}, D.~S. {Gurjar}, and S.~{Yadav}, ``{Two-way cooperative cognitive
  radio networks with nonlinear RF-energy harvester},'' in \emph{IEEE
  LATINCOM}, 2020, pp. 1--6.

\bibitem{pw2021app8}
W.~{Chen}, H.~{Ding}, S.~{Wang}, D.~B. {da Costa}, and F.~{Gong}, ``{Impartial
  SWIPT-assisted user cooperation schemes},'' \emph{IEEE Trans. Wireless
  Commun.}, vol.~19, no.~5, pp. 3361--3375, 2020.

\bibitem{pw2021app9}
K.~{Agrawal}, M.~{F. Flanagan}, and S.~{Prakriya}, ``{NOMA with
  battery-assisted energy harvesting full-duplex relay},'' \emph{IEEE Trans.
  Veh. Technol.}, vol.~69, no.~11, pp. 13\,952--13\,957, 2020.

\bibitem{pw2021app11}
Y.~Zhang, S.~Feng, and W.~Tang, ``{Performance analysis and optimization for
  power beacon-assisted wireless powered cooperative NOMA systems},''
  \emph{IEEE Access}, vol.~8, pp. 198\,436--198\,450, 2020.

\bibitem{pw2_2021_swipt3}
L.~Ma, E.~Li, and Q.~Yang, ``{On the performance of full-duplex cooperative
  NOMA with non-linear EH},'' \emph{IEEE Access}, vol.~9, pp.
  145\,968--145\,976, 2021.

\bibitem{pw2_2021_swipt4}
U.~Makhanpuri, K.~Agrawal, A.~Jee, and S.~Prakriya, ``{Performance of
  full-duplex cooperative NOMA network with nonlinear energy harvesting},'' in
  \emph{IEEE PIMRC}, 2021, pp. 495--500.

\bibitem{ref_pw2_app_3}
X.~{Xie}, J.~{Chen}, and Y.~{Fu}, ``{Outage performance and QoS optimization in
  full-duplex system with non-linear energy harvesting model},'' \emph{IEEE
  Access}, vol.~6, pp. 44\,281--44\,290, 2018.

\bibitem{ref_pw1_app_4}
J.~Zhang, G.~Pan, and Y.~Xie, ``{Secrecy analysis of wireless-powered
  multi-antenna relaying system with nonlinear energy harvesters and imperfect
  CSI},'' \emph{IEEE Trans. Green Commun. and Netw.}, vol.~2, no.~2, pp.
  460--470, 2017.

\bibitem{ref_pw1_app_8}
S.~{Pejoski}, Z.~{Hadzi-Velkov}, and R.~{Schober}, ``{Optimal power and time
  allocation for WPCNs with piece-wise linear EH Model},'' \emph{IEEE Wireless
  Commun. Lett.}, vol.~7, no.~3, pp. 364--367, 2018.

\bibitem{pw2_2022_wpcn1}
N.~Shanin, L.~Cottatellucci, and R.~Schober, ``{Optimal transmit strategy for
  multi-user MIMO WPT systems with non-linear energy harvesters},'' \emph{IEEE
  Trans. Commun.}, vol.~70, no.~3, pp. 1726--1741, 2022.

\bibitem{pw2_2021_wpcn1}
------, ``{Optimal transmit strategy for MIMO WPT systems with non-linear
  energy harvesting},'' in \emph{DCOSS}, 2021, pp. 520--527.

\bibitem{pw2_2022_wpcn2}
K.~Xiong, Y.~Liu, L.~Zhang, B.~Gao, J.~Cao, P.~Fan, and K.~B. Letaief, ``{Joint
  optimization of trajectory, task offloading, and CPU control in UAV-assisted
  wireless powered fog computing networks},'' \emph{IEEE Trans. Green Commun.
  Netw.}, vol.~6, no.~3, pp. 1833--1845, 2022.

\bibitem{pw2_2023app2}
L.~Zhang, J.~Zhang, N.~Hu, X.~Li, and G.~Pan, ``{Outage performance for
  NOMA-based FSO-RF systems with transmit antenna selection and nonlinear
  energy harvesting},'' \emph{IEEE Internet Things J.}, vol.~10, no.~7, pp.
  6491--6506, 2023.

\bibitem{pw2_2021app1}
L.~{Shi}, W.~{Cheng}, Y.~{Ye}, H.~{Zhang}, and R.~Q. {Hu}, ``{Heterogeneous
  power-splitting based two-way df relaying with non-linear energy
  harvesting},'' in \emph{IEEE GLOBECOM}, 2018, pp. 1--7.

\bibitem{ref_pw2_app_4}
G.~Lu, L.~Shi, and Y.~Ye, ``{Maximum throughput of TS/PS scheme in an AF
  relaying network with non-linear energy harvester},'' \emph{IEEE Access},
  vol.~6, pp. 26\,617--26\,625, 2018.

\bibitem{ref_pw2_app_6}
L.~Shi, Y.~Ye, R.~Q. Hu, and H.~Zhang, ``{Energy efficiency maximization for
  SWIPT enabled two-way DF relaying},'' \emph{IEEE Signal Process. Lett.},
  vol.~26, no.~5, pp. 755--759, 2019.

\bibitem{pw2_2021app3}
L.~{Shi}, Y.~{Ye}, R.~Q. {Hu}, and H.~{Zhang}, ``{Energy efficiency
  maximization for SWIPT enabled two-way DF relaying},'' \emph{IEEE Signal
  Process. Lett.}, vol.~26, no.~5, pp. 755--759, 2019.

\bibitem{pw2_2021app8}
P.~{Raut}, P.~K. {Sharma}, T.~A. {Tsiftsis}, and Y.~{Zou}, ``{Power-time
  splitting-based non-linear energy harvesting in FD short-packet
  communications},'' \emph{IEEE Trans. Veh. Technol.}, vol.~69, no.~8, pp.
  9146--9151, 2020.

\bibitem{pw2_2021app12}
S.~A.~A. {Kazmi} and S.~{Coleri}, ``{Optimization of full-duplex relaying
  system with non-linear energy harvester},'' \emph{IEEE Access}, vol.~8, pp.
  201\,566--201\,576, 2020.

\bibitem{pw2_2021app14}
Y.~{Liu}, F.~{Gao}, X.~{Deng}, T.~{Wu}, and X.~{Zhang}, ``{Performance analysis
  for incremental DF relaying networks with non-linear energy harvesting},'' in
  \emph{IEEE ICCT}, 2020, pp. 354--360.

\bibitem{pw2_2021app13}
S.~{Xu}, X.~{Song}, L.~{Xia}, J.~{Cao}, H.~{Qi}, and Z.~{Xie}, ``{Optimal power
  allocation for non-linear EH cooperative network with multiple
  eavesdroppers},'' in \emph{IEEE IES}, 2020, pp. 3913--3917.

\bibitem{pw3_2021_swipt2}
S.~Wang, Z.~He, and Y.~Rong, ``{Joint transceiver optimization for DF
  multicasting MIMO relay systems with wireless information and power
  transfer},'' \emph{IEEE Trans. Commun.}, vol.~69, no.~7, pp. 4953--4967,
  2021.

\bibitem{pw3_2021_swipt3}
Y.~Chen and F.~Gao, ``{Secrecy outage analysis for secure DF cooperative
  networks based relay selection scheme with non-linear energy harvester},'' in
  \emph{IEEE ICCT}.\hskip 1em plus 0.5em minus 0.4em\relax IEEE, 2021, pp.
  344--348.

\bibitem{pw3_2022_swipt5}
V.~Panse, T.~K. Jain, and A.~Kothari, ``{Relay selection in SWIPT-enabled
  cooperative networks},'' in \emph{PCEMS}, 2022, pp. 62--67.

\bibitem{pw2_2021app4}
J.~H. {Moon}, J.~J. {Park}, and D.~I. {Kim}, ``{Reconfigurable heterogeneous
  energy harvester with adaptive mode switching},'' in \emph{IEEE SPAWC}, 2019,
  pp. 1--5.

\bibitem{pw2_2021app11}
J.~H. {Moon}, J.~J. {Park}, K.~Y. {Lee}, and D.~I. {Kim}, ``{Heterogeneously
  reconfigurable energy harvester: An algorithm for optimal reconfiguration},''
  \emph{IEEE Internet Things J.}, vol.~8, no.~3, pp. 1437--1452, 2021.

\bibitem{pw2_2021app9}
J.~J. {Park}, J.~H. {Moon}, K.~Y. {Lee}, and D.~I. {Kim},
  ``{Transmitter-oriented dual-mode SWIPT with deep-learning-based adaptive
  mode switching for IoT sensor networks},'' \emph{IEEE Internet Things J.},
  vol.~7, no.~9, pp. 8979--8992, 2020.

\bibitem{pw3_2021_swipt1}
J.~H. Moon, J.~J. Park, K.-Y. Lee, and D.~I. Kim, ``{Heterogeneously
  reconfigurable energy harvester: An algorithm for optimal reconfiguration},''
  \emph{IEEE Internet Things J.}, vol.~8, no.~3, pp. 1437--1452, 2020.

\bibitem{pw3_2022_swipt4}
E.~Goudeli, C.~Psomas, I.~Krikidis, H.~Kiani, D.~Chatzichristodoulou, and
  S.~Nikolaou, ``{Detection schemes for integrated SWIPT receivers with
  non-linear energy harvesting},'' in \emph{IEEE VTC Spring}.\hskip 1em plus
  0.5em minus 0.4em\relax IEEE, 2022, pp. 1--5.

\bibitem{pw3_2021_swipt6}
R.~Gupta and I.~Krikidis, ``{Simultaneous wireless power transfer and
  modulation classification},'' in \emph{IEEE VTC Spring}, 2021, pp. 1--6.

\bibitem{pw2_2021app5}
Y.~{Liu}, Y.~{Ye}, H.~{Ding}, F.~{Gao}, and H.~{Yang}, ``{Outage performance
  analysis for SWIPT-based incremental cooperative NOMA networks with
  non-linear harvester},'' \emph{IEEE Commun. Lett.}, vol.~24, no.~2, pp.
  287--291, 2020.

\bibitem{pw3_2023app2}
P.~K. Sharma, N.~Sharma, S.~Dhok, and A.~Singh, ``{RIS-assisted FD short packet
  communication with non-linear EH},'' \emph{IEEE Communications Letters},
  vol.~27, no.~2, pp. 522--526, 2023.

\bibitem{pw3_2023app3}
S.~Kurma, P.~K. Sharma, K.~Singh, S.~Mumtaz, and C.-P. Li, ``{URLLC-based
  cooperative industrial IoT networks with nonlinear energy harvesting},''
  \emph{IEEE Transactions on Industrial Informatics}, vol.~19, no.~2, pp.
  2078--2088, 2023.

\bibitem{pw2_2021app6}
H.~{Yang}, Y.~{Ye}, X.~{Chu}, and M.~{Dong}, ``{Resource and power allocation
  in SWIPT-enabled device-to-device communications based on a nonlinear energy
  harvesting model},'' \emph{IEEE Internet Things J.}, vol.~7, no.~11, pp.
  10\,813--10\,825, 2020.

\bibitem{pw2_2021app7}
M.~{Babaei}, U.~{Ayg\"ol\"u}, M.~{Ba\c{s}aran}, and L.~{Durak-Ata}, ``{BER
  performance of full-duplex cognitive radio network with nonlinear energy
  harvesting},'' \emph{IEEE Trans. Green Commun. Netw.}, vol.~4, no.~2, pp.
  448--460, 2020.

\bibitem{pw2_2021app16}
G.~{Sacarelo} and Y.~H. {Kim}, ``{Rate-energy tradeoffs of wireless powered
  backscatter communication with power splitting and time switching},''
  \emph{IEEE Access}, vol.~9, pp. 10\,844--10\,857, 2021.

\bibitem{pw3_2022_wpcn1}
Z.~Hao, X.~Jia, and J.~Xu, ``{Age of information in wireless sensor networks
  with non-linear energy harvesting and outdated channel state information},''
  in \emph{IEEE VTC Spring}, 2022, pp. 1--5.

\bibitem{pw2_2021app2}
D.~{Mishra} and H.~{Johansson}, ``{Optimal channel estimation for hybrid energy
  beamforming under phase shifter impairments},'' \emph{IEEE Trans. Commun.},
  vol.~67, no.~6, pp. 4309--4325, 2019.

\bibitem{pw2_2021app10}
A.~{Hakimi}, M.~{Mohammadi}, Z.~{Mobini}, and Z.~{Ding}, ``{Full-duplex
  non-orthogonal multiple access cooperative spectrum-sharing networks with
  non-linear energy harvesting},'' \emph{IEEE Trans. Veh. Technol.}, vol.~69,
  no.~10, pp. 10\,925--10\,936, 2020.

\bibitem{pw2_2021app15}
L.~Shi, Y.~Ye, X.~Chu, and G.~Lu, ``{Computation energy efficiency maximization
  for a NOMA-based WPT-MEC network},'' \emph{IEEE Internet Things J.}, vol.~8,
  no.~13, pp. 10\,731--10\,744, 2021.

\bibitem{pw3_2022_wpcn2}
B.~Luo, P.~L. Yeoh, R.~Schober, and B.~S. Krongold, ``{Distributed multiantenna
  frequency-selective energy beamforming with joint total and individual power
  constraints},'' \emph{IEEE Trans. Green Commun. Netw.}, vol.~6, no.~4, pp.
  2100--2114, 2022.

\bibitem{pw3_2022_wpcn3}
H.~Alakoca, M.~Babaei, L.~Durak-Ata, and E.~Basar, ``{RIS-empowered non-linear
  energy harvesting communications over Nakagami-$m$ channels},'' \emph{IEEE
  Commun. Lett.}, vol.~26, no.~9, pp. 2215--2219, 2022.

\bibitem{pw3_2023app1}
A.~Olutayo, Y.~Dong, J.~Cheng, J.~F. Holzman, and V.~C.~M. Leung,
  ``{Performance of wireless powered communication systems over Beaulieu-Xie
  channels with nonlinear energy harvesters},'' \emph{IEEE open j. Commun.
  Soc.}, vol.~4, pp. 456--463, 2023.

\bibitem{ref_lg2_app_2}
S.~Wang, M.~Xia, and Y.-C. Wu, ``{Space-time signal optimization for SWIPT:
  Linear versus nonlinear energy harvesting model},'' \emph{IEEE Commun.
  Lett.}, vol.~22, no.~2, pp. 408--411, 2017.

\bibitem{ref_lg2_app_4}
K.-G. Nguyen, Q.-D. Vu, L.-N. Tran, and M.~Juntti, ``{Energy efficiency
  fairness for multi-pair wireless-powered relaying systems},'' \emph{IEEE J.
  Sel. Areas Commun.}, vol.~37, no.~2, pp. 357--373, 2018.

\bibitem{mlg_2021_swipt1}
D.~Gunasinghe and G.~A.~A. Baduge, ``{Performance analysis of SWIPT for
  intelligent reflective surfaces for wireless communication},'' \emph{IEEE
  Commun. Lett.}, vol.~25, no.~7, pp. 2201--2205, 2021.

\bibitem{mlg_2022_swipt2}
X.~Peng, P.~Wu, H.~Tan, and M.~Xia, ``{Optimization for IRS-assisted MIMO-OFDM
  SWIPT system with non-linear EH model},'' \emph{IEEE Internet Things J.},
  2022.

\bibitem{mlg_2022_swipt3}
D.~L. Galappaththige, R.~Shrestha, and G.~A.~A. Baduge, ``{Exploiting cell-free
  massive MIMO for enabling simultaneous wireless information and power
  transfer},'' \emph{IEEE Trans. Green Commun. Netw.}, vol.~5, no.~3, pp.
  1541--1557, 2021.

\bibitem{lg2_2021app1}
A.~M. {Almasoud} and A.~E. {Kamal}, ``{Wireless-powered machine-to-machine
  multicasting in cellular networks},'' \emph{IEEE Trans. Green Commun. Netw.},
  vol.~4, no.~2, pp. 515--528, 2020.

\bibitem{lg2_2021app2}
F.~{Zhou} and R.~Q. {Hu}, ``{Computation efficiency maximization in
  wireless-powered mobile edge computing networks},'' \emph{IEEE Trans.
  Wireless Commun.}, vol.~19, no.~5, pp. 3170--3184, 2020.

\bibitem{lg2_2021app3}
F.~{Zhou}, Y.~{Wu}, R.~Q. {Hu}, and Y.~{Qian}, ``{Computation efficiency in a
  wireless-powered mobile edge computing network with NOMA},'' in \emph{IEEE
  ICC}, 2019, pp. 1--7.

\bibitem{mlg_2021_wpcn1}
Y.~Xu, B.~Gu, and D.~Li, ``{Robust energy-efficient optimization for secure
  wireless-powered backscatter communications with a non-linear EH model},''
  \emph{IEEE Commun. Lett.}, vol.~25, no.~10, pp. 3209--3213, 2021.

\bibitem{lg_2023app3}
T.~S. Muratkar, A.~Bhurane, and A.~Kothari, ``{Effect of dynamic reflection
  coefficient on backscatter system in time-selective scenario},'' \emph{IEEE
  J. Radio Freq. Id.}, vol.~7, pp. 45--49, 2023.

\bibitem{lg_2023app5}
X.~Li, J.~Jiang, H.~Wang, C.~Han, G.~Chen, J.~Du, C.~Hu, and S.~Mumtaz,
  ``{Physical layer security for wireless-powered ambient backscatter
  cooperative communication networks},'' \emph{IEEE Trans. Cogn. Commun.
  Netw.}, pp. 1--1, to appear, 2023.

\bibitem{ref_2rd_app_1}
A.~{\"{O}z\c{c}elikkale}, M.~{Koseoglu}, and M.~{Srivastava}, ``{Optimization
  vs. reinforcement learning for wirelessly powered sensor networks},'' in
  \emph{IEEE SPAWC}, 2018, pp. 1--5.

\bibitem{2rd_2021app1}
A.~{\"{O}z\c{c}elikkale}, M.~{Koseoglu}, M.~{Srivastava}, and A.~{Ahl\'en},
  ``{Deep reinforcement learning based energy beamforming for powering sensor
  networks},'' in \emph{IEEE MLSP Wkshps}, 2019, pp. 1--6.

\bibitem{ref_frac_app_1}
M.~Maleki, A.~M.~D. Hoseini, and M.~Masjedi, ``{Performance analysis of SWIPT
  relay systems over Nakagami-$m$ fading channels with non-linear energy
  harvester and hybrid protocol},'' in \emph{Electrical Engineering (ICEE)},
  Mashhad, Iran, May 2018, pp. 610--615.

\bibitem{fr_2021app3}
G.~A. {Ropokis}, ``{Multi-relay cooperation with self-energy recycling and
  power consumption considerations},'' in \emph{WiMob}, 2019, pp. 268--275.

\bibitem{fr_2021app2}
J.~{Zhang}, G.~{Zheng}, I.~{Krikidis}, and R.~{Zhang}, ``{Specific absorption
  rate-aware beamforming in MISO downlink SWIPT systems},'' \emph{IEEE Trans.
  Commun.}, vol.~68, no.~2, pp. 1312--1326, 2020.

\bibitem{frc_2022_swipt1}
O.~M. El-Nakhla, M.~I. Obayya, and S.~E. Kishk, ``{Stable matching relay
  selection (SMRS) for TWR D2D network with RF/RE EH capabilities},''
  \emph{IEEE Access}, vol.~10, pp. 22\,381--22\,391, 2022.

\bibitem{fr_2021app4}
Y.~{Xu} and B.~{Liu}, ``{Disaster-recovery communications utilizing SWIPT-based
  D2D relay network},'' in \emph{IEEE ICCC}, 2019, pp. 1041--1046.

\bibitem{fr_2021app5}
R.~{Ma}, H.~{Wu}, J.~{Ou}, S.~{Yang}, and Y.~{Gao}, ``{Power splitting-based
  SWIPT systems with full-duplex jamming},'' \emph{IEEE Trans. Veh. Technol.},
  vol.~69, no.~9, pp. 9822--9836, 2020.

\bibitem{fr_2021app6}
M.~{Dimitropoulou}, C.~{Psomas}, and I.~{Krikidis}, ``{K-th best device
  selection for scheduling in wireless powered communication networks},'' in
  \emph{IEEE ICC}, 2020, pp. 1--6.

\bibitem{fr_2021app7}
Y.~{Ye}, L.~{Shi}, X.~{Chu}, and G.~{Lu}, ``{Throughput fairness guarantee in
  wireless powered backscatter communications with HTT},'' \emph{IEEE Wireless
  Commun. Lett.}, vol.~10, no.~3, pp. 449--453, 2021.

\bibitem{fr_2021app10}
Y.~{Liu}, Z.~{Zhang}, L.~{Shi}, and Y.~{Ye}, ``{Backscatter assisted wireless
  powered non-orthogonal multiple access systems},'' in \emph{IEEE ICCT}, 2020,
  pp. 339--343.

\bibitem{frc_2021_wpcn2}
M.~Dimitropoulou, C.~Psomas, and I.~Krikidis, ``{Generalized selection in
  wireless powered networks with non-linear energy harvesting},'' \emph{IEEE
  Trans. Commun.}, vol.~69, no.~8, pp. 5634--5648, 2021.

\bibitem{frc_2021_wpcn3}
Y.~Ye, L.~Shi, X.~Chu, and G.~Lu, ``{Total transmission time minimization in
  wireless powered hybrid passive-active communications},'' in \emph{IEEE VTC
  Spring}.\hskip 1em plus 0.5em minus 0.4em\relax IEEE, 2021, pp. 1--5.

\bibitem{fr_2021app9}
O.~T. {Demir} and E.~{Bj\"ornson}, ``{Joint power control and LSFD for
  wireless-powered Cell-Free massive MIMO},'' \emph{IEEE Trans. Wireless
  Commun.}, vol.~20, no.~3, pp. 1756--1769, 2021.

\bibitem{fr_2021app8}
L.~Shi, Y.~Ye, X.~Chu, and G.~Lu, ``{Computation bits maximization in a
  backscatter assisted wirelessly powered MEC network},'' \emph{IEEE Commun.
  Lett.}, vol.~25, no.~2, pp. 528--532, 2020.

\bibitem{frc_2021_wpcn1}
L.~Shi, Y.~Ye, G.~Zheng, and G.~Lu, ``{Computational EE fairness in
  backscatter-assisted wireless powered MEC networks},'' \emph{IEEE Wireless
  Commun. Lett.}, vol.~10, no.~5, pp. 1088--1092, 2021.

\bibitem{frc_2022_wpcn1}
Z.~Chu, P.~Xiao, D.~Mi, W.~Hao, Z.~Lin, Q.~Chen, and R.~Tafazolli, ``{Wireless
  powered intelligent radio environment with non-linear energy harvesting},''
  \emph{IEEE Internet Things J.}, 2022.

\bibitem{fr_2023app1}
Z.~Yang, J.~A. Hussein, P.~Xu, G.~Chen, Y.~Wu, and Z.~Ding, ``{A novel hybrid
  successive interference cancellation for Uplink wireless power transfer NOMA
  in Internet of Things},'' \emph{IEEE Trans. Veh. Technol.}, vol.~72, no.~5,
  pp. 6090--6102, 2023.

\bibitem{han2021joint_log_aap_1}
J.~Han, G.~H. Lee, S.~Park, and J.~K. Choi, ``{Joint subcarrier and
  transmission power allocation in OFDMA-based WPT system for mobile edge
  computing in IoT environment},'' \emph{IEEE Internet Things J.}, 2021.

\bibitem{ref_frac_app_2}
Y.~Ye, Y.~Li, L.~Shi, R.~Q. Hu, and H.~Zhang, ``{Improved hybrid relaying
  protocol for DF relaying in the presence of a direct link},'' \emph{IEEE
  Wireless Commun. Lett.}, vol.~8, no.~1, pp. 173--176, 2018.

\bibitem{fr_2021app1}
Y.~{Ye}, Y.~{Li}, L.~{Shi}, R.~Q. {Hu}, and H.~{Zhang}, ``{Improved hybrid
  relaying protocol for DF relaying in the presence of a direct link},''
  \emph{IEEE Wireless Commun. Lett.}, vol.~8, no.~1, pp. 173--176, 2019.

\bibitem{2022_all_mod1}
P.~N. Alevizos, G.~Vougioukas, and A.~Bletsas, ``{Nonlinear energy harvesting
  models in wireless information and power transfer},'' in \emph{IEEE SPAWC},
  2018, pp. 1--5.

\bibitem{2022_all_mod2}
B.~A. Mouris, H.~Ghauch, R.~Thobaben, and B.~L.~G. Jonsson, ``{Multi-tone
  signal optimization for wireless power transfer in the presence of wireless
  communication links},'' \emph{IEEE Trans. Wireless Commun.}, vol.~19, no.~5,
  pp. 3575--3590, 2020.

\bibitem{2022_all_mod3}
J.~B. Lee, Y.~Rong, L.~Gopal, and C.~W.~R. Chiong, ``{Mutual information
  maximization for SWIPT AF MIMO relay systems with non-linear EH models and
  imperfect channel state information},'' \emph{IEEE Trans. Veh. Technol.},
  vol.~71, no.~8, pp. 8503--8518, 2022.

\bibitem{2022_all_mod4}
J.~Wang and Y.~Ge, ``{A radio frequency energy harvesting-based multihop
  clustering routing protocol for cognitive radio sensor networks},''
  \emph{IEEE Sensors J.}, vol.~22, no.~7, pp. 7142--7156, 2022.

\bibitem{frac_opt_1}
J.~Y., \emph{{An efficient global optimization algorithm for nonlinear
  sum-of-ratios problem}}.

\bibitem{2004Convex}
S.~Boyd and L.~Vandenberghe, \emph{{Convex optimization}}.\hskip 1em plus 0.5em
  minus 0.4em\relax Convex Optimization, 2004.

\bibitem{2017Convex}
C.~Y. Chi, W.~C. Li, and C.~H. Lin, \emph{{Convex optimization for signal
  processing and communications: From fundamentals to applications}}.\hskip 1em
  plus 0.5em minus 0.4em\relax Convex Optimization for Signal Processing and
  Communications: From Fundamentals to Applications, 2017.

\bibitem{2008CVX}
M.~Grant, ``{CVX: MATLAB software for disciplined convex programming},''
  \emph{http://cvxr.com/cvx}, 2008.

\bibitem{du2021millimeter}
H.~Du, J.~Zhang, J.~Cheng, and B.~Ai, ``{Millimeter wave communications with
  reconfigurable intelligent surfaces: Performance analysis and
  optimization},'' \emph{IEEE Trans. Commun.}, vol.~69, no.~4, pp. 2752--2768,
  2021.

\bibitem{du2022performance}
H.~Du, D.~Niyato, Y.-A. Xie, Y.~Cheng, J.~Kang, and D.~I. Kim, ``{Performance
  analysis and optimization for jammer-aided multiantenna uav covert
  communication},'' \emph{IEEE J. Sel. Areas Commun.}, vol.~40, no.~10, pp.
  2962--2979, 2022.

\bibitem{du2022_2}
H.~Du, J.~Zhang, K.~Guan, D.~Niyato, H.~Jiao, Z.~Wang, and T.~K{\"u}rner,
  ``{Performance and optimization of reconfigurable intelligent surface aided
  THz communications},'' \emph{IEEE Trans. Commun.}, vol.~70, no.~5, pp.
  3575--3593, 2022.

\bibitem{ai_0}
L.~Wang, X.~Sun, R.~Jiang, W.~Jiang, Z.~Zhong, and D.~W. Kwan~Ng, ``{Optimal
  energy efficiency for Multi-MEC and blockchain empowered IoT: A deep learning
  approach},'' in \emph{IEEE ICC}, 2021, pp. 1--6.

\bibitem{du2023exploring}
H.~Du, R.~Zhang, D.~Niyato, J.~Kang, Z.~Xiong, D.~I. Kim, H.~V. Poor
  \emph{et~al.}, ``{Exploring collaborative distributed diffusion-based
  AI-generated content (AIGC) in wireless networks},'' \emph{arXiv preprint
  arXiv:2304.03446}, 2023.

\bibitem{2022_ai_0}
E.~Baccour, N.~Mhaisen, A.~A. Abdellatif, A.~Erbad, A.~Mohamed, M.~Hamdi, and
  M.~Guizani, ``{Pervasive AI for IoT applications: A survey on
  resource-efficient distributed artificial intelligence},'' \emph{IEEE Commun.
  Surveys Tuts.}, vol.~24, no.~4, pp. 2366--2418, 2022.

\bibitem{ai_2033app1}
Y.~Shi, K.~Yang, T.~Jiang, J.~Zhang, and K.~B. Letaief,
  ``{Communication-efficient edge AI: algorithms and systems},'' \emph{IEEE
  Commun. Surveys Tuts.}, vol.~22, no.~4, pp. 2167--2191, 2020.

\bibitem{du2023generative}
H.~Du, Z.~Li, D.~Niyato, J.~Kang, Z.~Xiong, H.~Huang, and S.~Mao, ``{Generative
  AI-aided optimization for AI-generated content (AIGC) services in edge
  networks},'' \emph{arXiv preprint arXiv:2303.13052}, 2023.

\bibitem{ai_2023aap2}
R.~Zhang, K.~Xiong, X.~Tian, Y.~Lu, P.~Fan, and K.~B. Letaief, ``{Inverse
  reinforcement learning meets power allocation in multi-user cellular
  networks},'' in \emph{IEEE INFOCOM WKSHPS}, 2022, pp. 1--2.

\bibitem{rostami2020using}
M.~Rostami, D.~Isele, and E.~Eaton, ``{Using task descriptions in lifelong
  machine learning for improved performance and zero-shot transfer},'' \emph{J.
  Artif. Intell. Res.}, vol.~67, pp. 673--704, 2020.

\bibitem{lg1_swipt_2021app25}
M.~{Varasteh}, J.~{Hoydis}, and B.~{Clerckx}, ``{Learning to communicate and
  energize: Modulation, coding, and multiple access designs for wireless
  information-power transmission},'' \emph{IEEE Trans. Commun.}, vol.~68,
  no.~11, pp. 6822--6839, 2020.

\bibitem{ai_eh_0}
J.~Jang and H.~J. Yang, ``{Deep learning-aided user association and power
  control with renewable energy sources},'' \emph{IEEE Trans. Commun.},
  vol.~70, no.~4, pp. 2387--2403, 2022.

\bibitem{ai_eh_1}
R.~Zhang, K.~Xiong, W.~Guo, X.~Yang, P.~Fan, and K.~B. Letaief,
  ``{Q-learning-based adaptive power control in wireless RF energy harvesting
  heterogeneous networks},'' \emph{IEEE Syst. J.}, vol.~15, no.~2, pp.
  1861--1872, 2021.

\bibitem{ai_eh_2}
Y.~Li, X.~Su, H.~Jiang, and C.~S. Chen, ``{Throughput maximization for wireless
  powered communication: Reinforcement learning approaches},'' in \emph{IEEE
  IWQOS}, 2021, pp. 1--10.

\bibitem{ai_eh_3}
Y.~Yu, J.~Tang, J.~Huang, X.~Zhang, D.~K.~C. So, and K.-K. Wong,
  ``{Multi-objective optimization for UAV-assisted wireless powered IoT
  networks based on extended DDPG algorithm},'' \emph{IEEE Trans. Commun.},
  vol.~69, no.~9, pp. 6361--6374, 2021.

\bibitem{ai_eh_4}
F.~Zhou, Y.~Wu, and Q.~Wu, ``{Resource allocation based on deep reinforcement
  learning for wideband cognitive radio networks},'' in \emph{URSI GASS}, 2021,
  pp. 01--04.

\bibitem{lg_2023app9}
M.~A. Ouamri, G.~Barb, D.~Singh, A.~B.~M. Adam, M.~S.~A. Muthanna, and X.~Li,
  ``{Nonlinear energy-harvesting for D2D networks underlaying UAV with SWIPT
  using MADQN},'' \emph{IEEE Wireless Commun. Lett.}, pp. 1--1, to appear,
  2023.

\bibitem{pw2_2023app1}
X.~Mi and H.~He, ``Multi-agent deep reinforcement learning for d2d-assisted mec
  system with energy harvesting,'' in \emph{ICACT}, 2023, pp. 145--153.

\bibitem{ai_convex_1}
A.~Omidkar, A.~Khalili, H.~H. Nguyen, and H.~Shafiei,
  ``{Reinforcement-learning-based resource allocation for
  energy-harvesting-aided D2D communications in IoT networks},'' \emph{IEEE
  Internet Things J.}, vol.~9, no.~17, pp. 16\,521--16\,531, 2022.

\bibitem{ai_convex_2}
Z.~Ding, R.~Schober, and H.~V. Poor, ``{No-pain no-gain: DRL assisted
  optimization in energy-constrained CR-NOMA networks},'' \emph{IEEE Trans.
  Commun.}, vol.~69, no.~9, pp. 5917--5932, 2021.

\bibitem{ai_convex_3}
L.~Li, H.~Xu, J.~Ma, A.~Zhou, and J.~Liu, ``{Joint EH time and transmit power
  optimization based on DDPG for EH communications},'' \emph{IEEE Commun.
  Lett.}, vol.~24, no.~9, pp. 2043--2046, 2020.

\bibitem{du2023spear}
H.~Du, D.~Niyato, J.~Kang, Z.~Xiong, K.-Y. Lam, Y.~Fang, and Y.~Li, ``{Spear or
  shield: leveraging generative AI to tackle security threats of intelligent
  network services},'' \emph{arXiv preprint arXiv:2306.02384}, 2023.

\bibitem{zhang2021joint}
R.~Zhang, K.~Xiong, Y.~Lu, B.~Gao, P.~Fan, and K.~B. Letaief, ``{Joint
  coordinated beamforming and power splitting ratio optimization in MU-MISO
  SWIPT-enabled HetNets: A multi-agent DDQN-based approach},'' \emph{IEEE J.
  Sel. Areas Commun.}, vol.~40, no.~2, pp. 677--693, 2021.

\bibitem{zhang2023energy}
R.~Zhang, K.~Xiong, Y.~Lu, P.~Fan, D.~W.~K. Ng, and K.~B. Letaief, ``{Energy
  efficiency maximization in RIS-assisted SWIPT networks with RSMA: A PPO-based
  approach},'' \emph{IEEE J. Sel. Areas Commun.}, vol.~41, no.~5, pp.
  1413--1430, 2023.

\end{thebibliography}

\end{document}